\documentclass[aps,eqsecnum,preprint,floats,epsf,epsfig,nofootinbib]{revtex4}
\usepackage{graphicx}
\usepackage{epsfig}

\begin{document}
\def\be{\begin{eqnarray}}
\def\en{\end{eqnarray}}
\def\non{\nonumber}
\def\la{\langle}
\def\ra{\rangle}
\def\pp{{\prime\prime}}
\def\nc{N_c^{\rm eff}}
\def\vp{\varepsilon}
\def\hep{\hat{\varepsilon}}
\def\a{{\cal A}}
\def\B{{\cal B}}
\def\c{{\cal C}}
\def\d{{\cal D}}
\def\e{{\cal E}}
\def\p{{\cal P}}
\def\t{{\cal T}}
\def\B{{\cal B}}
\def\P{{\cal P}}
\def\S{{\cal S}}
\def\T{{\cal T}}
\def\C{{\cal C}}
\def\A{{\cal A}}
\def\E{{\cal E}}
\def\V{{\cal V}}
\def\CP{$CP$~}
\def\up{\uparrow}
\def\dw{\downarrow}
\def\vma{{_{V-A}}}
\def\vpa{{_{V+A}}}
\def\smp{{_{S-P}}}
\def\spp{{_{S+P}}}
\def\lrpartial{\buildrel\leftrightarrow\over\partial}
\def\J{{J/\psi}}
\def\3bar{{\bf \bar 3}}
\def\6bar{{\bf \bar 6}}
\def\10bar{{\bf \ov{10}}}
\def\ov{\overline}
\def\Lqcd{{\Lambda_{\rm QCD}}}
\def\pr{{Phys. Rev.}~}
\def\prl{{ Phys. Rev. Lett.}~}
\def\pl{{ Phys. Lett.}~}
\def\np{{ Nucl. Phys.}~}
\def\zp{{ Z. Phys.}~}
\def\lsim{ {\ \lower-1.2pt\vbox{\hbox{\rlap{$<$}\lower5pt\vbox{\hbox{$\sim$}
}}}\ } }
\def\gsim{ {\ \lower-1.2pt\vbox{\hbox{\rlap{$>$}\lower5pt\vbox{\hbox{$\sim$}
}}}\ } }

\font\el=cmbx10 scaled \magstep2{\obeylines\hfill BNL-HET-04/17}

\font\el=cmbx10 scaled \magstep2{\obeylines\hfill December, 2004}

\vskip 1.5 cm

\centerline{\large\bf Final State Interactions in Hadronic $B$
Decays}
\bigskip
\centerline{\bf Hai-Yang Cheng,$^1$ Chun-Khiang Chua$^1$ and
Amarjit Soni$^2$}
\medskip
\centerline{$^1$ Institute of Physics, Academia Sinica}
\centerline{Taipei, Taiwan 115, Republic of China}
\medskip

\medskip
\centerline{$^2$ Physics Department, Brookhaven National
Laboratory} \centerline{Upton, New York 11973}
\medskip

\bigskip
\bigskip
\centerline{\bf Abstract}
\bigskip
\small

There exist many experimental indications that final-state
interactions (FSIs) may play a prominent role not only in charmful
$B$ decays but also in charmless $B$ ones. We examine the
final-state rescattering effects on the hadronic $B$ decay rates
and their impact on direct \CP violation. The color-suppressed
neutral modes such as $B^0\to
D^0\pi^0,\pi^0\pi^0,\rho^0\pi^0,K^0\pi^0$ can be substantially
enhanced by long-distance rescattering effects. The direct
$CP$-violating partial rate asymmetries in charmless $B$ decays to
$\pi\pi/\pi K$ and $\rho\pi$ are significantly affected by
final-state rescattering and their signs are generally different
from that predicted by the short-distance approach. For example,
direct \CP asymmetry in $B^0\to\rho^0\pi^0$ is increased to around
60\% due to final state rescattering effects whereas the
short-distance picture gives about 1\%. Evidence of direct \CP
violation in the decay $\ov B^0\to K^-\pi^+$ is now established,
while the combined BaBar and Belle measurements of $\ov
B^0\to\rho^\pm\pi^\mp$ imply a $3.6\sigma$ direct \CP asymmetry in
the $\rho^+\pi^-$ mode. Our predictions for \CP violation agree
with experiment in both magnitude and sign, whereas the QCD
factorization predictions (especially for $\rho^+\pi^-$) seem to
have some difficulty with the data. Direct \CP violation in the
decay $B^-\to\pi^-\pi^0$ is very small ($\lsim 1\%$) in the
Standard Model even after the inclusion of FSIs. Its measurement
will provide a nice way to search for New Physics as in the
Standard Model QCD penguins cannot contribute (except by isospin
violation). Current data on $\pi K$ modes seem to violate the
isospin sum rule relation, suggesting the presence of electroweak
penguin contributions. We have also investigated whether a large
transverse polarization in $B\to \phi K^*$ can arise from the
final-state rescattering of $D^{(*)}\bar D_s^{(*)}$ into $\phi
K^*$. While the longitudinal polarization fraction can be reduced
significantly from short-distance predictions due to such FSI
effects, no sizable perpendicular polarization is found owing
mainly to the large cancellations occurring in the processes
$\overline B\to D_s^* \bar D\to\phi\ov K^*$ and $\overline B\to
D_s\bar D^*\to\phi\ov K^*$ and this can be understood as a
consequence of \CP and SU(3) [CPS] symmetry. To fully account for
the polarization anomaly (especially the perpendicular
polarization) observed in $B\to \phi K^*$, FSI from other states
or other mechanism, e.g. the penguin-induced annihilation, may
have to be invoked. Our conclusion is that the small value of the
longitudinal polarization in $VV$ modes cannot be regarded as a
clean signal for New Physics.

\eject
\section{Introduction}

The importance of final-state interactions (FSIs) has long been
recognized in hadronic charm decays since some resonances are
known to exist at energies close to the mass of the charmed meson.
As for hadronic $B$ decays, the general folklore is that FSIs are
expected to play only a minor role there as the energy release in
the energetic $B$ decay is so large that the final-state particles
are moving fast and hence they do not have adequate time for
getting involved in final-state rescattering. However, from the
data accumulated at $B$ factories and at CLEO, there are growing
indications that soft final-state rescattering effects do play an
essential role in $B$ physics.

Some possible hints at FSIs in the $B$ sector are:

\begin{enumerate}
 \item There exist some decays that do not receive any factorizable
contributions, for example $B\to K\chi_{0c}$, owing to the
vanishing matrix element of the $(V-A)$ current,
$\la\chi_{0c}|\bar c\gamma_\mu(1-\gamma_5)c|0\ra=0$.
Experimentally, it was reported by both Belle \cite{BelleKchi} and
BaBar \cite{BaBarKchi} that this decay mode has a sizable
branching ratio, of order $(2\sim 6)\times 10^{-4}$. This implies
that the nonfactorizable correction is important and/or the
rescattering effect is sizable. Studies based on the light-cone
sum rule approach indicate that nonfactorizable contributions to
$B\to\chi_{c0}K$ due to soft gluon exchanges is too small to
accommodate the data \cite{Melic,Wang}. In contrast, it has been
shown that the rescattering effect from the intermediate charmed
mesons is able to reproduce the observed large branching ratio
\cite{Colangelo2002}.

 \item The color-suppressed modes $\ov B^0\to D^{(*)0}\pi^0$
have been measured by  Belle \cite{BelleBDpi}, CLEO
\cite{CLEOBDpi} and BaBar \cite{BaBarBDpi}. Their branching ratios
are all significantly larger than theoretical expectations based
on naive factorization. When combined with the color-allowed $\ov
B\to D^{(*)}\pi$ decays, it indicates non-vanishing relative
strong phases among various $\ov B\to D^{(*)}\pi$ decay
amplitudes. Denoting $\T$ and $\C$ as the color-allowed tree
amplitude and color-suppressed $W$-emission amplitude,
respectively, it is found that $\C/\T\sim 0.45\,{\rm exp}(\pm
i60^\circ)$ (see e.g. \cite{ChengBDpi}), showing a non-trivial
relative strong phase between $\C$ and $\T$ amplitudes. The large
magnitude and phase of $\C/\T$ compared to naive expectation
implies the importance of long-distance FSI contributions to the
color-suppressed internal $W$-emission via final-state
rescattering of the color-allowed tree amplitude.

\item A model independent analysis of charmless $B$ decay data
based on the topological quark diagrammatic approach yields a
larger value of $|\C/\T|$ and a large strong relative phase
between the $\C$ and $\T$ amplitudes \cite{Chiang}. For example,
one of the fits in \cite{Chiang} leads to
$\C/\T=(0.46^{+0.43}_{-0.30})\,{\rm
exp}([-i(94^{+43}_{-52})]^\circ$, which is indeed consistent with
the result inferred from $B\to D\pi$ decays. This means that FSIs
in charmless $B$ decays are as important as in $B\to D\pi$ ones.
The presence of a large color-suppressed amplitude is somewhat a
surprise from the current model calculations such as those in the
QCD factorization approach and most likely indicates a prominent
role for final-state rescattering.

\item Both BaBar \cite{BaBar2pi0} and Belle \cite{Belle2pi0} have
reported a sizable branching ratio of order $2\times 10^{-6}$ for
the decay $B^0\to\pi^0\pi^0$. This cannot be explained by either
QCD factorization or the PQCD approach and it again calls for a
possible rescattering effect to induce $\pi^0\pi^0$. Likewise, the
color-suppressed decay $B^0\to\rho^0\pi^0$ with the branching
ratio $(1.9\pm1.2)\times 10^{-6}$ averaged from the Belle and
BaBar measurements (see Table \ref{tab:Brhopi}) is substantially
higher than the prediction based on PQCD \cite{CDLu} or the
short-distance approach. Clearly, FSI is one of the prominent
candidates responsible for this huge enhancement.

\item Direct \CP violation in the decay $B^0\to K^+\pi^-$ with the
magnitude $\A_{K\pi}=-0.133\pm0.030\pm0.009$ at $4.2\sigma$ was
recently announced by BaBar \cite{BaBarKpi}, which agrees with the
previous Belle measurement of $\A_{K\pi}=-0.088\pm0.035\pm0.013$
at $2.4\sigma$ based on a 140\,fb$^{-1}$ data sample
\cite{BelleKpi}. A further evidence of this direct \CP violation
at $3.9\sigma$ was just reported by Belle with 253\,fb$^{-1}$ of
data \cite{BelleKpi2}. In the calculation of QCD factorization
\cite{BBNS,BN}, the predicted \CP asymmetry
$\A_{K\pi}=(4.5^{+9.1}_{-9.9})\%$ is apparently in conflict with
experiment. It is conceivable that FSIs via rescattering will
modify the prediction based on the short-distance interactions.
Likewise, a large direct \CP asymmetry in the decay
$B^0\to\pi^+\pi^-$ was reported by Belle \cite{Bellepipi},  but it
has not been confirmed by BaBar \cite{BaBarpipi}. The weighted
average of Belle and BaBar gives
$\A_{\pi\pi}=0.31\pm0.24$~\cite{ichep} with the PDG scale factor
of $S=2.2$ on the error. Again, the central value of the QCD
factorization prediction \cite{BN} has a sign opposite to the
world average.

\item Under the factorization approach, the color-suppressed mode
$\ov B^0\to D_s^+ K^-$  can only proceed via $W$-exchange. Its
sizable branching ratio of order $4\times 10^{-5}$ observed by
Belle \cite{BelleDsK} and BaBar \cite{BaBarDsK} will need a large
final-state rescattering contribution if the short-distance
$W$-exchange effect is indeed small according to the existing
model calculations.

\item The measured longitudinal fractions for $B\to\phi K^*$ by
both  BaBar \cite{BaBarphiK*} and  Belle \cite{BellephiK*} are
close to 50\%. This is in sharp contrast to the general argument
that factorizable amplitudes in $B$ decays to light vector meson
pairs give a longitudinal polarization satisfying the scaling law:
$1-f_L={\cal O}(1/m_b^2)$.~\footnote{More recently Kagan has
suggested that the penguin-induced annihilation can cause an
appreciable deviation from this expectation numerically, though
the scaling law is still respected {\it formally}
\cite{Kagan1,Kagan2}. However, it is difficult to make a reliable
estimation of this deviation.}
This law remains formally true even when nonfactorizable graphs
are included in QCD factorization. Therefore, in order to obtain a
large transverse polarization in $B\to\phi K^*$, this scaling law
valid at short-distance interactions must be violated. The effect
of long-distance rescattering on this scaling law should be
examined.

\end{enumerate}

The presence of FSIs can have interesting impact on the direct \CP
violation phenomenology. As stressed in \cite{AS98}, traditional
discussions have centered around the absorptive part of the
penguin graph in $b\to s$ transitions \cite{BSS} and as a result
causes ``simple" \CP violation; long-distance final state
rescattering effects, in general, will lead to a different pattern
of \CP violation, namely, ``compound" \CP violation. Predictions
of simple \CP violation are quite distinct from that of compound
\CP violation. The sizable \CP asymmetry observed in  $\ov B^0\to
K^-\pi^+$ decays is a strong indication for large direct \CP
violation driven by long-distance rescattering effects. Final
state rescattering phases in $B$ decays are unlikely to be small
possibly causing large compound $CP$-violating partial rate
asymmetries in these modes. As shown below, the sign of \CP
asymmetry can be easily flipped by long-distance rescattering
effects. Hence, it is important to explore the compound \CP
violation.

Of course, it is notoriously difficult to study FSI effects in a
systematic way as it is nonperturbative in nature. Nevertheless,
we can gain some control on rescattering effects by studying them
in a phenomenological way. More specifically, FSIs can be modelled
as the soft rescattering of certain intermediate two-body hadronic
channels, e.g. $B\to D\ov D\to\pi\pi$, so that they can be treated
as the one-particle-exchange processes at the hadron level. That
is, we shall study long-distance rescattering effects by
considering one-particle-exchange graphs. As the exchanged
particle is not on-shell, form factors must be introduced to
render the whole calculation meaningful in the perturbative sense.
This approach has been applied to the study of FSIs in charm
decays for some time \cite{FSIinD,Ablikim}. In the context of $B$
physics, the so-called ``charming penguin" contributions to
charmless hadronic $B$ decays have been treated as long-distance
effects manifesting in the rescattering processes such as $B\to
D_s\bar D\to K\pi$ \cite{charmpenguin1,Isola2003,charmpenguin}.
Likewise, the dynamics for final-state interactions is assumed to
be dominated by the mixing of the final state with $D^{(*)}\bar
D^{(*)}$ in \cite{Barshay,Barshay2}. Final-state rescattering
effects were also found in $B$ to charmonium final states, e.g.
$B^-\to K^-\chi_{c0}$, a process prohibited in the naive
factorization approach \cite{Colangelo2002}. Effects of
final-state rescattering on $K\pi$ and $\pi\pi$ final states have
been discussed extensively in the literature
\cite{FSIKpi,AS98,FSIChua2}.

The approach of modelling FSIs as soft rescattering processes of
intermediate two-body states has been criticized on several
grounds \cite{BBNS01}. First, there are many more intermediate
multi-body channels in $B$ decays and systematic cancellations
among them are predicted to occur in the heavy quark limit. This
effect of cancellation will be missed if only a few intermediate
states are taken into account. Second, the hadronic dynamics of
multi-body decays is very complicated and in general not under
theoretical control. Moreover, the number of channels and the
energy release in $B$ decays are large. We wish to stress that the
$b$ quark mass ($\sim4.5$ GeV) is not very large and far from
infinity in reality. The aforementioned cancellation may not occur
or may not be very effective for the finite $B$ mass. For
intermediate two-body states, we always consider those channels
that are quark-mixing-angle most favored so that they give the
dominant long-distance contributions. Whether there exist
cancellations between two-body and multi-body channels is not
known. Following \cite{Smith}, we may assume that two-body
$\rightleftharpoons$ $n$-body rescatterings are negligible either
justified from the $1/N_c$ argument \cite{Hooft} or suppressed by
large cancellations. We view our treatment of the two-body
hadronic model for FSIs as a working tool. We work out the
consequences of this tool to see if it is empirically working. If
it turns out to be successful, then it will imply the possible
dominance of intermediate two-body contributions. In other
approaches such as QCD factorization \cite{BBNS}, the complicated
hadronic $B$ decays in principle can be treated systematically as
$1/m_b$ power corrections which vanish in the heavy quark limit.
However, one should recognize that unless the coefficients of
power corrections are known or calculable in a model independent
manner, the short-distance picture without a control of the
$1/m_b$ coefficients and without being able to include the effects
of FSIs have their limitation as well. When the short-distance
scenario is tested against experiment and failings occur such as
the many examples mentioned before, then it does not necessarily
mean that New Physics has been discovered and an examination of
rescattering effects can be helpful.

The layout of the present paper is as follows. In Sec. II we give
an overview of the role played by the final-state interactions in
hadronic $B$ decays. As a warm-up, we begin in Sec. III with a
study of final-state rescattering contributions to $B\to D\pi$
decays which proceed only through tree diagrams. We then proceed
to the penguin-dominated $B\to K\pi$ decays in Sec. IV,
tree-dominated $B\to \pi\pi$ decays in Sec. V and $B\to \rho\pi$
decays in Sec. VI. Effects of FSIs on the branching ratios and
direct \CP asymmetries are studied. Sec. VII is devoted to the
polarization anomaly discovered recently in $B\to \phi K^*$
decays. Conclusion and discussion are given in Sec. VIII. Appendix
A gives some useful formula for the phase-space integration, while
the theoretical input parameters employed in the present paper are
summarized in Appendix B.

\section{Final-State Interactions}

In the diagrammatic approach, all two-body nonleptonic weak decays
of heavy mesons can be expressed in terms of six distinct quark
diagrams \cite{Chau,CC86,CC87}: $\t$, the color-allowed external
$W$-emission tree diagram; $\c$, the color-suppressed internal
$W$-emission diagram; $\e$, the $W$-exchange diagram; $\a$, the
$W$-annihilation diagram; ${\cal P}$, the penguin diagram; and
${\cal V}$, the vertical $W$-loop diagram. It should be stressed
that these quark diagrams are classified according to the
topologies of weak interactions with all strong interaction
effects included and hence they are {\it not} Feynman graphs. All
quark graphs used in this approach are topological with all the
strong interactions included, i.e. gluon lines are included in all
possible ways.

\begin{figure}[ht]
\vspace{0cm}
 \centerline{\epsfxsize4 in \epsffile{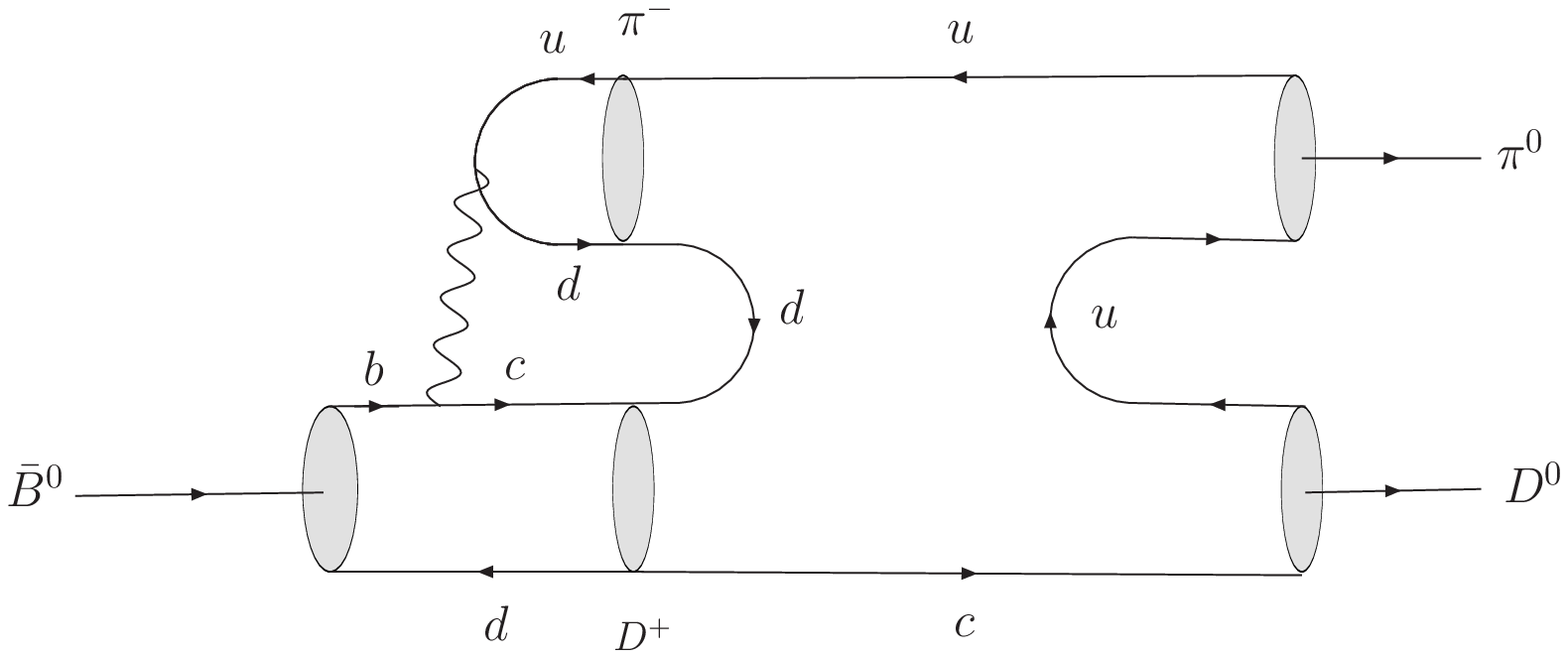}}
\centerline{(a)}
\smallskip
 \centerline{\epsfxsize4 in \epsffile{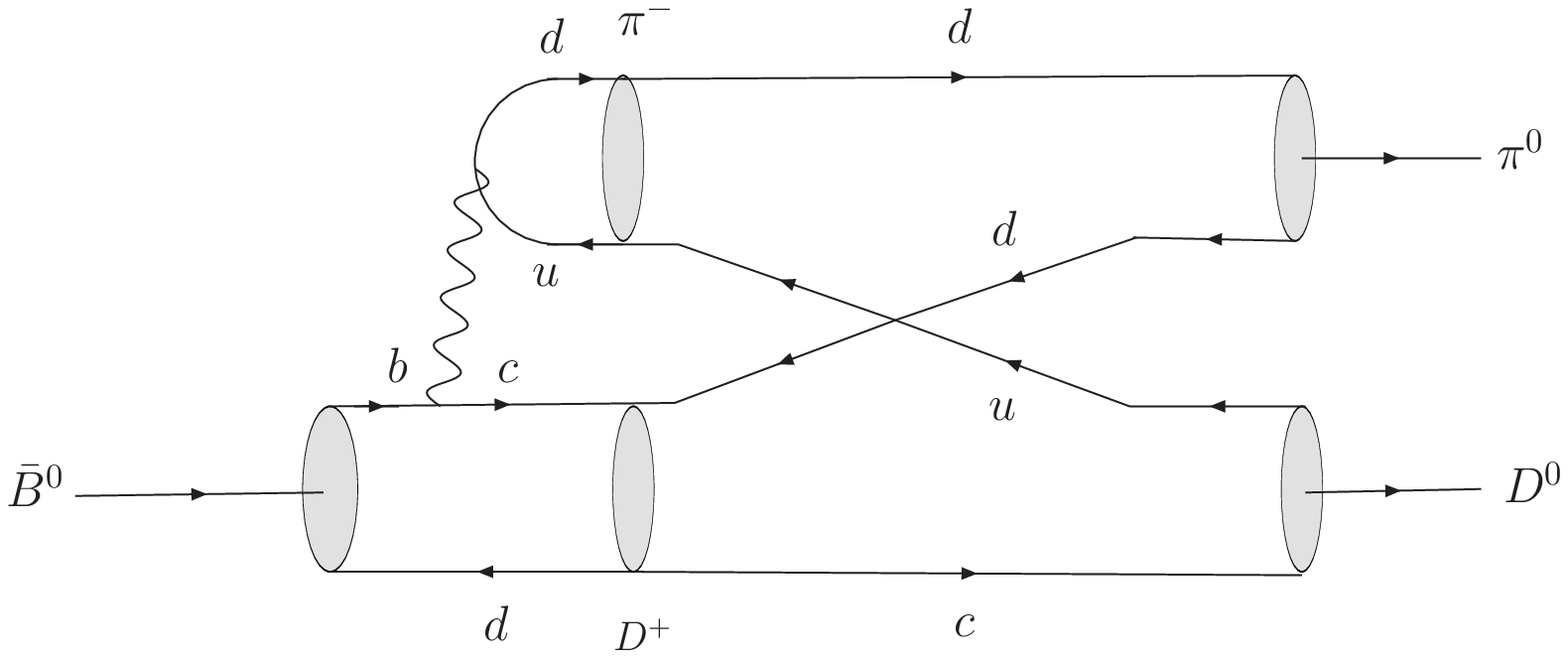}}
\centerline{(b)}
\smallskip
 \centerline{\epsfxsize4 in \epsffile{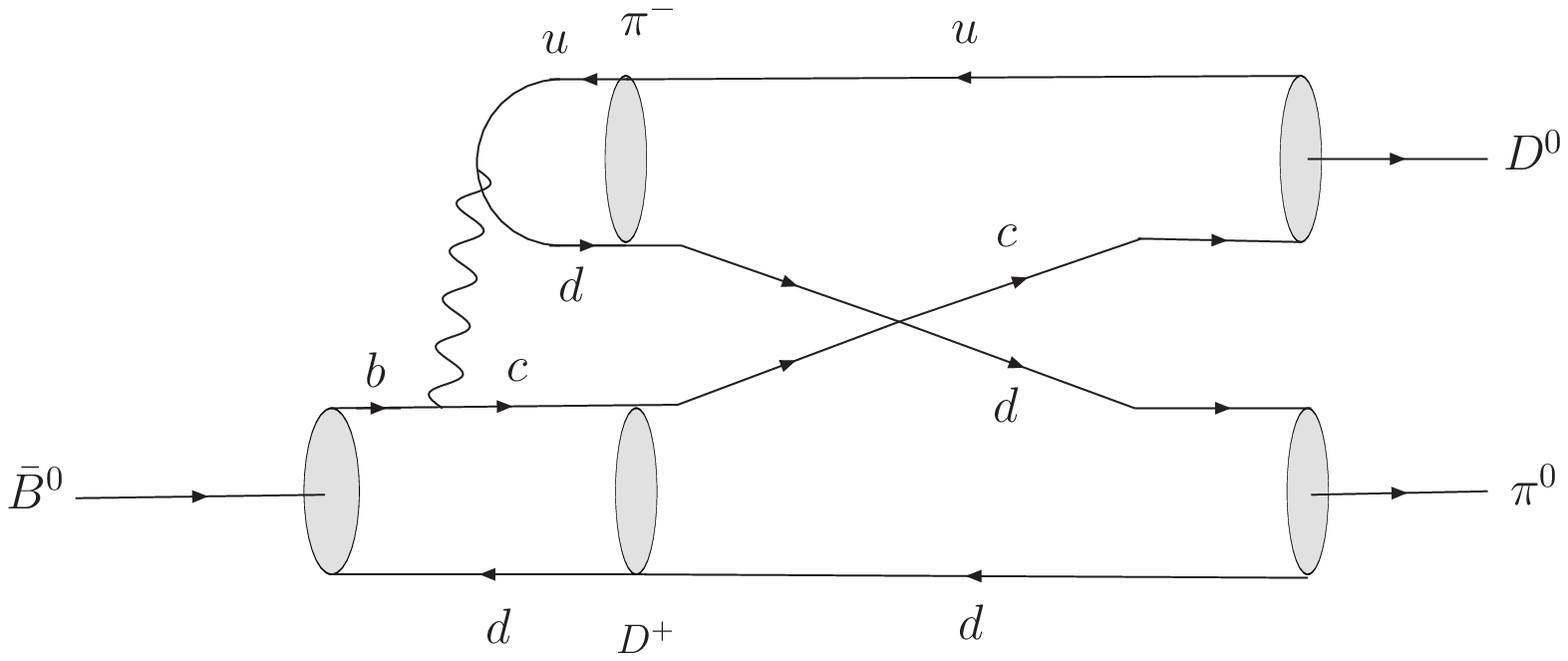}}
 \centerline{(c)}
 \caption[]{\small Contributions to $\ov B^0\to D^0\pi^0$ from
    the color-allowed weak decay $\ov B^0\to D^+\pi^-$ followed by a
    resonant-like rescattering (a) and quark exchange (b) and (c). While (a)
    has the same topology as the $W$-exchange graph, (b) and (c) mimic
    the color-suppressed internal $W$-emission graph.}
\end{figure}

\begin{figure}[ht]
\vspace{-1cm}
  \centerline{\psfig{figure=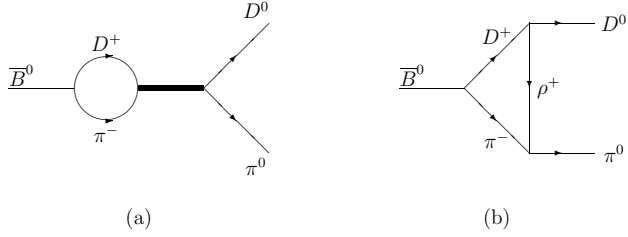,width=14cm}}
\vspace{-14.5cm}
    \caption[]{\small Manifestation of Fig. 1(a) as the
    long-distance $s$- and $t$-channel contributions to the
    $W$-exchange amplitude in $\ov B^0\to D^0\pi^0$. The thick
    line in (a) represents a resonance.}
\end{figure}

\begin{figure}[h]
\vspace{-1cm}
  \centerline{\psfig{figure=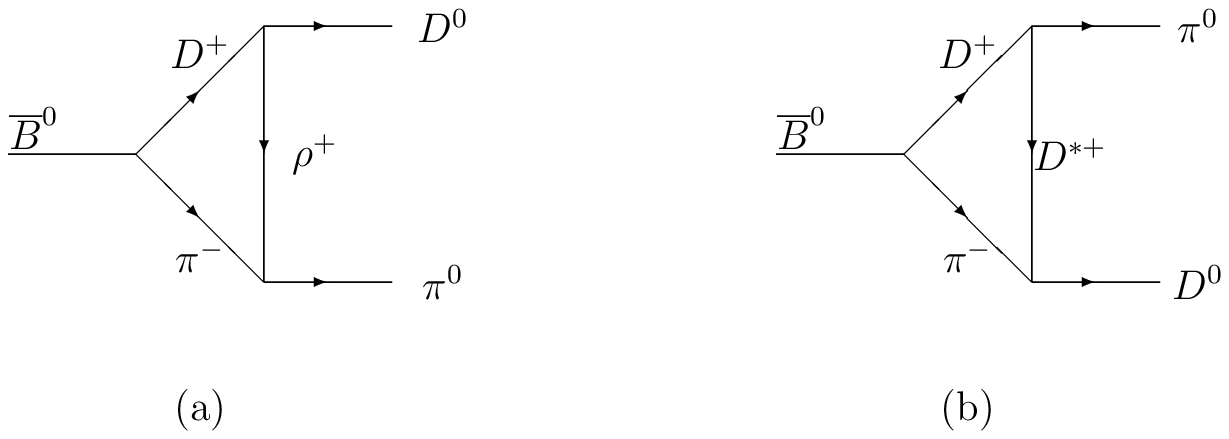,width=16cm}}
\vspace{-17cm}
    \caption[]{\small Manifestation of Figs. 1(b) and 1(c) as the long-distance
    $t$-channel contributions to the
    color-suppressed internal $W$-emission amplitude in $\ov B^0\to D^0\pi^0$.}
\end{figure}

As stressed above, topological graphs can provide information on
final-state interactions (FSIs). In general, there are several
different forms of FSIs: elastic scattering and inelastic
scattering such as quark exchange, resonance formation,$\cdots$,
etc. Take the decay $\ov B^0\to D^0\pi^0$ as an illustration. The
topological amplitudes $\c,~\e,~\a$ can receive contributions from
the tree amplitude $\t$ of e.g. $\ov B^0\to D^+\pi^-$ via
final-state rescattering, as illustrated in Fig. 1: Fig. 1(a) has
the same topology as $W$-exchange, while 1(b) and 1(c) mimic the
internal $W$-emission amplitude $\c$. Therefore, even if the
short-distance $W$-exchange vanishes, a long-distance $W$-exchange
can be induced via inelastic FSIs. Historically, it was first
pointed out in \cite{Donoghue} that rescattering effects required
by unitarity can produce the reaction $D^0\to\ov K^0\phi$, for
example, even in the absence of the $W$-exchange diagram. Then it
was shown in \cite{CC87} that this rescattering diagram belongs to
the generic $W$-exchange topology.

Since FSIs are nonperturbative in nature, in principle it is
extremely difficult to calculate their effects. It is customary to
evaluate the long-distance $W$-exchange contribution, Fig. 1(a),
at the hadron level manifested as Fig. 2
\cite{FSIinD,aaoud,Ablikim}. Fig. 2(a) shows the resonant
amplitude coming from $\ov B^0\to D^+\pi^-$ followed by a
$s$-channel $J^P=0^+$ particle exchange with the quark content
$(c\bar u)$, which couples to $\ov D^0\pi^0$ and $D^+\pi^-$. Fig.
2(b) corresponds to the $t$-channel contribution with a $\rho$
particle exchange. The relative phase between $\t$ and $\c$
indicates some final-state interactions responsible for this.
Figs. 1(b) and 1(c) show that final-state rescattering via quark
exchange has the same topology as the color-suppressed internal
$W$-emission amplitude. At the hadron level, Figs. 1(b) and 1(c)
are manifested in the rescattering processes with one particle
exchange in the $t$ channel \cite{FSIinD} (see Fig. 3). Note that
Figs. 3(a) and 2(b) are the same at the meson level even though at
the quark level they correspond to different processes, namely,
annihilation and quark exchange, respectively. In contrast, Fig.
3(b) is different than Fig. 3(a) in the context of the $t$-channel
exchanged particle.

For charm decays, it is expected that the long-distance
$W$-exchange is dominated by resonant FSIs as shown in Fig. 2(a).
That is, the resonance formation of FSI via $q\bar q$ resonances
is probably the most important one due to the fact that an
abundant spectrum of resonances is known to exist at energies
close to the mass of the charmed meson. However, a direct
calculation of this diagram is subject to many theoretical
uncertainties. For example, the coupling of the resonance to
$D\pi$ states is unknown and the off-shell effects in the chiral
loop should be properly addressed \cite{aaoud}. Nevertheless, as
emphasized in \cite{Zen,Weinberg}, most of the properties of
resonances follow from unitarity alone, without regard to the
dynamical mechanism that produces the resonance. Consequently, as
shown in \cite{Zen,Chenga1a2}, the effect of resonance-induced
FSIs [Fig. 2(a)] can be described in a model-independent manner in
terms of the mass and width of the nearby resonances. It is found
that the $\E$ amplitude is modified by resonant FSIs by (see e.g.
\cite{Chenga1a2})
 \be \label{E}
 \e=e+(e^{2i\delta_r}-1)\left(e+{\T\over 3}\right),
 \en
with
  \be \label{phase}
e^{2i\delta_r}=1-i\,{\Gamma\over m_D-m_R+i\Gamma/2},
 \en
where the reduced $W$-exchange amplitude $\e$ before resonant FSIs
is denoted by $e$. Therefore, resonance-induced FSIs amount to
modifying the $W$-exchange amplitude and leaving the other
quark-diagram amplitudes $\t$ and $\c$ intact. We thus see that
even if the short-distance $W$-exchange vanishes (i.e. $e=0$), as
commonly asserted, a long-distance $W$-exchange contribution still
can be induced from the tree amplitude $\t$ via FSI rescattering
in resonance formation.

In $B$ decays, in contrast to $D$ decays, the resonant FSIs will
be expected to be suppressed relative to the rescattering effect
arising from quark exchange owing to the lack of the existence of
resonances at energies close to the $B$ meson mass. This means
that one can neglect the $s$-channel contribution from Fig. 2(a).

As stressed before, the calculation of the meson level Feynman
diagrams in Fig. 2 or Fig. 3 involves many theoretical
uncertainties. If one naively calculates the diagram, one obtains
an answer which does not make sense in the context of perturbation
theory since the contributions become so large that perturbation
theory is no longer trustworthy. For example, consider the loop
contribution to $B^0\to \pi^+\pi^-$ via the rescattering process
$D^+D^-\to\pi^+\pi^-$. Since $B^0\to\pi^+\pi^-$ is CKM suppressed
relative to $B^0\to D^+D^-$, the loop contribution is larger than
the initial $B\to\pi\pi$ amplitude. Because the $t$-channel
exchanged particle is not on-shell, as we shall see later, a
form-factor cutoff must be introduced to the vertex to render the
whole calculation meaningful.

\section{$B\to D\pi$ Decays}

The color-suppressed decays of $\ov B^0$ into
$D^{(*)0}\pi^0,D^0\eta,D^0\omega$ and $D^0\rho^0$ have been
observed by Belle~\cite{BelleBDpi} and $\ov B^0$ decays into
$D^{(*)0}\pi^0$ have been measured by CLEO \cite{CLEOBDpi}.
Recently, BaBar \cite{BaBarBDpi} has presented the measurements of
$\ov B^0$ decays into $D^{(*)0}(\pi^0,\eta,\omega)$ and
$D^0\eta'$. All measured color-suppressed decays have similar
branching ratios with central values between $1.7\times 10^{-4}$
and $4.2\times 10^{-4}$. They are all significantly larger than
theoretical expectations based on naive factorization. For
example, the measurement $\B(\ov B^0\to D^0\pi^0)=2.5\times
10^{-4}$ (see below) is larger than the theoretical prediction,
$(0.58\sim 1.13)\times 10^{-4}$ \cite{ChengBDpi}, by a factor of
$2\sim 4$. Moreover, the three $B\to D\pi$ amplitudes form a
non-flat triangle, indicating nontrivial relative strong phases
between them. In this section we will focus on $B\to D\pi$ decays
and illustrate the importance of final-state rescattering effects.

In terms of the quark-diagram topologies $\T$, $\C$ and $\E$,
where $\T$ is the color-allowed external $W$-emission tree
amplitude, $\C,~\E$ are color-suppressed internal $W$-emission and
$W$-exchange amplitudes, respectively, the $\ov B\to D\pi$
amplitudes can be expressed as
 \be \label{eq:BDpiamp}
 A(\ov B^0\to D^+\pi^-) &=& \T+\E,\non\\
 A(B^-\to D^0\pi^-) &=& \T+\C,  \\
 A(\ov B^0\to D^0\pi^0) &=& {1\over\sqrt{2}}(-\C+\E),  \non
 \en
and they satisfy the isospin triangle relation
 \be \label{eq:isoBDpi}
 A(\ov B^0\to D^+\pi^-)=\sqrt{2}A(\ov B^0\to D^0\pi^0)+A(B^-\to D^0\pi^-).
 \en
Using the data \cite{PDG}
 \be \label{eq:exptBDpi}
\B(\ov B^0\to D^+\pi^-) &=& (2.76\pm0.25)\times 10^{-3}, \non \\
\B(B^-\to D^0\pi^-) &=& (4.98\pm0.29)\times 10^{-3},
 \en
and the world average $\B(\ov B^0\to D^0\pi^0)=(2.5\pm0.2)\times
10^{-4}$ from the measurements
 \be \label{eq:exptBD0pi0}
\B(\ov B^0\to D^0\pi^0)=\cases{(2.9\pm0.2\pm0.3)\times 10^{-4}, &
BaBar \cite{BaBarBDpi} \cr (2.31\pm0.12\pm0.23)\times 10^{-4}, &
Belle \cite{BelleBDpi} \cr (2.74^{+0.36}_{-0.32}\pm0.55)\times
10^{-4}, & CLEO \cite{CLEOBDpi}}
 \en
we find (only the central values for phase angles are shown here)
 \be \label{eq:BDpiQDvalue}
 \left.{\C-\E\over \T+\E}\right|_{D\pi}=(0.41\pm0.05)\, e^{\pm i53^\circ}, && \qquad\quad
  \left.{\C-\E\over \T+\C}\right|_{D\pi}=(0.29\pm0.03)\, e^{\pm i38^\circ},
 \en
where we have used the relation, for example,
 \be
 \cos\theta_{\{ \C-\E,\T+\E \}}={\B(\ov B^0\to
 D^+\pi^-)+2\B(\ov B^0\to D^0\pi^0)-{\tau(B^0)\over\tau(B^+)}\B(B^-\to D^0\pi^-)\over
 2\sqrt{\B(\ov B^0\to D^+\pi^-)}\,\sqrt{2\B(\ov B^0\to D^0\pi^0)}}
 \en
to extract the strong phases. For the numerical results in Eq.
(\ref{eq:BDpiQDvalue}), we have employed the $B$ meson lifetimes
$\tau(B^0)=(1.536\pm0.014)\times 10^{-12}s$ and
$\tau(B^+)=(1.671\pm0.018)\times 10^{-12}s$ \cite{PDG}.

It is known that in Cabibbo-allowed $D\to PP$ decays, the
topological amplitudes $\T,\C,\E$ can be {\it individually}
extracted from the data with the results \cite{Rosner}
 \be
 \left.{\C\over\T}\right|_{D\to PP}=(0.73\pm0.05)\,e^{-i152^\circ},
 \qquad
 \left.{\E\over\T}\right|_{D\to PP}=(0.59\pm0.05)\,e^{i114^\circ}.
 \en
Therefore, in charm decays the color-suppressed amplitude $\C$ is
not color suppressed and the $W$-exchanged amplitude is quite
sizable ! The large phase of $\E$ is suggestive of nearby
resonance effects. The three amplitudes $\T,\C$ and $\E$ in $B\to
D\pi$ decays can be individually determined from the measurements
of $D\pi$ in conjunction with the data of $D_s^+K^-$ and $D^0\eta$
\cite{ChiangDpi}.

In the factorization approach, the short-distance factorizable
amplitudes read
 \be \label{eq:T,C,E}
 \T_{\rm SD} &=& i{G_F\over\sqrt{2}}\,V_{cb}V_{ud}^*\,a_1(D\pi)(m_B^2-m_D^2)f_\pi
F_0^{BD}(m_\pi^2), \non \\
{\cal C}_{\rm SD} &=& i
 {G_F\over\sqrt{2}}\,V_{cb}V_{ud}^*\,a_2(D\pi)(m_B^2-m_\pi^2)f_D
F_0^{B\pi}(m_D^2),  \\
 {\cal E}_{\rm SD} &=& i
{G_F\over\sqrt{2}}\,V_{cb}V_{ud}^*\,a_2(D\pi)(m_D^2-m_\pi^2)f_B
F_0^{0\to D\pi}(m_B^2), \non
 \en
where $a_{1,2}$ are the parameters related to the Wilson
coefficients and some calculable short-distance nonfactorizable
effects. The annihilation form factor $F_0^{0\to D\pi}(m_B^2)$ is
expected to be suppressed at large momentum transfer, $q^2=m_B^2$,
corresponding to the conventional helicity suppression. Based on
the argument of helicity and color suppression, one may therefore
neglect short-distance (hard) $W$-exchange contributions. In the
QCD factorization approach, contrary to the parameter $a_1(D\pi)$,
$a_2(D\pi)$ is not calculable owing to the presence of infrared
divergence caused by the gluon exchange between the emitted $D^0$
meson and the $(\ov B^0\pi^0)$ system. In other words, the
nonfactorizable contribution to $a_2$ is dominated by
nonperturbative effects. For charmless $B$ decays, a typical value
of $a_2\approx 0.22\,{\rm exp}(-i30^\circ)$ is obtained in the QCD
factorization approach \cite{BBNS}. Recall that the experimental
value for $B\to\J K$ is $|a_2(\J K)|=0.26\pm 0.02$ \cite{a1a2}.
However, neglecting the $W$-exchange contribution, a direct fit to
the $D\pi$ data requires that $a_2/a_1\approx (0.45-0.65)e^{\pm
i60^\circ}$ \cite{ChengBDpi,Xing,NP,Lee}. The question is then why
the magnitude and phase of $a_2/a_1$ are so different from the
model expectation. To resolve this difficulty, we next turn to
long-distance rescattering effects. An early effort along this
direction is based on a quasi-elastic scattering model
\cite{FSIChua1}. For a recent study of the color-suppressed $B\to
D^{(*)0}M$ decays in the perturbative QCD approach based on $k_T$
factorization theorem and in soft-collinear effective theory, see
\cite{KeumDpi} and \cite{Mantry} respectively.

\subsection{Long-distance contributions to $B\to D\pi$}

\begin{figure}[t]
\vspace{-1cm}
  \centerline{\psfig{figure=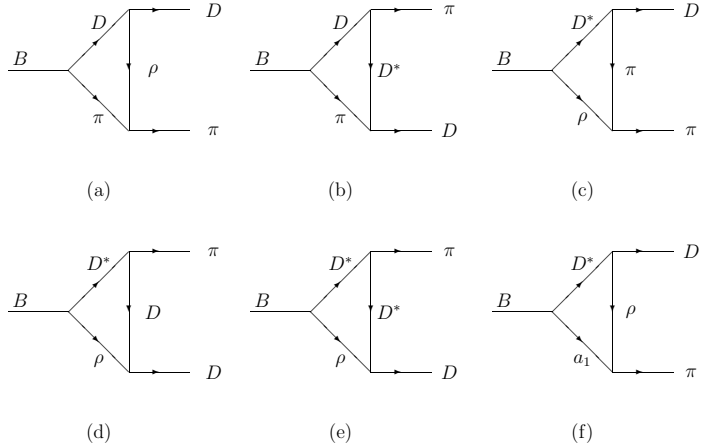,width=13cm}}
\vspace{-11cm}
    \caption[]{\small Long-distance $t$-channel rescattering contributions to $B\to D\pi$. } \label{fig:BDpi}
\end{figure}


Some possible leading long-distance FSI contributions to $B\to
D\pi$ are depicted in Fig. \ref{fig:BDpi}. Apart from the $D\pi$
intermediate state contributions as shown in Figs. 2 and 3, here
we have also included rescattering contributions from the
$D^*\rho$ and $D^*a_1$ intermediate states. As noted in passing,
the $s$-channel contribution is presumably negligible owing to the
absence of nearby resonances. Hence, we will focus only on the
$t$-channel long-distance contributions. For each diagram in Fig.
\ref{fig:BDpi}, one should consider all the possible isospin
structure and draw all the possible sub-diagrams at the quark
level \cite{Ablikim}. While all the six diagrams contribute to
$B^-\to D^0\pi^-$, only the diagrams \ref{fig:BDpi}(a),
\ref{fig:BDpi}(c) and \ref{fig:BDpi}(f) contribute to $\ov B^0\to
D^+\pi^-$ and \ref{fig:BDpi}(b), \ref{fig:BDpi}(d) and
\ref{fig:BDpi}(e) to $\ov B^0\to D^0\pi^0$. To see this, we
consider the contribution of Fig. 4(a) to $\ov B^0\to D^0\pi^0$ as
an example. The corresponding diagrams of Fig. 4(a) at the quark
level are Figs. 1(a) and 1(b). At the meson level, Fig. 4(a)
contains Figs. 2(b) and 3(a). Owing to the wave function
$\pi^0=(\bar uu-\bar dd)/\sqrt{2}$, Fig. 1(a) and hence Fig. 2(b)
has an isospin factor of $1/\sqrt{2}$, while Fig. 1(b) and hence
Fig. 3(a) has a factor of $-1/\sqrt{2}$. Consequently, there is a
cancellation between Figs. 2(b) and 3(a).  Another way for
understanding this cancellation is to note that Fig. 2(b)
contributes to $\E$, while Fig. 3(a) to $\C$. From Eq.
(\ref{eq:BDpiamp}), it is clear that there is a cancellation
between them.

Given the weak Hamiltonian in the form $H_{\rm W}=\sum_i\lambda_i
Q_i$, where $\lambda_i$ is the combination of the quark mixing
matrix elements and $Q_i$ is a $T$-even local operator ($T$: time
reversal), the absorptive part of Fig. 4 can be obtained by using
the optical theorem and time-reversal invariant weak decay
operator $Q_i$. From the time reversal invariance of $Q~(=U_T Q^*
U^\dagger_T$), it follows that
\begin{equation}
\langle i;{\rm out}| Q|B;{\rm in}\rangle^* =\sum_j S^*_{ji}
\langle j;{\rm out}|Q|B;{\rm in}\rangle,
\label{eq:timerev}
\end{equation}
where $S_{ij}\equiv\langle i;{\rm out}|\,j;{\rm in}\rangle$ is the
strong interaction $S$-matrix element, and we have used $U_T|{\rm
out\,(in)}\rangle^*=|{\rm in\,(out)}\rangle$ to fix the phase
convention.\footnote{Note that in the usual phase convention we
have $T|P(\vec p)\ra=-|P(-\vec p)\ra,~T|V(\vec
p,\lambda)\ra=-(-)^\lambda|V(-\vec p,\lambda)\ra$ with $\lambda$
being the helicity of the vector meson. For the $B\to PP,VP,VV$
decays followed by two-particle to two-particle rescatterings, the
$|PP\ra$ and $S$,$D$-wave $|VV\ra$ states satisfy the $U_T|{\rm
out\,(in)}\rangle^*=|{\rm in\,(out)}\rangle$ relation readily,
while for $|B\ra$ and $P$-wave $|VV\ra$ as well as $|VP\ra$ states
we may assign them an additional phase $i$ to satisfy the above
relation. We shall return to the usual phase convention once the
optical theorem for final-state rescattering, namely,
Eq.~(\ref{eq:ImA}), is obtained.}
Eq. (\ref{eq:timerev}) implies an identity related to the optical
theorem. Noting that $S=1+i T$, we find
\begin{equation}
2\,\A bs\, \langle i;{\rm out}|Q|B;{\rm in}\rangle =\sum_j
T^*_{ji} \langle j;{\rm out}|Q|B;{\rm in}\rangle, \label{eq:ImA}
\end{equation}
where use of the unitarity of the $S$-matrix has been made.
Specifically, for two-body $B$ decays, we have
 \be
 \A bs\, M(p_B\to p_1p_2) &=& {1\over 2}\sum_j\left(\Pi_{k=1}^j\int {d^3\vec
 q_k\over (2\pi)^32E_k}\right)(2\pi)^4 \non \\
 &\times & \delta^4(p_1+p_2-\sum_{k=1}^j q_k)M(p_B\to
 \{q_k\})T^*(p_1p_2\to\{q_k\}).
 \en
Thus the optical theorem relates the absorptive part of the
two-body decay amplitude to the sum over all possible $B$ decay
final states $\{q_k\}$, followed by strong $\{q_k\}\to p_1p_2$
rescattering.

Neglecting the dispersive parts for the moment, the FSI
corrections to the topological amplitudes are
 \be
 \T &=& \T_{\rm SD},  \non \\
 \C &=& \C_{\rm SD}+i\A bs\,(4a+4b+4c+4d+4e+4f),\\
 \E &=& \E_{\rm SD}+i\A bs\,(4a+4c+4f). \non
 \en
The color-allowed amplitude $\T$ does receive contributions from,
for example, the $s$-channel $\ov B^0\to D^+\pi^-\to D^+\pi^-$ and
the $t$-channel rescattering process $\ov B^0\to D^0\pi^0\to
D^+\pi^-$. However, they are both suppressed: The first one is
subject to ${\cal O}(1/m_b^2)$ suppression while the weak decay in
the second process is color suppressed. Therefore, it is safe to
neglect long-distance corrections to $\T$. As a result,
 \be
 A(\ov B^0\to D^+\pi^-) &=& \T_{\rm SD}+\E_{\rm SD}+i\A bs\,(4a+4c+4f),  \non \\
 A(B^-\to D^0\pi^-) &=& \T_{\rm SD}+\C_{\rm SD}+i\A bs\,(4a+4b+4c+4d+4e+4f),  \\
 A(\ov B^0\to D^0\pi^0) &=& {1\over\sqrt{2}}(-\C+\E)_{\rm
 SD}-{i\over\sqrt{2}}\A bs\,(4b+4d+4e). \non
 \en
Note that the isospin relation (\ref{eq:isoBDpi}) is still
respected, as it should be.

To proceed we write down the relevant Lagrangian
\cite{Yan92,Casalbuoni}\footnote{In the chiral and heavy quark
limits, the effective Lagrangian of (\ref{eq:LDDV}) can be recast
compactly in terms of superfields \cite{Casalbuoni}
 \be \label{eq:LwithH}
 {\cal L} =&& i\la H_b v^\mu {\cal D}_{\mu ba}\ov H_a\ra+ig\la
 H_b\gamma_\mu\gamma_5 A_{ba}^\mu\ov H_a\ra \non \\
 && +i\beta\la H_b v^\mu(V_\mu-\rho_\mu)_{ba}\ov H_a\ra+i\lambda\la
 H_b\sigma^{\mu\nu}F_{\mu\nu}(\rho)_{ba}\ov H_a\ra,
 \en
where the superfield $H$ is given by $H={1+v\!\!\!/\over
2}(D^{*\mu}\gamma_\mu-i\gamma_5 D)$, and $(V_\mu)_{ba}$ and
$(A_\mu)_{ba}$ are the matrix elements of vector and axial
currents, respectively, constructed from Goldstone bosons.
 }
 \be
 {\cal L} &=&
 -ig_{\rho\pi\pi}\Big(\rho^+_\mu\pi^0\lrpartial{}^{\!\mu}\pi^-
 +\rho^-_\mu\pi^+\lrpartial{}^{\!\mu}\pi^0+\rho^0_\mu\pi^-\lrpartial{}^{\!\mu}\pi^+\Big)
 \non \\
 &-& ig_{D^*DP}(D^i\partial^\mu P_{ij}
 D_\mu^{*j\dagger}-D_\mu^{*i}\partial^\mu P_{ij}D^{j\dagger})
 +{1\over 2}g_{D^*D^*P}
 \vp_{\mu\nu\alpha\beta}\,D_i^{*\mu}\partial^\nu P^{ij}
 \lrpartial{}^{\!\alpha} D^{*\beta\dagger}_j \non \\
 &-& ig_{DDV} D_i^\dagger \lrpartial_{\!\mu} D^j(V^\mu)^i_j
 -2f_{D^*DV} \epsilon_{\mu\nu\alpha\beta}
 (\partial^\mu V^\nu)^i_j
 (D_i^\dagger\lrpartial{}^{\!\alpha} D^{*\beta j}-D_i^{*\beta\dagger}\lrpartial{}{\!^\alpha} D^j)
 \non\\
 &+& ig_{D^*D^*V} D^{*\nu\dagger}_i \lrpartial_{\!\mu} D^{*j}_\nu(V^\mu)^i_j
 +4if_{D^*D^*V} D^{*\dagger}_{i\mu}(\partial^\mu V^\nu-\partial^\nu
 V^\mu)^i_j D^{*j}_\nu,
 \label{eq:LDDV}
 \en
with the convention $\vp^{0123}=1$, where $P$ and $V_\mu$ are
$3\times 3$ matrices for the octet pseudoscalar and nonet vector
mesons, respectively
 \be
 P &=& \left(\matrix{{\pi^0\over\sqrt{2}}+{\eta\over\sqrt{6}} & \pi^+ & K^+ \cr
 \pi^- & -{\pi^0\over\sqrt{2}}+{\eta\over\sqrt{6}} & K^0  \cr
 K^- & \ov K^0 & -\sqrt{2\over 3}\eta }\right), \quad
 V = \left(\matrix{{\rho^0\over\sqrt{2}}+{\omega\over\sqrt{2}} & \rho^+ & K^{*+} \cr
 \rho^- & -{\rho^0\over\sqrt{2}}+{\omega\over\sqrt{2}} & K^{*0}  \cr
 K^{*-} & \ov K^{*0} & \phi }\right).
 \en
Note that our phase convention on fields is fixed by
 \be
 \la 0|A_\mu|P\ra=i f_P p_\mu,
 \qquad
 \la 0|V_\mu|V\ra=m_V \vp_\mu,
 \en
and is different from~\cite{Casalbuoni}. In the chiral and heavy
quark limits, we have~\cite{Casalbuoni}
 \be
 g_{D^*D^*\pi}={g_{D^*D\pi}\over \sqrt{m_Dm_{D^*}}}={2\over f_\pi}g,
 \quad g_{DDV}=g_{D^*D^*V}=\frac{\beta g_V}{\sqrt2},
 \quad
 f_{D^*DV}=\frac{f_{D^*D^*V}}{m_{D^*}}=\frac{\lambda g_V}{\sqrt2},
 \en
with $f_\pi=132$ MeV. The parameters $g_V$, $\beta$ and $\lambda$
thus enter into the effective chiral Lagrangian describing the
interactions of heavy mesons with low momentum light vector mesons
(see e.g. \cite{Casalbuoni}). The parameter $g_V$ respects the
relation $g_V=m_\rho/f_\pi$ \cite{Casalbuoni}. We shall follow
\cite{Isola2003} to use $\beta=0.9$ and $\lambda=0.56$~GeV$^{-1}$.
Instead of writing down the Feynman rules for the vertices, we
work out the corresponding matrix elements:
 \be
 \la\rho^+(\vp)\pi^-(p_2)|i{\cal L}|\pi^0(p_1)\ra &=& -ig_{\rho\pi\pi}\vp\cdot (p_1+p_2), \non \\
 \la D(p_2)\pi(q)|i{\cal L}|D^*(\vp,p_1)\ra &=&   -ig_{D^*D\pi}\,\vp\cdot q, \non \\
 \la D^*(\vp_2,p_2)\pi(q)|i{\cal L}|D^*(\vp_1,p_1)\ra &=& -ig_{D^*D^*\pi}
 \epsilon_{\mu\nu\alpha\beta}\vp_1^\mu\vp_2^{*\nu}  q^\alpha p^\beta_1, \non \\
 \la D(p_2)\rho(\vp)|i{\cal L}|D(p_1)\ra &=& -{i\over \sqrt{2}}\,\beta g_V\,\vp\cdot (p_1+p_2), \non \\
 \la D^*(\vp_2,p_2)\rho(\vp_\rho,q)|i{\cal L}|D(p_1)\ra &=& -i
 {2\sqrt{2}}\lambda\, g_V\,\epsilon_{\mu\nu\alpha\beta}\vp^\mu_\rho\vp^{*\nu}_{D^*}p_1^\alpha
 q^\beta.
 \en

Figs. \ref{fig:BDpi}(a) and \ref{fig:BDpi}(b) arise from the weak
decay $B\to D\pi$ followed by the rescattering of $D\pi$ to
$D\pi$. Denoting the momenta by $B(p_B)\to D(p_1)\pi(p_2)\to
D(p_3)\pi(p_4)$, it follows that the absorptive part of
\ref{fig:BDpi}(a) is given by \footnote{As noticed in
\cite{Ablikim}, the isospin factor of $1/\sqrt{2}$ or
$-1/\sqrt{2}$ should be dropped for the intermediate state
$\rho^0$ or $\pi^0$.}
 \be \label{eq:abs4a}
 \A bs\,(4a) &=& {1\over 2}\int {d^3\vec p_1\over (2\pi)^32E_1}\,{d^3 \vec p_2\over
(2\pi)^3 2E_2}\,(2\pi)^4\delta^4(p_B-p_1-p_2)A(\ov B^0\to D^+
\pi^-) \non \\
 &\times&  i{1\over\sqrt{2}}\,\beta g_V\,{F^2(t,m_\rho)\over
 t-m_{\rho}^2+im_\rho\Gamma_\rho}\, (-i)g_{\rho\pi\pi}\,(p_1+p_3)^\mu(p_2+p_4)^\nu
 \left(-g_{\mu\nu}+{k_\mu k_\nu\over m_\rho^2}\right)  \non \\
 &=& \int_{-1}^1\,{|\vec p_1|d\cos\theta\over 16\pi m_B}{1\over
 \sqrt{2}}\,g_V g_{\rho\pi\pi}\beta\,A(\ov B^0\to D^+\pi^-){F^2(t,m_\rho)\over
 t-m_\rho^2+im_\rho\Gamma_\rho} H_1,
 \en
where $\theta$ is the angle between $\vec p_1$ and $\vec  p_3$,
$k$ is the momentum of the exchanged $\rho$ meson ($k^2=t$), and
 \be
 t &\equiv&
 (p_1-p_3)^2 = m_1^2+m_3^2-2E_1E_3+2|\vec p_1||\vec
 p_3|\cos\theta, \non \\
 H_1 &=& -(p_1\cdot p_2+p_3\cdot p_4+p_1\cdot p_4+p_2\cdot
 p_3)-{(m_1^2-m_3^2)(m_2^2-m_4^2)\over m_\rho^2}.
 \en
The form factor $F(t,m)$ in Eq. (\ref{eq:abs4a}) takes care of the
off-shell effect of the exchanged particle, which is usually
parametrized as
 \be \label{eq:FF}
 F(t,m)=\,\left({\Lambda^2-m^2\over \Lambda^2-t}\right)^n,
 \en
normalized to unity at $t=m^2$. The monopole behavior of the form
factor (i.e. $n=1$) is preferred as it is consistent with the QCD
sum rule expectation \cite{Gortchakov}. However, we shall return
back to this point when discussing the FSI effects in $B\to \phi
K^*$ decays (see Sec. VII.C).

Likewise, the absorptive part of \ref{fig:BDpi}(b) is given by
 \be
 \A bs\,(4b) &=& {1\over 2}\int {d^3\vec p_1\over (2\pi)^32E_1}\,{d^3\vec p_2\over
(2\pi)^3 2E_2}\,(2\pi)^4\delta^4(p_B-p_1-p_2)A(\ov B^0\to D^+
\pi^-)
\non \\
 &\times& ig_{D^*D\pi}p_4^\mu\,{F^2(t,m_{D^*})\over
 t-m_{D^*}^2}\,(-i)g_{D^*D\pi}(-p_2^\nu)
 \left(-g_{\mu\nu}+{k_\mu k_\nu\over m_{D^*}^2}\right) \non \\
 &=& -\int_{-1}^1\,{|\vec p_1|d\cos\theta\over 16\pi m_B}
 g_{D^*D\pi}^2\,A(\ov B^0\to D^+\pi^-)\,{F^2(t,m_{D^*})\over
 t-m_{D^*}^2}\, H_2,
 \en
with
 \be
 H_2=-p_2\cdot p_4+{(p_2\cdot p_3-m_2^2)(p_1\cdot p_4-m_4^2)\over
 m_{D^*}^2}.
 \en

Figs. \ref{fig:BDpi}(c)-\ref{fig:BDpi}(e) come from the
rescattering process $B\to D^*(p_1)\rho(p_2)\to D(p_3)\pi(p_4)$.
The absorptive part of \ref{fig:BDpi}(c) reads
 \be
 \A bs\,(4c) &=& {1\over 2}\int {d^3\vec p_1\over (2\pi)^32E_1}\,{d^3\vec p_2\over
(2\pi)^3
2E_2}\,(2\pi)^4\delta^4(p_B-p_1-p_2)\sum_{\lambda_1,\lambda_2}A(\ov
B^0\to D^{*+} \rho^-)   \non \\
 &\times&  (-i)g_{D^*D\pi}\,\vp_1\cdot (-p_3)\,{F^2(t,m_\pi)\over
 t-m_\pi^2}\, (-i)g_{\rho\pi\pi}\,2\vp_2\cdot p_4.
 \en
To proceed, we note that the factorizable amplitude of $B\to
V_1V_2$ is given by
 \be \label{eq:BVV}
 A(B\to V_1V_2) &=& {G_F\over\sqrt{2}}V_{\rm CKM}\la V_2 | (\bar{q}_2 q_3)_\vma|0\ra\la
V_1|(\bar{q}_1b)_\vma|\ov B \ra \non \\
&=& - if_{V_2}m_2\Bigg[ (\vp^*_1\cdot\vp^*_2) (m_{B}+m_{1})A_1^{
BV_1}(m_{2}^2)- (\vp^*_1\cdot p_{_{B}})(\vp^*_2 \cdot
p_{_{B}}){2A_2^{ BV_1}(m_{2}^2)\over m_{B}+m_{1} } \non \\ &-&
i\epsilon_{\mu\nu\alpha\beta}\vp^{*\mu}_2\vp^{*\nu}_1p^\alpha_{_{B}}
p^\beta_1\,{2V^{ BV_1}(m_{2}^2)\over m_{B}+m_{1} }\Bigg],
 \en
with $(\bar q_1q_2)_\vma\equiv \bar q_1\gamma_\mu(1-\gamma_5)q_2$.
Therefore,
 \be
 \A bs\,(4c) &=& -i2\int_{-1}^1\,{|\vec p_1|d\cos\theta\over 16\pi m_B}
 \,g_{D^*D\pi}g_{\rho\pi\pi}\,{F^2(t,m_\pi)\over
 t-m_\pi^2} \non \\
 &\times& f_\rho m_\rho\Big[
 (m_B+m_{D^*})A_1^{BD^*}(m_\rho^2)H_3-{2A_2^{BD^*}(m_\rho^2)\over
 m_B+m_{D^*}}H'_3\Big],
 \en
where
 \be
 H_3 &=& \left(p_3\cdot p_4-{(p_1\cdot p_3)(p_1\cdot p_4)\over
m_1^2}-{(p_2\cdot p_3) (p_2\cdot p_4)\over m_2^2}+{(p_1\cdot p_2)
(p_1\cdot p_3)(p_2\cdot p_4)\over m_1^2m_2^2}\right), \non \\
 H'_3 &=& \left(p_2\cdot p_3-{(p_1\cdot p_2)(p_1\cdot p_3)\over
 m_1^2}\right)\left(p_1\cdot p_4-{(p_1\cdot p_2)(p_2\cdot p_4)\over
 m_2^2}\right).
 \en
Likewise,
  \be
  \A bs\,(4d)&=& {1\over 2}\int {d^3\vec p_1\over (2\pi)^32E_1}\,{d^3\vec
p_2\over (2\pi)^3
2E_2}\,(2\pi)^4\delta^4(p_B-p_1-p_2)\sum_{\lambda_1,\lambda_2}A(\ov
B^0\to D^{*+} \rho^-)   \non \\
 &\times& (-i)g_{D^*D\pi}\vp_1\cdot p_4 \,{F^2(t,m_{D^*})\over
 t-m_{D^*}^2}(-i){\beta\over\sqrt{2}}\,g_V\,\vp_2\cdot(k+p_3) \non
 \\
 &=& i\sqrt{2}\int_{-1}^1\,{|\vec p_1|d\cos\theta\over 16\pi m_B}
 \,\beta\,g_V\,g_{D^*D\pi}\,{F^2(t,m_D)\over
 t-m_D^2} \non \\
 &\times& f_\rho m_\rho \Big[
 (m_B+m_{D^*})A_1^{BD^*}(m_\rho^2)H_4-{2A_2^{BD^*}(m_\rho^2)\over
 m_B+m_{D^*}}H'_4\Big],
 \en
where $H_4$ and $H'_4$ can be obtained from $H_3$ and $H'_3$,
respectively, by interchanging $p_3$ and $p_4$, and
 \be
 \A bs\,(4e) &=& {1\over 2}\int {d^3\vec p_1\over (2\pi)^32E_1}\,{d^3\vec p_2\over
(2\pi)^3
2E_2}\,(2\pi)^4\delta^4(p_B-p_1-p_2)\sum_{\lambda_1,\lambda_2}A(\ov
B^0\to D^{*+}\rho^-) \non \\ &\times&
ig_{D^*D^*\pi}\,\epsilon_{\mu\nu\alpha\beta}\vp^\mu_1\, p_4^\alpha
p_1^\beta\,{F^2(t,m_{D^*}) \over
 t-m_{D^*}^2}\,i2\sqrt{2}\,\lambda\, g_V\,
 \epsilon_{\rho\sigma\xi\eta}\vp^\rho_2 p_3^\xi
 (-p_2)^\eta\left(-g^{\nu\sigma}+{k^\nu k^\sigma\over m_{D^*}^2}\right)   \non \\
 &=& i2\sqrt{2}\int_{-1}^1\,{|\vec p_1|d\cos\theta\over 16\pi m_B}
 g_V\lambda\,g_{D^*D^*\pi}\,{F^2(t,m_{D^*})\over
 t-m_{D^*}^2} \non \\
 &\times& f_\rho m_\rho\Big[
 (m_B+m_{D^*})A_1^{BD^*}(m_\rho^2)H_5-{2A_2^{BD^*}(m_\rho^2)\over
 m_B+m_{D^*}}H'_5\Big],
 \en
where
 \be
 H_5 &=& 2(p_1\cdot p_2) (p_3\cdot p_4)-2(p_1\cdot p_3)(p_2\cdot
 p_4),  \non \\
 H'_5 &=& m_B^2\Big[(p_1\cdot p_2)(p_3\cdot p_4)-(p_1\cdot p_4)(p_2\cdot
 p_4)\Big]+(p_2\cdot p_B)(p_4\cdot p_B)(p_1\cdot p_3)  \\
 &+& (p_1\cdot
 p_B)(p_3\cdot p_B)(p_2\cdot p_4)-(p_1\cdot
p_B)(p_2\cdot p_B)(p_3\cdot p_4)-(p_3\cdot p_B)(p_4\cdot
p_B)(p_1\cdot p_2).  \non
 \en

Fig. \ref{fig:BDpi}(f) comes from the weak decay $\ov B^0\to
D^{*+}a_1^-$ followed by a strong scattering. We obtain
 \be \label{eq:(f)}
 \A bs\,(4f) &=& {1\over 2}\int {d^3\vec p_1\over (2\pi)^32E_1}\,{d^3\vec p_2\over
(2\pi)^3
2E_2}\,(2\pi)^4\delta^4(p_B-p_1-p_2)\sum_{\lambda_1,\lambda_2}A(\ov
B^0\to D^{*+}a_1^-)  \\ &\times& i2\sqrt{2}g_V\lambda
 \epsilon_{\mu\nu\alpha\beta}\vp^\mu_1\,p_3^\alpha
k^\beta\,{F^2(t,m_\rho) \over
 t-m_\rho^2+im_\rho\Gamma_\rho}\,(gG_{\rho\sigma}+\ell L_{\rho\sigma})\vp_2^\rho
 \left(-g^{\nu\sigma}+{k^\nu k^\sigma\over m_{D^*}^2}\right) \non
 \en
where
 \be
 G^{\rho\sigma} &=& \delta^{\rho\sigma}-{1\over Y}\Big[
 m_2^2k^\rho k^\sigma+k^2 p_2^\rho p^\sigma_2+p_2\cdot k(p_2^\rho
 k^\sigma+k^\rho p_2^\sigma)\Big] \non \\
 L^{\rho\sigma} &=& {p_2\cdot k\over Y}\left(
 p_2^\rho+k^\rho{m_2^2\over p_2\cdot k}\right)\left(
 k^\sigma+p_2^\sigma{k^2\over p_2\cdot k}\right),
 \en
and $Y=(p_2\cdot k)^2-k^2m_2^2$. The parameters $g$ and $\ell$
appearing in Eq. (\ref{eq:(f)}) also enter into the strong decay
amplitude of $a_1\to \rho\pi$ parametrized as
 \be \label{eq:a1torhopi}
 A(a_1\to  \rho\pi)=(gG_{\mu\nu}+\ell
 L_{\mu\nu})\vp_{a_1}^\mu\vp^\nu_\rho
 \en
and hence they can determined from the measured decay rate and the
ratio of $D$ and $S$ waves:
 \be
 \Gamma(a_1\to \rho\pi) &=& {p_c\over 12\pi
 m_{a_1}^2}\left(2|g|^2+{m_{a_1}^2m_\rho^2\over (p_{a_1}\cdot
 p_\rho)^2}|\ell|^2\right), \non \\
 {D\over S} &=& -\sqrt{2}\,{(E_\rho-m_\rho)g+p_c^2 m_{a_1}h\over
 (E_\rho+2m_\rho)g+p_c^2 m_{a_1}h },
 \en
with $p_c$ being the c.m. momentum of the $\rho$ or $\pi$ in the
$a_1$ rest frame, and
 \be
 h={p_{a_1}\cdot
 p_\rho\over Y}\left(-g+\ell {m_{a_1}^2m_\rho^2\over (p_{a_1}\cdot
 p_\rho)^2}\right).
 \en
Hence, Eq. (\ref{eq:(f)}) can be recast to
 \be
 \A bs\,(4f) &=& i2\sqrt{2}\int_{-1}^1\,{|\vec p_1|d\cos\theta\over 16\pi m_B}
 \,g_V\lambda\,f_{a_1}m_{a_1}\,{F^2(t,m_\rho)\over
 t-m_\rho^2+im_\rho\Gamma_\rho}\,{2V^{BD^*}(m_{a_1}^2)\over
 m_B+m_{D^*}} H_6,
 \en
with
 \be
 H_6 &=& \left({g\over Y}p_2\cdot k-{\ell\over Y}\,{m_2^2k^2\over
 p_2\cdot k}\right)\Big[m_1^2(p_2\cdot p_4)(p_2\cdot p_3)+ m_2^2(p_1\cdot p_3)(p_1\cdot p_4)
 +(p_1\cdot p_2)^2p_3\cdot p_4  \non \\
 &-& m_1^2m_2^2\, p_3\cdot p_4-(p_2\cdot p_4)(p_1\cdot
 p_3)(p_1\cdot p_2)-(p_1\cdot p_2)(p_2\cdot
 p_3)(p_1\cdot p_4)\Big] \non \\
 &-& 2g\Big(m_1^2p_2\cdot p_3-(p_1\cdot p_2)(p_1\cdot
 p_3)\Big).
 \en

It should be emphasized that attention must be paid to the
relative sign between $B\to PP$ and $B\to VV$ decay amplitudes
when calculating long-distance contributions from various
rescattering processes. For the one-body matrix element defined by
$\la 0|\bar q\gamma_\mu\gamma_5|P\ra=if_Pq_\mu$, the signs of
$B\to PP$ and $B\to VV$ amplitudes, respectively, are fixed as in
Eqs. (\ref{eq:T,C,E}) and (\ref{eq:BVV}). This can be checked
explicitly via heavy quark symmetry (see e.g. \cite{CCH}).

The dispersive part of the rescattering amplitude can be obtained
from the absorptive part via the dispersion relation
 \be \label{eq:dispersive}
 {\cal D}is\,A(m_B^2)={1\over \pi}\int_s^\infty { {\A bs}\,A(s')\over
 s'-m_B^2}ds'.
 \en
Unlike the absorptive part, it is known that the dispersive
contribution suffers from the large uncertainties due to some
possible subtractions and the complication from integrations. For
this reason, we will assume the dominance of the rescattering
amplitude by the absorptive part and ignore the dispersive part in
the present work except for the decays $B^0\to\pi^+\pi^-$ and
$B^0\to\pi^0\pi^0$ where a dispersive contribution arising from
$D\bar D\to\pi\pi$ and $\pi\pi\to\pi\pi$ rescattering via
annihilation may play an essential role.

\subsection{Numerical results}
To estimate the contributions from rescattering amplitudes we need
to specify various parameters entering into the vertices of
Feynman diagrams. The on-shell strong coupling $g_{\rho\pi\pi}$ is
determined from the $\rho\to\pi\pi$ rate to be
$g_{\rho\pi\pi}=6.05\pm0.02$. The coupling $g_{D^*D\pi}$ has been
extracted by CLEO to be $17.9\pm0.3\pm1.9$ from the measured
$D^{*+}$ width \cite{CLEOg}. The parameters relevant for the
$D^*D^{(*)}\rho$ couplings are $g_V=5.8$, $\beta=0.9$ and
$\lambda=0.56\,{\rm GeV}^{-1}$.

For form factors we use the values determined from the covariant
light-front model \cite{CCH}. For the coefficients $g$ and $\ell$
in Eq. (\ref{eq:a1torhopi}), one can use the experimental results
of $D/S=-0.1\pm0.028$ and $\Gamma(a_1\to\rho\pi)=250-600$ MeV
\cite{PDG} to fix them. Specifically, we use $D/S=-0.1$ and
$\Gamma(a_1\to\rho\pi)=400$ MeV to obtain $g=4.3$ and
$\ell=5.8$\,. For decay constants, we use $f_\pi=132$ MeV,
$f_D=200$ MeV, $f_{D^*}=230$ MeV and $f_{a_1}=-205$
MeV.\footnote{The sign of $f_{a_1}$ is opposite to $f_\pi$ as
noticed in \cite{CCH}.}

Since the strong vertices are determined for physical particles
and since the exchanged particle is not on-shell, it is necessary
to introduce the form factor $F(t)$ to account for the off-shell
effect of the $t$-channel exchanged particle. For the cutoff
$\Lambda$ in the form factor $F(t)$ [see Eq. (\ref{eq:FF})],
$\Lambda$ should be not far from the physical mass of the
exchanged particle. To be specific, we write
 \be \label{eq:Lambda}
 \Lambda=m_{\rm exc}+\eta\Lambda_{\rm QCD},
 \en
where the parameter $\eta$ is expected to be of order unity and it
depends not only on the exchanged particle but also on the
external particles involved in the strong-interaction vertex.
Since we do not have first-principles calculations of $\eta$, we
will determine it from the measured branching ratios given in
(\ref{eq:exptBDpi}) and (\ref{eq:exptBD0pi0}). Taking
$\Lambda_{\rm QCD}=220$ MeV, we find $\eta=2.2$ for the exchanged
particles $D^*$ and $D$ and $\eta=1.1$ for $\rho$ and $\pi$. As
noted in passing, the loop corrections will be larger than the
initial $B\to D\pi$ amplitudes if the off-shell effect is not
considered. Although the strong couplings are large in the
magnitude, the rescattering amplitude is suppressed by a factor of
$F^2(t)\sim m^2\Lambda_{\rm QCD}^2/t^2$. Consequently, the
off-shell effect will render the perturbative calculation
meaningful.

Numerically,  we obtain (in units of GeV)
 \be \label{eq:numBDpi}
 A(\ov B^0\to D^+\pi^-) &=& i6.41\times 10^{-7}-(0.90+i0.09)\times
 10^{-7}, \non \\
 A(\ov B^0\to D^0\pi^0) &=& -i0.91\times 10^{-7}-1.56\times
 10^{-7}, \non \\
 A(B^-\to D^0\pi^-) &=& i7.69\times 10^{-7}+(1.30-i0.09)\times
 10^{-7}
 \en
for $a_1=0.90$ and $a_2=0.25$,\footnote{As mentioned before,
$a_2(D\pi)$ is {\it a priori} not calculable in the QCD
factorization approach. Therefore, we choose $a_2(D\pi)\approx
0.25$ for the purpose of illustration. Nevertheless, a rough
estimate of $a_2(D\pi)$ has been made in \cite{BBNS} by treating
the charmed meson as a light meson while keeping its highly
asymmetric distribution amplitude. It yields $a_2(D\pi)\approx
0.25\exp(-i40^\circ)$.}
where the first term on the right hand side of each amplitude
arises from the short-distance factorizable contribution and the
second term comes from absorptive parts of the final-state
rescattering processes. Note that the absorptive part of $\ov
B^0\to D^+\pi^-$ and $B^-\to D^0\pi^-$ decay amplitudes are
complex owing to a non-vanishing $\rho$ width. To calculate the
branching ratios, we will take into account the uncertainty from
the $\Lambda_{\rm QCD}$ scale. Recall that $\Lambda_{\rm
QCD}=217^{+25}_{-21}$ MeV is quoted in PDG \cite{PDG}. Therefore,
allowing 15\% error in $\Lambda_{\rm QCD}$, the flavor-averaged
branching ratios are found to be
 \be
 \B(\ov B^0\to  D^+\pi^-) &=& (3.1^{+0.0}_{-0.0})\times 10^{-3}~ (3.2\times 10^{-3}), \non \\
 \B(B^-\to D^0\pi^-)&=& (5.0^{+0.1}_{-0.0})\times 10^{-3}~(4.9\times 10^{-3}), \non \\
 \B(\ov B^0\to D^0\pi^0)&=& (2.5^{+1.1}_{-0.8})\times 10^{-4}~(0.6\times  10^{-4}),
 \en
where the branching ratios for decays without FSIs are shown in
parentheses. We see that the $D^0\pi^0$ mode is sensitive to the
cutoff scales $\Lambda_{D^*}$ and $\Lambda_D$, whereas the other
two color-allowed channels are not. It is clear that the naively
predicted $\B(\ov B^0\to D^0\pi^0)$ before taking into account
rescattering effects is too small compared to experiment. In
contrast, $\ov B^0\to D^+\pi^-$ and $B^-\to D^0\pi^-$ are almost
not affected by final-state rescattering. Moreover,
 \be \label{eq:C/TinBDpi}
 \left.{\C\over \T}\right|_{D\pi} = \cases{ 0.28\,e^{-i48^\circ}, &
 \cr 0.20, & } \quad
 \left.{\E\over \T}\right|_{D\pi} = \cases{ 0.14\,e^{i96^\circ}, & with~FSI \cr
 0, & without~FSI}
 \en
where we have assumed negligible short-distance $W$-exchange and
taken the central values of the cutoffs. Evidently, even if the
short-distance weak annihilation vanishes, a long-distance
contribution to $W$-exchange can be induced from rescattering.
Consequently,
 \be \label{eq:CE/TEinBDpi}
 \left.{\C-\E\over \T+\E}\right|_{D\pi} = \cases{ 0.40\,e^{-i67^\circ}, & \cr
 0.20,& } \quad
 \left.{\C-\E\over \T+\C}\right|_{D\pi} = \cases{ 0.33\,e^{-i50^\circ}, & with~FSI \cr
 0.17, & without~FSI}
 \en
Therefore, our results are consistent with  experiment
(\ref{eq:BDpiQDvalue}). This indicates that rescattering will
enhance the ratios and provide the desired strong phases. That is,
the enhancement of $R\equiv(\C-\E)/(\T+\E)$ relative to the naive
expectation is ascribed to the enhancement of color-suppressed
$W$-emission and the long-distance $W$-exchange, while the phase
of $R$ is accounted for by the strong phases of $\C$ and $\E$
topologies.

Because the rescattering long-distance amplitudes have the same
weak phases as the short-distance factorizable ones, it is clear
that there is no direct \CP violation induced from rescattering
FSIs.

Since the decay $\ov B^0\to D_s^+K^-$ proceeds only through the
topological $W$-exchange diagram, the above determination of $\E$
allows us to estimate its decay rate. From Eq.
(\ref{eq:C/TinBDpi}) it follows that $|\E/(\T+\E)|^2=0.020$ in the
presence of FSIs. Therefore, we obtain
 \be
 {\Gamma(\ov B^0\to D_s^+K^-)\over \Gamma(\ov B^0\to
 D^+\pi^-)}=0.019\,,
 \en
which is indeed consistent with the experimental value of
$0.014\pm0.005$ \cite{PDG}. The agreement will be further improved
after taking into account SU(3) breaking in the $\E$ amplitude of
$\ov B^0\to D_s^+K^-$.

\section{$B\to\pi K$ Decays}

\begin{table}[h]
\caption{Experimental data for \CP averaged branching ratios (top,
in units of $10^{-6}$) and \CP asymmetries (bottom) for $B\to\pi
K$ \cite{HFAG,PDG}.} \label{tab:piK}
\begin{ruledtabular}
\begin{tabular}{c c c c c}
Mode & BaBar & Belle & CLEO & Average  \\
\hline
 $B^-\to \pi^-\ov K^0$ & $26.0\pm1.3\pm1.0$ & $22.0\pm1.9\pm1.1$ &
 $18.8^{+3.7+2.1}_{-3.3-1.8}$ & $24.1\pm1.3$ \\
 $\ov B^0\to \pi^+K^-$ & $17.9\pm0.9\pm0.7$ & $18.5\pm1.0\pm0.7$ &
 $18.0^{+2.3+1.2}_{-2.1-0.9}$ & $18.2\pm0.8$ \\
 $B^-\to \pi^0K^-$ & $12.0\pm0.7\pm0.6$ &
 $12.0\pm1.3^{+1.3}_{-0.9}$ & $12.9^{+2.4+1.2}_{-2.2-1.1}$ &
 $12.1\pm0.8$ \\
 $\ov B^0\to\pi^0\ov K^0$ & $11.4\pm0.9\pm0.6$ &
 $11.7\pm2.3^{+1.2}_{-1.3}$ & $12.8^{+4.0+1.7}_{-3.3-1.4}$ &
 $11.5\pm1.0$  \\
 \hline
 $\A_{\pi^-K_S}$ & $-0.087\pm0.046\pm0.010$ &
 $0.05\pm0.05\pm0.01$ & $0.18\pm0.24\pm0.02$ &
 $-0.020\pm0.034$ \\
 $\A_{\pi^+K^-}$ & $-0.133\pm0.030\pm0.009$ &
 $-0.101\pm0.025\pm0.005$ & $-0.04\pm0.16\pm0.02$ & $-0.113\pm0.019$
 \\
 $\A_{\pi^0K^-}$ & $0.06\pm0.06\pm0.01$ &
 $0.04\pm0.05\pm0.02$ & $-0.29\pm0.23\pm0.02$ & $0.04\pm0.04$
 \\
 $\A_{\pi^0K_S}$ & $-0.06\pm0.18\pm0.06$ & $-0.12\pm0.20\pm0.07$ &
 &  $-0.09\pm0.14$ \\
 ${\cal S}_{\pi^0K_S}$ & $0.35^{+0.30}_{-0.33}\pm0.04$ & $0.30\pm0.59\pm0.11$ & &
 $0.34^{+0.27}_{-0.29}$\\
\end{tabular}
\end{ruledtabular}
\end{table}

The penguin dominated $B\to \pi K$ decay amplitudes have the
general expressions

 \be
 A(\ov B^0\to \pi^+K^-) &=& \T+\P+{2\over 3}\P'_{\rm EW}+\P_A, \non \\
 A(\ov B^0\to \pi^0\ov K^0) &=& {1\over \sqrt{2}}(\C-\P+\P_{\rm
 EW}+{1\over 3}\P'_{\rm EW}-\P_A),  \\
 A(B^-\to\pi^-\ov K^0) &=& \P-{1\over 3}\P'_{\rm EW}+\A+\P_A,  \non \\
 A(B^-\to\pi^0 K^-) &=& {1\over\sqrt{2}}(\T+\C+\P+\P_{\rm
 EW}+{2\over 3}\P'_{\rm EW}+\A+\P_A), \non
 \en
where $\P_{\rm EW}$ and $\P'_{\rm EW}$ are color-allowed and
color-suppressed electroweak penguin amplitudes, respectively, and
$\P_A$ is the penguin-induced weak annihilation amplitude. The
decay amplitudes satisfy the isospin relation
 \be
 A(\ov B^0\to\pi^+K^-)+\sqrt{2}A(\ov B^0\to\pi^0\ov
 K^0)=-A(B^-\to\pi^-\ov K^0)+\sqrt{2}A(B^-\to\pi^0 K^-).
 \en
Likewise, a similar isospin relation holds for charge conjugate
fields
 \be
 A(B^0\to\pi^-K^+)+\sqrt{2}A(B^0\to\pi^0\ov
 K^0)=-A(B^+\to\pi^+K^0)+\sqrt{2}A(B^+\to\pi^0 K^+).
 \en

In the factorization approach \cite{Ali,CCTY},
 \be
 \T_{\rm SD} &=& \kappa_1\lambda_ua_1, \qquad  \C_{\rm SD} = \kappa_2\lambda_u a_2,
 \qquad  \P_{\rm SD}=\kappa_1\Big[\lambda_u(a_4^u+a_6^u
 r_\chi^K)+\lambda_c(a_4^c+a_6^c r_\chi^K)\Big], \non \\  \P^{\rm SD}_{\rm
 EW} &=& {3\over 2}\kappa_2(\lambda_u+\lambda_c)(-a_7+a_9),  \quad \P'^{\rm SD}_{\rm
 EW}={3\over 2}\kappa_1\Big[\lambda_u(a_8^u r_\chi^K+a_{10}^u)+\lambda_c(a_8^c
 r_\chi^K+a_{10}^c)\Big],
 \en
with $\lambda_q\equiv V_{qs}^*V_{qb}$,
$r_\chi^K=2m_K^2/[m_b(\mu)(m_s+m_q)(\mu)]$, and
 \be
 \kappa_1=i{G_F\over\sqrt{2}}f_K F_0^{B\pi}(m_K^2)(m_B^2-m_\pi^2),
 \qquad \kappa_2=i{G_F\over\sqrt{2}}f_\pi
 F_0^{BK}(m_\pi^2)(m_B^2-m_K^2).
 \en
The parameters $a_i^{u,c}$ can be calculated in the QCD
factorization approach \cite{BBNS}. They are basically the Wilson
coefficients in conjunction with short-distance nonfactorizable
corrections such as vertex corrections and hard spectator
interactions. Formally, $a_i(i\neq 6,8)$ and $a_{6,8}r_\chi^K$
should be renormalization scale and scheme independent. In
practice, there exists some residual scale dependence in
$a_i(\mu)$ to finite order. At the scale $\mu=2.1$ GeV, the
numerical results are
 \be \label{eq:ai}
  &&a_1 = 0.9921 + i0.0369, \qquad
  a_2 = 0.1933 - i0.1130, \qquad
  a_3 = -0.0017 + i0.0037, \non \\
  && a_4^u = -0.0298 - i0.0205, \qquad
  a_4^c = -0.0375 - i0.0079, \qquad
  a_5 = 0.0054 - i0.0050, \non \\
  && a_6^u = -0.0586 -i 0.0188, \qquad
  a_6^c = -0.0630 -i 0.0056, \qquad
  a_7 = i5.4\times 10^{-5}, \non \\
  && a_8^u = (45.0 - i5.2)\times 10^{-5}, \qquad
  a_8^c = (44.2 + i3.1)\times 10^{-5}, \qquad
  a_9 = -(953.9 + i24.5 )\times 10^{-5}, \non \\
  && a_{10}^u = (-58.3+i 86.1  )\times 10^{-5}, \qquad
  a_{10}^c = (-60.3 + i88.8 )\times 10^{-5}.
  \en
For current quark masses, we use $m_b(m_b)=4.4$ GeV,
$m_c(m_b)=1.3$ GeV, $m_s(2.1\,{\rm GeV})=90$ MeV and
$m_q/m_s=0.044$.

Using the above coefficients $a_i^{u,c}$ leads to
 \be
 \B(B^-\to \pi^-\ov K^0)_{\rm SD} &=& 17.8\times 10^{-6}, \qquad
 \A^{\rm SD}_{\pi^-K_S}=0.01,  \non \\
 \B(\ov B^0\to \pi^+K^-)_{\rm SD} &=& 13.9\times 10^{-6}, \qquad
 \A^{\rm SD}_{\pi^+K^-}=0.04, \non \\
 \B(B^-\to \pi^0K^-)_{\rm SD} &=& 9.7\times 10^{-6}, \qquad~
 \A^{\rm SD}_{\pi^0K^-}=0.08,  \non \\
 \B(\ov B^0\to \pi^0\ov K^0)_{\rm SD} &=& 6.3\times 10^{-6}, \qquad~
 \A^{\rm SD}_{\pi^0K_S}=-0.04\,,
 \en
where we have used $|V_{cb}|=0.041$, $|V_{ub}/V_{cb}|=0.09$ and
$\gamma=60^\circ$ for quark mixing matrix elements.\footnote{The
notations $(\alpha,\beta,\gamma)$ and $(\phi_1,\phi_2,\phi_3)$
with $\alpha\equiv\phi_2,\beta\equiv \phi_1,\gamma\equiv \phi_3$
are in common usage for the angles of the unitarity triangle.}
From Table \ref{tab:piK} it is clear that the predicted branching
ratios are slightly smaller than experiment especially for $\ov
B^0\to\pi^0 \ov K^0$ where the prediction is about two times
smaller than the data. Moreover, the predicted \CP asymmetry for
$\pi^+K^-$ is opposite to the experimental measurement in sign. By
now, direct \CP violation in $\ov B^0\to\pi^+K^-$ has been
established by both BaBar \cite{BaBarKpi} and Belle
\cite{BelleKpi}.

\begin{figure}[t]
\vspace{-1cm}
  \centerline{\psfig{figure=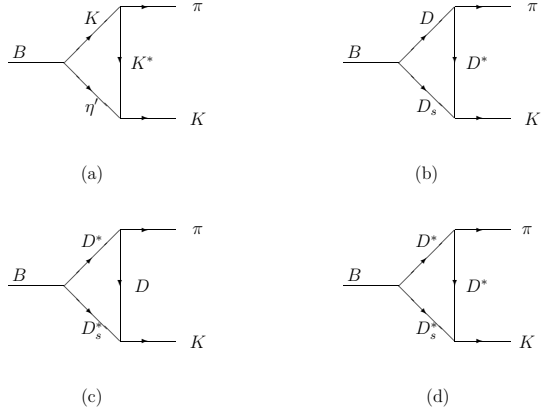,width=12cm}}
\vspace{-10cm}
    \caption[]{\small Long-distance $t$-channel rescattering contributions
    to $B\to K\pi$.  }\label{fig:BKpi}
\end{figure}

Leading long-distance rescattering contributions to $B\to K\pi$
are shown in Fig. \ref{fig:BKpi}. The absorptive amplitudes are
  \be
 \A bs\,(5a) &=& -\int_{-1}^1 {|\vec p_1|d\cos\theta\over 16\pi
 m_B}\,g_{K^*K\pi}\,g_{K^*K\eta'}
 A(\ov B^0\to \ov K^0\eta')\,{F^2(t,m_\rho)\over
 t-m_{K^*}^2}\,H_1, \non \\
 \A bs\,(5b) &=&
 \int_{-1}^1 {|\vec p_1|d\cos\theta\over 16\pi
 m_B}\,g_{D^*D\pi}\,g_{D^*D_sK}
 A(\ov B^0\to D^+D_s^-){F^2(t,m_{D^*})\over t-m_{D^*}^2}\,J_2, \\
  \A bs\,(5c)&=& -i{G_F\over\sqrt{2}}V_{cb}V_{cs}^*f_{D_s^*} m_{D_s^*}\int_{-1}^1
{|\vec p_1|d\cos\theta\over 16\pi m_B}\,g_{D^*D\pi}\,g_{D_s^*DK}\,
{F^2(t,m_D) \over t-m_D^2} \non
\\ &\times& \Big[(m_B+m_{D^*})A_1^{BD^*}(m_{D_s^*}^2)H_4-{2A_2^{BD^*}(m_{D_s^*}^2)\over
  m_B+m_{D^*}}H'_4\Big], \non \\
 \A bs\,(5d)&=& i{G_F\over\sqrt{2}}V_{cb}V_{cs}^*f_{D_s^*}m_{D_s^*}\int_{-1}^1
{|\vec p_1|d\cos\theta\over 16\pi
m_B}\,g_{D^*D^*\pi}\,g_{D_s^*DK}\, {F^2(t,m_{D^*}) \over
t-m_{D^*}^2} \non
\\ &\times& \Big[(m_B+m_{D^*})A_1^{BD^*}(m_{D_s^*}^2)J_5-{2A_2^{BD^*}(m_{D_s^*}^2)\over
m_B+m_{D^*}}J'_5\Big],  \non
 \en
with
 \be
 J_2=-p_3\cdot p_4-{(p_1\cdot p_3-m_3^2)(p_2\cdot p_4-m_4^2)\over
 m_{D^*}^2},
 \en
and $J_5$, $J'_5$ can be obtained from $H_5$ and $H'_5$,
respectively, under the replacement of $p_3\leftrightarrow p_4$.

Hence,
 \be
 A(B^-\to \pi^-\ov K^0) &=& A(B^-\to \pi^-\ov K^0)_{\rm SD}+i\A
 bs(5a+5b+5c+5d),  \non \\
 A(\ov B^0\to \pi^+K^-) &=& A(\ov B^0\to \pi^+K^-)_{\rm SD}+i\A
 bs(5a+5b+5c+5d),  \non \\
 A(B^-\to \pi^0 K^-) &=& A(B^-\to \pi^0 K^-)_{\rm SD}+{i\over\sqrt{2}}\A
 bs(5a+5b+5c+5d),  \non \\
 A(\ov B^0\to \pi^0\ov K^0) &=& A(\ov B^0\to \pi^0\ov K^0)_{\rm SD}-{i\over\sqrt{2}}\A
 bs(5a+5b+5c+5d).
 \en
As noticed before, the rescattering diagrams 5(b)-5(d) with charm
intermediate states contribute only to the topological penguin
graph. Indeed, it has been long advocated that charming-penguin
long-distance contributions increase significantly the $B\to K\pi$
rates and yield better agreement with experiment
\cite{charmpenguin}. For a recent work along this line, see
\cite{Isola2003}.

To proceed with the numerical calculation, we use the experimental
value $A(\ov B^0\to\ov K^0\eta')=(-7.8+i4.5)\times 10^{-8}$ GeV
with the phase determined from the topological approach
\cite{Chiang}. For strong couplings, $g_{K^*K\pi}$ defined for
$K^{*+}\to K^0\pi^+$ is fixed to be 4.6 from the measured $K^*$
width, and the $K^*K\eta'$ coupling is fixed to be
 \be
 g_{K^{*+}\to  K^+\eta'}={1\over\sqrt{3}}g_{K^{*+}\to
 K^+\pi^0}={1\over \sqrt{6}}g_{K^{*+}\to K^0\pi^+}
 \en
by approximating the $\eta'$ wave function by
$\eta'={1\over\sqrt{6}}(u\bar u+d\bar d+2s\bar s)$. A $\chi^2$ fit
to the measured decay rates yields $\eta_{D^{(*)}}=0.69$ for the
exchanged particle $D$ or $D^*$, while $\eta_{K^*}$ is less
constrained as the exchanged $K^*$ particle contribution [Fig.
5(a)] is suppressed relative to the $D$ or $D^*$ contribution.
Since the decay $B\to D\ov D_s$ is quark-mixing favored over $B\to
D\pi$, the cutoff scale here thus should be lower than that in
$B\to D\pi$ decays where $\eta_{D^{(*)}}=2.2$ as it should be. As
before, we allow 15\% uncertainty for the $\Lambda_{\rm QCD}$
scale and obtain the branching ratios and \CP asymmetries as shown
in Table \ref{tab:comparepiK}, where use has been made of
$f_{D^*_s}\approx f_{D_s}=230$ MeV and the heavy quark symmetry
relations
 \be
 g_{D^*D_sK}=\sqrt{m_{D_s}\over m_D}\,g_{D^*D\pi}, \qquad
 g_{D_s^*DK}=\sqrt{m_{D^*_s}\over m_{D^*}}\,g_{D^*D\pi}.
 \label{eq:SU3breaking}
 \en
It is found that the absorptive contribution from the $K\eta'$
intermediate states is suppressed relative to $D^{(*)}\ov
D^{(*)}_s$ as the latter are quark-mixing-angle most favored, i.e.
$B\to D^{(*)}\ov D_s^{(*)}\gg B\to K\eta'$. Evidently, all $\pi K$
modes are sensitive to the cutoffs $\Lambda_{D^*}$ and
$\Lambda_D$. We see that the sign of the direct \CP partial rate
asymmetry for $\ov B^0\to\pi^+K^-$ is flipped after the inclusion
of rescattering and the predicted
$\A_{\pi^+K^-}=-0.14^{+0.01}_{-0.03}$ is now in good agreement
with the world average of $-0.11\pm0.02$. Note that the branching
ratios for all $\pi K$ modes are enhanced by $(30\sim 40)\%$ via
final-state rescattering.

\begin{table}[t]
\caption{ \label{tab:comparepiK} Comparison of experimental data
and fitted outputs for \CP averaged branching ratios (top, in
units of $10^{-6}$) and the theoretical predictions of \CP
asymmetries (bottom) for $B\to\pi K$.}
\begin{ruledtabular}
\begin{tabular}{c r r r}
Mode & Expt. & SD & SD+LD \\
\hline
 ${\mathcal B}(B^-\to \pi^-\ov K^0)$
   & $24.1\pm1.3$
   & 17.8
   & $23.3^{+4.6}_{-3.7}$
   \\
 ${\mathcal B}(\ov B^0\to \pi^+K^-)$
   & $18.2\pm0.8$
   & 13.9
   & $19.3^{+5.0}_{-3.1}$
   \\
 ${\mathcal B}(B^-\to \pi^0K^-)$
   & $12.1\pm0.8$
   & 9.7
   & $12.5^{+2.6}_{-1.6}$
   \\
 ${\mathcal B}(\ov B^0\to\pi^0\ov K^0)$
   & $11.5\pm1.0$
   & 6.3
   & $9.1^{+2.5}_{-1.6}$
   \\
 \hline
 $\A_{\pi^-K_S}$
   & $-0.02\pm0.03$
   & 0.01
   & $0.024^{+0.0}_{-0.001}$
   \\
 $\A_{\pi^+K^-}$
   & $-0.11\pm0.02$
   & 0.04
   & $-0.14^{+0.01}_{-0.03}$
   \\
 $\A_{\pi^0K^-}$
   & $0.04\pm0.04$
   & 0.08
   & $-0.11^{+0.02}_{-0.04}$
   \\
 $\A_{\pi^0K_S}$
   & $-0.09\pm0.14$
   & $-0.04$
   & $0.031^{+0.008}_{-0.014}$
   \\
 ${\cal S}_{\pi^0K_S}$
   & $0.34\pm0.28$
   & 0.79
   & $0.78\pm0.01$
   \\
\end{tabular}
\end{ruledtabular}
\end{table}

Note that there is a sum-rule relation \cite{AS98}
 \be
 2\Delta\Gamma(\pi^0K^-)-\Delta\Gamma(\pi^-\ov
 K^0)-\Delta\Gamma(\pi^+K^-)+2\Delta\Gamma(\pi^0\ov K^0)=0,
 \en
based on isospin symmetry. Hence, a violation of the above
relation would provide an important test for the presence of
electroweak penguin contributions. It is interesting to check how
good the above relation works in light of the present data. We
rewrite it as
 \be
 \frac{\A_{\pi^+ K^-} \B(\pi^+ K^-)-2 \A_{\pi^0 \ov K^0} \B(\pi^0 \ov K^0)}
      {2 \A_{\pi^0 K^-} \B(\pi^0 K^-)-\A_{\pi^- \ov K^0} \B(\pi^- \ov K^0)}
 =\frac{\tau_{\overline B^0}}{\tau_{B^-}}.
 \en
The left hand side of the above equation yields $0.07\pm2.28$,
while the right hand side is $0.923\pm0.014$~\cite{PDG}. Although
the central values of the data seem to imply the need of
non-vanishing electroweak penguin contributions and/or some New
Physics, it is not conclusive yet with present data.
For further implications of this sum-rule
relation, see \cite{AS98}.

\section{$B\to \pi\pi$ Decays}

\subsection{Short-distance contributions and Experimental Status}

The experimental results of \CP averaged branching ratios and \CP
asymmetries for charmless $B\to\pi\pi$ decays are summarized in
Table \ref{tab:Bpipi}. For a neutral $B$ meson decay into a \CP
eigenstate $f$, \CP asymmetry is defined by
 \be
 {\Gamma(\ov B(t)\to f)-\Gamma(B(t)\to f)\over
 \Gamma(\ov B(t)\to f)+\Gamma(B(t)\to
 f)}={\cal S}_f\sin(\Delta mt)-\C_f\cos(\Delta mt),
 \en
where $\Delta m$ is the mass difference of the two neutral $B$
eigenstates, ${\cal S}_f$ is referred to as mixing-induced \CP
asymmetry and $\A_f=-\C_f$ is the direct \CP asymmetry. Note that
the Belle measurement \cite{Belle2pi0} gives an evidence of \CP
violation in $B^0\to \pi^+\pi^-$ decays at the level of 5.2
standard deviations, while this has not been confirmed by BaBar
\cite{BaBarpipi}. As a result, we follow Particle Data Group
\cite{PDG} to use a scale factor of 2.2 and 2.5, respectively, for
the error-bars in $\A_{\pi^+\pi^-}$ and $S_{\pi^+\pi^-}$ (see
Table \ref{tab:Bpipi}). The $CP$-violating parameters $\C_f$ and
$\S_f$ can be expressed as
 \be
 \C_f={1-|\lambda_f|^2\over 1+|\lambda_f|^2}, \qquad \S_f={2\,{\rm
 Im}\lambda_f\over 1+|\lambda_f|^2},
 \en
where
 \be
 \lambda_f={q\over p}\,{A(\ov B^0\to f)\over A(B^0\to f)}.
 \en
For $\pi\pi$ modes, $q/p=e^{-i2\beta}$ with $\sin
2\beta=0.726\pm0.037$ \cite{Ligeti}.

\begin{table}[t]
\caption{Experimental data for \CP averaged branching ratios (top,
in units of $10^{-6}$) and \CP asymmetries (bottom) for
$B\to\pi\pi$ \cite{HFAG,PDG}.} \label{tab:Bpipi}
\begin{ruledtabular}
\begin{tabular}{c c c c c}
Mode & BaBar & Belle & CLEO & Average  \\
\hline
 $\ov B^0\to \pi^+\pi^-$ & $4.7\pm0.6\pm0.2$ & $4.4\pm0.6\pm0.3$ &
 $4.5^{+1.4+0.5}_{-1.2-0.4}$ & $4.6\pm0.4$ \\
 $\ov B^0\to\pi^0\pi^0$ &  $1.17\pm0.32\pm0.10$ & $2.32^{+0.44+0.22}_{-0.48-0.18}$ &
 $<4.4$ & $1.51\pm0.28$ \\
 $B^-\to \pi^-\pi^0$ & $5.8\pm0.6\pm0.4$ &
 $5.0\pm1.2\pm0.5$ & $4.6^{+1.8+0.6}_{-1.6-0.7}$ &
 $5.5\pm0.6$\\
 \hline
  $\A_{\pi^+\pi^-}$ & $0.09\pm0.15\pm0.04$ & $0.58\pm0.15\pm0.07$ & &
  $0.31\pm0.24$\footnotemark[1]    \\
 ${\cal S}_{\pi^+\pi^-}$ & $-0.30\pm0.17\pm0.03$ & $-1.00\pm0.21\pm0.07$
 & & $-0.56\pm0.34$\footnotemark[1] \\
 $\A_{\pi^0\pi^0}$ & $0.12\pm0.56\pm0.06$ &
 $0.43\pm0.51^{+0.17}_{-0.16}$ & & $0.28\pm0.39$
 \\
 $\A_{\pi^-\pi^0}$ & $-0.01\pm0.10\pm0.02$ &
 $-0.02\pm0.10\pm0.01$ & & $-0.02\pm0.07$ \\
\end{tabular}
\footnotetext[1]{Error-bars of $\A_{\pi^+\pi^-}$ and
$S_{\pi^+\pi^-}$ are scaled by $S=2.2$ and 2.5, respectively. When
taking into account the measured correlation between $\A$ and $S$,
the averages are cited by Heavy Flavor Averaging Group \cite{HFAG}
to be $\A_{\pi^+\pi^-}=0.37\pm0.11$ and
$S_{\pi^+\pi^-}=-0.61\pm0.14$ with errors not being scaled up.}
\end{ruledtabular}
\end{table}

The $B\to \pi\pi$ decay amplitudes have the general expressions
 \be \label{eq:ampBpipi}
 A(\ov B^0\to\pi^+\pi^-) &=& \T+\P+{2\over 3}\P'_{\rm EW}+\E+\P_A+\V, \non \\
 A(\ov B^0\to\pi^0\pi^0) &=& -{1\over\sqrt{2}}(\C-\P+\P_{\rm EW}+{1\over 3}\P'_{\rm
 EW}-\E-\P_A-\V), \\
 A(B^-\to \pi^-\pi^0) &=& {1\over\sqrt{2}}(\T+\C+\P_{\rm EW}+\P'_{\rm
 EW}), \non
 \en
where $\V$ is the vertical $W$-loop topological diagram (see Sec.
II). In the factorization approach, the quark diagram amplitudes
are given by \cite{Ali,CCTY}
 \be \label{eq:qcdfBpipi}
 \T_{\rm SD} &=& \kappa\lambda_ua_1, \qquad  \C_{\rm SD} = \kappa\lambda_u
 a_2,\qquad  \P_{\rm SD}=\kappa\Big[\lambda_u(a_4^u+a_6^u
 r_\chi^\pi)+\lambda_c(a_4^c+a_6^c r_\chi^\pi)\Big], \non \\
  \P_{\rm
 EW}^{\rm SD} &=& {3\over 2}\kappa(\lambda_u+\lambda_c)(-a_7+a_9),  \quad \P'^{\rm SD}_{\rm
 EW}={3\over 2}\kappa\Big[\lambda_u(a_8^u r_\chi^\pi+a_{10}^u)+\lambda_c(a_8^c
 r_\chi^\pi+a_{10}^c)\Big],
 \en
where $\lambda_q\equiv V^*_{qd}V_{qb}$, $\kappa\equiv
i{G_F\over\sqrt{2}}f_\pi F_0^{B\pi}(m_\pi^2)(m_B^2-m_\pi^2)$ and
the chiral enhancement factor
$r_\chi^\pi=2m_\pi^2/[m_b(\mu)(m_u+m_d)(\mu)]$ arises from the
$(S-P)(S+P)$ part of the penguin operator $Q_6$. The explicit
expressions of the weak annihilation amplitude $\E$ in QCD
factorization can be found in \cite{BBNS}. It is easily seen that
the decay amplitudes satisfy the isospin relation
%
 \be \label{eq:isospinBpipi}
 A(\ov B^0\to \pi^+\pi^-)=\sqrt{2}\,\Big[A(B^-\to
 \pi^-\pi^0)+A(\ov B^0\to\pi^0\pi^0)\Big].
 \en
Note that there is another isospin relation for charge conjugate
fields
  \be
 A(B^0\to \pi^+\pi^-)=\sqrt{2}\,\Big[A(B^+\to
 \pi^+\pi^0)+A(B^0\to\pi^0\pi^0)\Big].
 \en

Substituting the coefficients $a_i^{u,c}$ given in Eq.
(\ref{eq:ai}) into Eqs. (\ref{eq:ampBpipi}) and
(\ref{eq:qcdfBpipi}) leads to\footnote{Charge conjugate modes are
implicitly included in the flavor-averaged branching ratios
throughout this paper.}
 \be \label{eq:BpipiSD}
 \B(\ov B^0\to\pi^+\pi^-)_{\rm SD} &=& 7.6\times 10^{-6}, \qquad
 \A_{\pi^+\pi^-}^{\rm SD}=-0.05, \non \\
 \B(\ov B^0\to\pi^0\pi^0)_{\rm SD} &=& 2.7\times 10^{-7}, \qquad
 \A_{\pi^0\pi^0}^{\rm SD}=0.61,  \\
 \B(B^-\to\pi^-\pi^0)_{\rm SD} &=& 5.1\times 10^{-6}, \qquad
 \A_{\pi^-\pi^0}^{\rm SD}=5\times 10^{-5}. \non
 \en
To obtain the above branching ratios and \CP asymmetries we have
applied the form factor $F_0^{B\pi}(0)=0.25$ as obtained from the
covariant light-front approach \cite{CCH} and neglected the
$W$-annihilation contribution. Nevertheless, the numerical results
are similar to that in \cite{BN} (see Tables 9 and 10 of
\cite{BN}) except for $\A_{\pi^-\pi^0}$ for which we obtain a sign
opposite to \cite{BN}.

We see that the predicted $\pi^+\pi^-$ rate is too large, whereas
$\pi^0\pi^0$ is too small compared to experiment (see Table
\ref{tab:Bpipi}). Furthermore, the predicted value of direct \CP
asymmetry, $A_{\pi^+ \pi^-}^{SD} = -0.05$, is in disagreement with
the experimental average of $0.31 \pm 0.24$. \footnote{In the
so-called perturbative QCD approach for nonleptonic $B$ decays
\cite{Keum}, the signs of $\pi^+\pi^-$ and $K^-\pi^+$ \CP
asymmetries are correctly predicted and the calculated magnitudes
are compatible with experiment.}
In the next subsection, we will study the long-distance
rescattering effect to see its impact on \CP violation.

In the topological analysis in \cite{Chiang}, it is found that in
order to describe $B\to\pi\pi$ and $B\to\pi K$ branching ratios
and \CP asymmetries, one must introduce a large value of $|\C/\T|$
and a non-trivial phase between $\C$ and $\T$. Two different fit
procedures have been adopted in \cite{Chiang} to fit $\pi\pi$ and
$\pi K$ data points by assuming negligible weak annihilation
contributions. One of the fits yields\footnote{The other fit
yields $|\C/\T|=1.43^{+0.40}_{-0.31}$ which is unreasonably too
large. This ratio is substantially reduced once the up and top
quark mediated penguins are taken into account \cite{Chiang}.}
 \be \label{eq:C/TinQDA}
 \gamma=(65^{+13}_{-35})^\circ, \qquad \left.{\C\over \T}
 \right|_{\pi\pi,\pi K}=(0.46^{+0.43}_{-0.30})
 \,{\rm exp}[-i(94^{+43}_{-52})]^\circ.
 \en
The result for the ratio $\C/\T$ in charmless $B$ decays is thus
consistent with (\ref{eq:BDpiQDvalue}) extracted from $B\to D\pi$.

\subsection{Long-distance contributions to $B\to \pi\pi$}

\begin{figure}[t]
\vspace{-1cm}
  \centerline{\psfig{figure=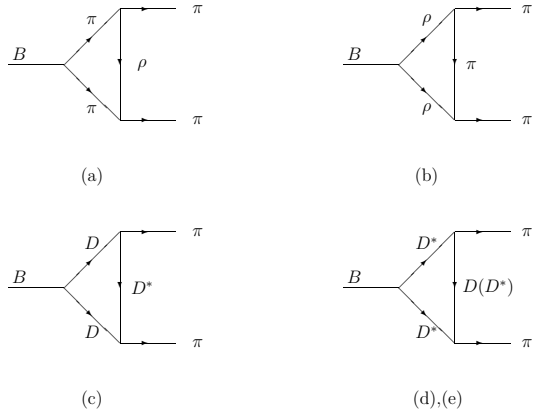,width=12cm}}
\vspace{-10cm}
    \caption[]{\small Long-distance $t$-channel rescattering contributions
    to $B\to \pi\pi$. Graphs (d) and (e) correspond to the exchanged
    particles $D$ and $D^*$, respectively. }\label{fig:Bpipi}
\end{figure}

\begin{figure}[t]

\centerline{
            {\epsfxsize3 in \epsffile{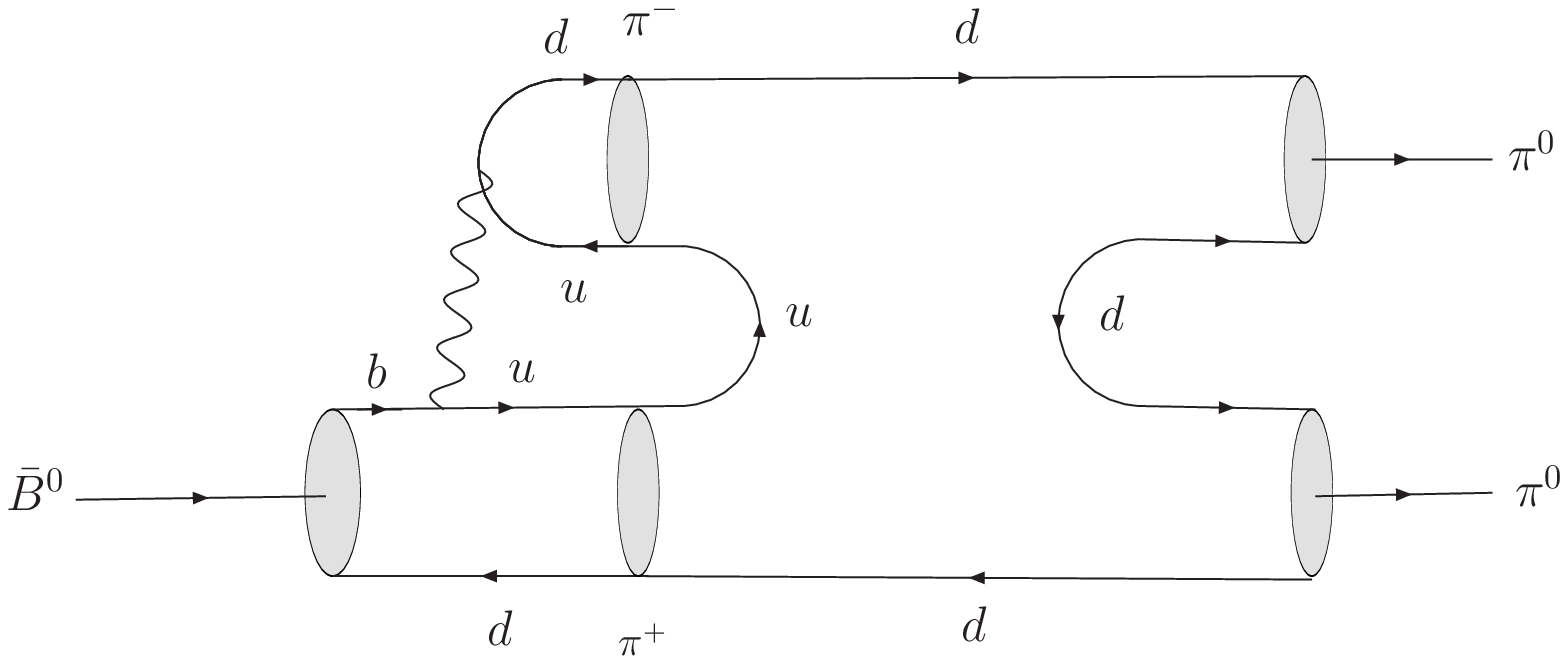}}{\epsfxsize3 in \epsffile{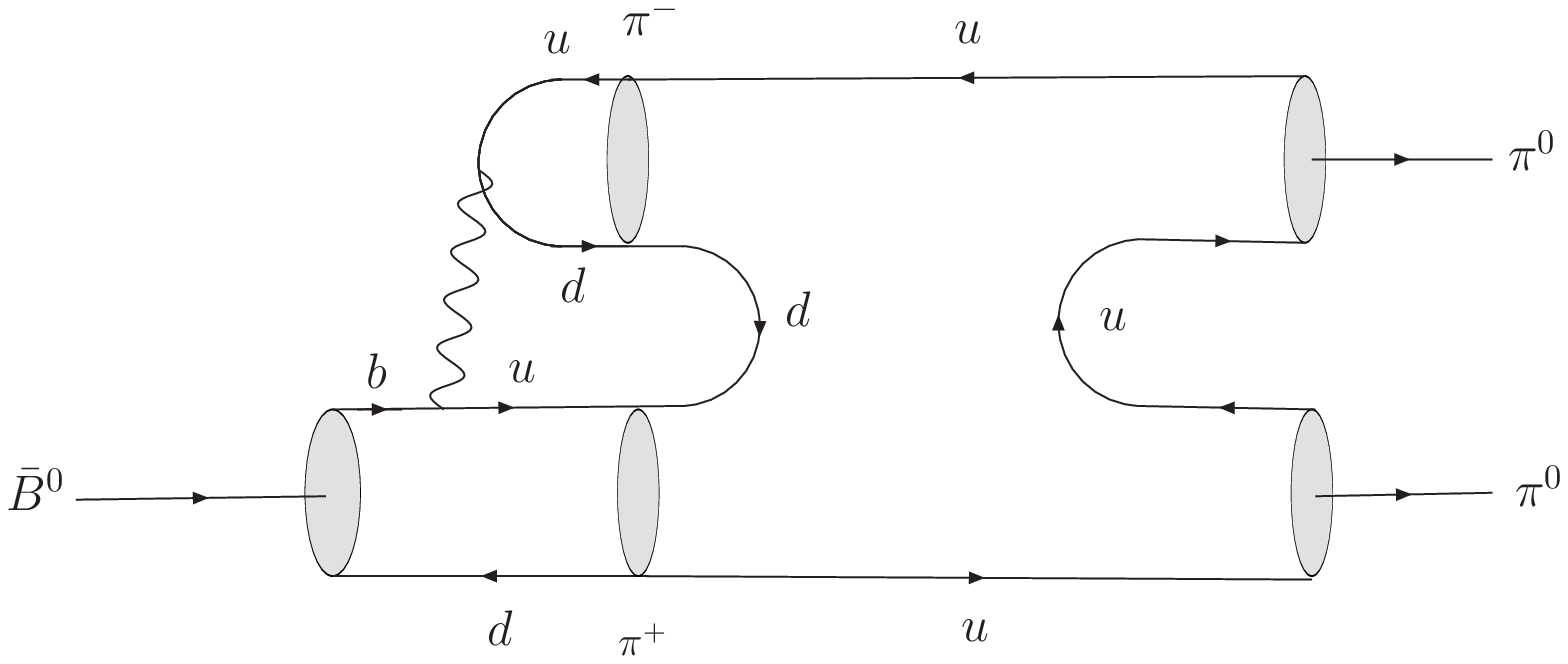}}
            }
\centerline{(a)\hspace{7.5cm} (b)\hspace{-1cm}}
 \caption{Contributions to $\ov B^0\to \pi^0\pi^0$ from
    the color-allowed weak decay $\ov B^0\to \pi^+\pi^-$ followed
    by quark annihilation processes (a) and (b). They have the same topologies
    as the penguin and $W$-exchange graphs, respectively.} \label{fig:Bpipis}
\end{figure}

Some leading rescattering contributions to $B\to\pi\pi$ are shown
in Figs. \ref{fig:Bpipi}(a)-\ref{fig:Bpipi}(e). While all the five
diagrams contribute to $\ov B^0\to \pi^+\pi^-$ and $\ov
B^0\to\pi^0\pi^0$, only the diagrams \ref{fig:Bpipi}(a) and
\ref{fig:Bpipi}(b) contribute to $ B^-\to \pi^-\pi^0$. Since the
$\pi^-\pi^0$ final state in $B^-$ decay must be in $I=2$, while
the intermediate $D\ov D$ state has $I=0$ or $I=1$, it is clear
that $B^-\to\pi^-\pi^0$ cannot receive rescattering from $B^-\to
D^-D^0\to\pi^-\pi^0$. It should be stressed that all the graphs in
Fig. \ref{fig:Bpipi} contribute to the penguin amplitude. To see
this, let us consider Fig. \ref{fig:Bpipis} which is one of the
manifestations of Fig. \ref{fig:Bpipi}(a) at the quark level.
(Rescattering diagrams with quark exchange are not shown in Fig.
\ref{fig:Bpipis}.) It is easily seen that Figs.
\ref{fig:Bpipis}(a) and \ref{fig:Bpipis}(b) can be redrawn as the
topologies $\P$ and $\E$, respectively. Likewise, Figs.
\ref{fig:Bpipi}(c)-(e) correspond to the topological $\P$.
Consequently,
 \be
 \T &=& \T_{\rm SD}, \non \\
 \C &=& \C_{\rm SD}+i\A bs\,(6a+6b), \\
 \E &=& \E_{\rm SD}+i\A bs\,(6a+6b), \non \\
 \P &=& \P_{\rm SD}+i\A bs\,(6a+6b+6c+6d+6e). \non
 \en
Hence,\footnote{In a similar context, it has been argued in
\cite{Ablikim} that the analogous Figs. \ref{fig:Bpipi}(a)-(b) do
not contribute to the decay $D^0\to\pi^0\pi^0$ owing to the
cancellation between the quark annihilation and quark exchange
diagrams at the quark level. However, a careful examination
indicates that the quark annihilation processes gain an additional
factor of 2 as noticed in \cite{FSIChua2}. Consequently, the
intermediate $\pi^+\pi^-$ state does contribute to
$B^0\to\pi^0\pi^0$ and $D^0\to\pi^0\pi^0$ via final-state
rescattering.}
 \be
 A(\ov B^0\to\pi^+\pi^-) &=& A(\ov B^0\to\pi^+\pi^-)_{\rm SD}+2i\A
 bs\,(6a+6b)+i\A bs(6c+6d+6e),  \non \\
 A(\ov B^0\to\pi^0\pi^0) &=& A(\ov B^0\to\pi^0\pi^0)_{\rm SD}+{1\over\sqrt{2}}i\A
 bs\,(6a+6b+6c+6d+6e),   \\
 A(B^-\to\pi^-\pi^0) &=& A(B^-\to\pi^-\pi^0)_{\rm SD}+{1\over\sqrt{2}}i\A
 bs\,(6a+6b),  \non
 \en

The absorptive parts of $B\to M_1(p_1)M_2(p_2)\to
\pi(p_3)\pi(p_4)$ with $M_{1,2}$ being the intermediate states in
Fig. \ref{fig:Bpipi} are
 \be
 \A bs\,(6a) &=& -\int_{-1}^1 {|\vec p_1|d\cos\theta\over 16\pi m_B}\,g_{\rho\pi\pi}^2
 A(\ov B^0\to\pi^+\pi^-)\,{F^2(t,m_\rho)\over
 t-m_\rho^2+im_\rho\Gamma_\rho}\,H_1, \non \\
   \A bs\,(6b) &=& -i{G_F\over\sqrt{2}}V_{ub}V_{ud}^*f_\rho m_\rho\int_{-1}^1
{|\vec p_1|d\cos\theta\over 16\pi m_B}\,g_{\rho\pi\pi}^2\,
{F^2(t,m_\pi) \over t-m_\pi^2} \non
\\ &\times& \Big[(m_B+m_\rho)A_1^{B\rho}(m_\pi^2)4H_3-{2A_2^{B\rho}(m_\pi^2)\over
m_B+m_\rho}4H'_3\Big], \non\\
 \A bs\,(6c) &=&
 \int_{-1}^1 {|\vec p_1|d\cos\theta\over 16\pi m_B}\,g_{D^*D\pi}^2
 A(\ov B^0\to D^+D^-){F^2(t,m_{D^*})\over t-m_{D^*}^2}\,J_2, \\
  \A bs\,(6d)&=& -i{G_F\over\sqrt{2}}V_{cb}V_{cd}^*f_{D^*} m_{D^*}\int_{-1}^1
{|\vec p_1|d\cos\theta\over 16\pi m_B}\,g_{D^*D\pi}^2\,
{F^2(t,m_D) \over t-m_D^2} \non
\\ &\times& \Big[(m_B+m_{D^*})A_1^{BD^*}(m_{D^*}^2)H_4-{2A_2^{BD^*}(m_{D^*}^2)\over
  m_B+m_{D^*}}H'_4\Big], \non \\
 \A bs\,(6e)&=& i{G_F\over\sqrt{2}}V_{cb}V_{cd}^*f_{D^*}m_{D^*}\int_{-1}^1
{|\vec p_1|d\cos\theta\over 16\pi m_B}\,g_{D^*D^*\pi}^2\,
{F^2(t,m_{D^*}) \over t-m_{D^*}^2} \non
\\ &\times& \Big[(m_B+m_{D^*})A_1^{BD^*}(m_{D^*}^2)J_5-{2A_2^{BD^*}(m_{D^*}^2)\over
m_B+m_{D^*}}J'_5\Big].  \non
 \en

Since in SU(3) flavor limit the final-state rescattering parts of
Figs. \ref{fig:Bpipi}(c)-(e) should be the same as that of Figs.
\ref{fig:BKpi}(b)-(d), it means that the cutoff scales
$\Lambda_{D}$ and $\Lambda_{D^*}$ appearing in the rescattering
diagrams of $B\to \pi\pi$ are identical to that in $B\to \pi K$
decays in the limit of SU(3) symmetry. Allowing 20\% SU(3)
breaking and recalling that $\eta_{D^{(*)}}=0.69$ in $B\to \pi K$
decays, we choose $\eta_{D^{(*)}}=\eta_\rho=0.83$ for $B\to
\pi\pi$ decays. Because the decay $\ov B^0\to D^+D^-$ is Cabibbo
suppressed relative to $\ov B^0\to D^+D_s^-$, it is evident that
the absorptive part of the charm loop diagrams, namely, Figs.
\ref{fig:Bpipi}(c)-(e), cannot enhance the $\pi^0\pi^0$ rate
sizably. However, since the branching ratio of the short-distance
induced $B^0\to\pi^+\pi^-$, namely $7.6\times 10^{-6}$ [see Eq.
(\ref{eq:BpipiSD})], already exceeds the measured value
substantially, this means that a dispersive part of the
long-distance rescattering amplitude must be taken into account.
This dispersive part also provides the main contribution to the
$\pi^0\pi^0$ rate. There is a subtle point here: If the dispersive
contribution is fixed by fitting to the measured $B\to \pi\pi$
rates, the $\chi^2$ value of order 0.2 is excellent and the
predicted direct \CP violation for $\pi^+\pi^-$ agrees with
experiment. Unfortunately, the calculated mixing-induced
$CP$-violating parameter $S$ will have a wrong sign. Therefore, we
need to accommodate the data of branching ratios and the parameter
$S$ simultaneously. We find (in units of GeV)
 \be \label{eq:pipiDis}
 {\cal D}is\,A=1.5\times 10^{-6}V_{cb}V^*_{cd}-6.7\times
 10^{-7}V_{ub}V_{ud}^*,
 \en
for $\eta_\pi=\eta_\rho=1.2$. It should be stressed that these
dispersive contributions cannot arise from the rescattering
processes in Figs. \ref{fig:Bpipi}(a)-(e) because they will also
contribute to $B\to K\pi$ decays via SU(3) symmetry and modify all
the previous predictions very significantly. As pointed out in
\cite{FSIChua2}, there exist $\pi\pi\to \pi\pi$ and $D\bar
D\to\pi\pi$ meson annihilation diagrams in which the two initial
quark pairs in the zero isospin configuration are destroyed and
then created. Such annihilation diagrams have the same topology as
the vertical $W$-loop diagram ${\cal V}$ as mentioned in Sec. II.
Hence, this FSI mechanism occurs in $\pi^+\pi^-$ and $\pi^0\pi^0$
but not in the $K\pi$ modes. That is, the dispersive term given in
Eq. (\ref{eq:pipiDis}) does not contribute to $B\to K\pi$ decays.

The resultant amplitudes are then given by (in units of
GeV)\footnote{Note that $B\to PP$ and $B\to VV$ decay amplitudes
are accompanied by a factor of $i$ in our phase convention [see
Eq. (\ref{eq:T,C,E})].}
 \be
 A(\ov B^0\to\pi^+\pi^-) &=& (2.03+i2.02)\times
 10^{-8}+(1.40-i0.77)\times 10^{-8}-i1.64\times 10^{-8}, \non \\
 A(B^0\to\pi^+\pi^-) &=& (-2.21+i2.04)\times
 10^{-8}-(2.21-i2.05)\times 10^{-8}-i1.64\times 10^{-8}, \non \\
 A(\ov B^0\to\pi^0\pi^0) &=& (-6.15+i3.46)\times
 10^{-9}+(7.89-i2.73)\times 10^{-9}-i1.07\times 10^{-8}, \non \\
 A(B^0\to\pi^0\pi^0) &=& (3.32+i1.11)\times
 10^{-9}+(3.32+i1.11)\times 10^{-9}-i1.07\times 10^{-8}, \non \\
 A(B^-\to\pi^-\pi^0) &=& (2.05+i1.08)\times
 10^{-8}+(2.07-i2.73)\times 10^{-9}, \non \\
 A(B^+\to\pi^+\pi^0) &=& (-1.90+i1.34)\times
 10^{-8}-(1.90-i1.34)\times 10^{-8},
 \en
where the numbers in the first parentheses on the right hand side
are due to short-distance contributions, while the second term in
parentheses and the third term arise from the absorptive and
dispersive parts, respectively, of long-distance rescattering.
(The dispersive contribution to $B^-\to\pi^-\pi^0$ is too small
and can be neglected.) Form the above equation, it is clear that
the long-distance dispersive amplitude gives the dominant
contribution to $\pi^0\pi^0$ and contributes destructively with
the short-distance $\pi^+\pi^-$ amplitude. The corresponding \CP
averaged branching ratios and direct \CP asymmetries are shown in
Table \ref{tab:comparepipi}, where the ranges indicate the
uncertainties of the QCD scale $\Lambda_{\rm QCD}$ by 15\%.  We
only show the ranges due to the cut-off as numerical predictions
depend sensitively on it. It is evident that the $\pi^-\pi^0$ rate
is not significantly affected by FSIs, whereas the $\pi^+\pi^-$
and $\pi^0\pi^0$ modes receive sizable long-distance
corrections.\footnote{A different mechanism for understanding the
$\pi\pi$ data is advocated in \cite{FSIChua2} where FSI is studied
based on a simple two parameter model by considering a
quasi-elastic scattering picture. A simultaneous fit to the
measured rates {\it and} \CP asymmetries indicates that
$\pi^+\pi^-$ is suppressed by the dispersive part of
$\pi\pi\to\pi\pi$ rescattering, while $\pi^0\pi^0$ is enhanced via
the absorptive part. This requires a large SU(3) rescattering
phase difference of order $90^\circ\sim 100^\circ$. The fitted
$\pi^-\pi^0$ rate is somewhat below experiment. In the present
work, it is the dispersive part of the rescattering from $D\bar D$
to $\pi\pi$ via annihilation that plays the role for the
enhancement of $\pi^0\pi^0$ and suppression of $\pi^+\pi^-$.}

Since the branching ratio of $B^0\to D^+D^-$ is about 50 times
larger than that of $B^0\to \pi^+\pi^-$, it has been proposed in
\cite{Barshay} that a small mixing of the $\pi\pi$ and $D\ov D$
channels can account for the puzzling observation of $B^0\to
\pi^0\pi^0$. However, the aforementioned dispersive part of
final-state rescattering was not considered in \cite{Barshay} and
consequently the predicted branching ratio
$\B(B^0\to\pi^+\pi^-)\sim 11\times 10^{-6}$ is too large compared
to experiment. In the present work, we have shown that it is the
dispersive part of long-distance rescattering from $D\bar
D\to\pi\pi$ that accounts for the suppression of the $\pi^+\pi^-$
mode and the enhancement of $\pi^0\pi^0$.

As far as \CP rate asymmetries are concerned, the long-distance
effect will generate a sizable direct \CP violation for
$\pi^+\pi^-$ with a correct sign. Indeed, the signs of direct \CP
asymmetries in $B\to\pi\pi$ decays are all flipped by the
final-state rescattering effects. It appears that FSIs lead to \CP
asymmetries for $\pi^+\pi^-$ and $\pi^0\pi^0$ opposite in sign.
The same conclusion is also reached in the $\pi\pi$ and $D\ov D$
mixing model \cite{Barshay} and in an analysis of charmless $B$
decay data \cite{FSIChua2} based on a quasi-elastic scattering
model \cite{FSIChua1} (see also \cite{Smith}). However, the
perturbative QCD approach based on the $k_T$ factorization theorem
predicts the same sign for $\pi^+\pi^-$ and $\pi^0\pi^0$ \CP
asymmetries, namely, $\A_{\pi^+\pi^-}=0.23\pm0.07$ and
$\A_{\pi^0\pi^0}=0.30\pm0.10$ \cite{Keum}. Moreover, a global
analysis of charmless $B\to PP$ decays based on the topological
approach yields $\A_{\pi^+\pi^-}\sim 0.33$ and
$\A_{\pi^0\pi^0}\sim 0.53$ \cite{Chiang} (see also \cite{Buras}
for the same conclusion reached in a different context). It is
worth mentioning that the isospin analyses of $B\to\pi\pi$ decay
data in \cite{CKMfitter} and \cite{Pivk} all indicate that given
the allowed region of the $\pi^0\pi^0$ branching ratio, there is
practically no constraints on $\pi^0\pi^0$ direct \CP violation
from BaBar and/or Belle data. And hence the sign of
$\A_{\pi^0\pi^0}$ is not yet fixed. At any rate, even a sign
measurement of direct \CP asymmetry in $\pi^0\pi^0$ will provide a
nice testing ground for discriminating between the FSI and PQCD
approaches for \CP violation.

\begin{table}[t]
\caption{ Same as Table \ref{tab:comparepiK} except for
$B\to\pi\pi$ decays. } \label{tab:comparepipi}
\begin{ruledtabular}
\begin{tabular}{c r r c}
Mode
   & Expt.
   & SD
   & SD+LD
   \\
\hline ${\mathcal B}(\ov B^0\to \pi^+\pi^-)$
    & $4.6\pm0.4$
    & 7.6
    & $4.6^{+0.2}_{-0.1}$
    \\
${\mathcal B}(\ov B^0\to\pi^0\pi^0)$
    & $1.5\pm0.3$
    & 0.3
    & $1.5^{+0.1}_{-0.0}$
    \\
${\mathcal B}(B^-\to \pi^-\pi^0)$
    & $5.5\pm0.6$
    & 5.1
    & $5.4\pm0.0$
    \\
 \hline
$\A_{\pi^+\pi^-}$
    & $0.31\pm0.24$
    & $-0.05$
    & $0.35^{+0.15}_{-0.14}$
    \\
${\cal S}_{\pi^+\pi^-}$
    & $-0.56\pm0.34$
    & $-0.66$
    & $-0.16^{+0.15}_{-0.16}$
    \\
$\A_{\pi^0\pi^0}$
    & $0.28\pm0.39$
    & $0.56$
    & $-0.30^{+0.01}_{-0.04}$
    \\
$\A_{\pi^-\pi^0}$
    & $-0.02\pm0.07$
    & $5\times 10^{-5}$
    & $-0.009^{+0.002}_{-0.001}$
    \\
\end{tabular}
\end{ruledtabular}
\end{table}

It is interesting to notice that based on SU(3) symmetry, there
are some model-independent relations for direct \CP asymmetries
between $\pi\pi$ and $\pi K$ modes \cite{Deshpande}
 \be
 \Delta \Gamma(\pi^+\pi^-)=-\Delta\Gamma(\pi^+ K^-), \qquad\quad
 \Delta \Gamma(\pi^0\pi^0)=-\Delta\Gamma(\pi^0\ov  K^0),
 \en
with $\Delta\Gamma(f_1f_2)\equiv \Gamma(\ov B\to
f_1f_2)-\Gamma(B\to f_1f_2)$.  From the measured rates, it follows
that
 \be
 \A_{\pi^+\pi^-}\approx -4.0\, \A_{\pi^+K^-}, \qquad
 \A_{\pi^0\pi^0}\approx -7.7\, \A_{\pi^0\ov K^0}.
 \en
It appears that the first relation is fairly satisfied by the
world averages of \CP asymmetries. Since direct \CP violation in
the $\pi^+ K^-$ mode is now established by $B$ factories, the
above relation implies that the $\pi^+\pi^-$ channel should have
$\A_{\pi^+\pi^-}\sim {\cal O}(0.4)$ with a positive sign. For the
neutral modes more data are clearly needed as the current
experimental results have very large errors.

Finally we present the results for the ratios $\C/\T$, $\E/\T$ and
$\V/\T$:
  \be \label{eq:ratios}
 \left.{\C\over \T}\right|_{\pi\pi} = 0.36\,e^{-i55^\circ},
 \qquad
 \left.{\E\over \T}\right|_{\pi\pi} = 0.19\,e^{-i85^\circ}, \qquad
  \left.{\V\over \T}\right|_{\pi\pi} = 0.56\,e^{i230^\circ},
 \en
to be compared with the short-distance result
$\C/\T|_{\pi\pi}^{\rm SD}=a_2/a_1=0.23\,{\rm exp}(-i32^\circ)$
from Eq. (\ref{eq:ai}). Evidently, the ratio of $\C/\T$ in $B\to
\pi\pi$ decays is similar to that in $B\to D\pi$ decays [see Eq.
(\ref{eq:C/TinBDpi})]. To analyze the $B\to\pi\pi$ data, it is
convenient to define $\T_{\rm eff}=\T+\E+\V$ and $\C_{\rm
eff}=\C-\E-\V$. It follows from Eq. (\ref{eq:ratios}) that
 \be
 \left.{\C_{\rm eff}\over \T_{\rm eff}}\right|_{\pi\pi} =
 0.71\,e^{i72^\circ}.
 \en
This is consistent with Eq. (\ref{eq:C/TinQDA}) obtained from the
global analysis of $B\to\pi\pi$ and $\pi K$ data.

\subsection{\CP violation in $B^-\to\pi^-\pi^0$ decays}
We see from Table \ref{tab:comparepipi} that for
$B^-\to\pi^-\pi^0$, even including FSIs $CP$-violating
partial-rate asymmetry in the SM is likely to be very small; the
SD contribution $\sim 5\times 10^{-4}$ may increase to the level
of one percent. Although this SM \CP asymmetry is so small that it
is difficult to measure experimentally, it provides a nice place
for detecting New Physics. This is because the isospin of the
$\pi^-\pi^0$ ( also $\rho^-\rho^0$) state in the $B$ decay is
$I=2$ and hence it does not receive QCD penguin contributions and
receives only the loop contributions from electroweak penguins.
Since these decays are tree dominated, SM predicts an almost null
\CP asymmetry. Final-state rescattering can enhance the
$CP$-violating effect at most to one percent level. Hence, a
measurement of direct \CP violation in $B^-\to \pi^-\pi^0$
provides a nice test of the Standard Model and New Physics. This
should be doable experimentally in the near future. If the
measured partial rate asymmetry turns out to be larger than, say,
2\%, this will mostly likely imply some New Physics beyond the
Standard Model.

\section{$B\to\rho\pi$ Decays}

The experimental results of \CP averaged branching ratios and \CP
asymmetries for $B\to\rho\pi$ decays are summarized in Table
\ref{tab:Brhopi}. The experimental determination of direct \CP
asymmetries for $\rho^+\pi^-$ and $\rho^-\pi^+$ is more
complicated as $B^0\to \rho^\pm\pi^\mp$ is not a \CP eigenstate.
The time-dependent \CP asymmetries are given by
 \be
 \A(t) &\equiv & {\Gamma(\ov
B^0(t)\to\rho^\pm\pi^\mp)-\Gamma(B^0(t)\to \rho^\pm\pi^\mp)\over
\Gamma(\ov B^0(t)\to\rho^\pm\pi^\mp)+\Gamma(B^0(t)\to
\rho^\pm\pi^\mp)} \non \\
 &=& (S_{\rho\pi}\pm \Delta S_{\rho\pi})\sin(\Delta
 mt)-(C_{\rho\pi}\pm \Delta C_{\rho\pi})\cos(\Delta mt),
 \en
where $\Delta m$ is the mass difference of the two neutral $B$
eigenstates, ${\cal S}_{\rho\pi}$ is referred to as mixing-induced
\CP asymmetry and $\C_{\rho\pi}$ is the direct \CP asymmetry,
while $\Delta S_{\rho\pi}$ and $\Delta C_{\rho\pi}$ are {\it
CP}-conserving quantities. Next consider the time- and
flavor-integrated charge asymmetry
 \be
 \A_{\rho\pi}\equiv {N(\rho^+\pi^-)-N(\rho^-\pi^+)\over
 N(\rho^+\pi^-)+N(\rho^-\pi^+)},
 \en
where $N(\rho^+\pi^-)$ and $N(\rho^-\pi^+)$ are the sum of of
yields for $B^0$ and $\ov B^0$ decays to $\rho^+\pi^-$ and
$\rho^-\pi^+$, respectively. Then,
 \be
 \A_{\rho^+\pi^-} &\equiv& {\Gamma(\ov B^0\to\rho^+\pi^-)-\Gamma(B^0\to
 \rho^-\pi^+)\over \Gamma(\ov B^0\to\rho^+\pi^-)+\Gamma(B^0\to
 \rho^-\pi^+)}={\A_{\rho\pi}-C_{\rho\pi}-\A_{\rho\pi}\Delta C_{\rho\pi}\over 1-\Delta
 C_{\rho\pi}-\A_{\rho\pi} C_{\rho\pi}},   \non \\
 \A_{\rho^-\pi^+} &\equiv& {\Gamma(\ov B^0\to\rho^-\pi^+)-\Gamma(B^0\to
 \rho^+\pi^-)\over \Gamma(\ov B^0\to\rho^-\pi^+)+\Gamma(B^0\to
 \rho^+\pi^-)}=-{\A_{\rho\pi}+C_{\rho\pi}+\A_{\rho\pi}\Delta C_{\rho\pi}\over 1+\Delta
 C_{\rho\pi}+\A_{\rho\pi} C_{\rho\pi}}.
 \en
Note that the quantities $\A_{\rho^\pm\pi^\mp}$ here correspond to
$\A_{\rho\pi}^{\mp\pm}$ defined in \cite{CKMfitter}. Hence, direct
\CP asymmetries $\A_{\rho^+\pi^-}$ and $\A_{\rho^-\pi^+}$ are
determined from the above equations together with the measured
correlation coefficients between the parameters in the
time-dependent fit to $B^0\to\rho^\pm\pi^\mp$. From Table
\ref{tab:Brhopi}, it is evident that the combined BaBar and Belle
measurements of $\ov B^0\to\rho^\pm\pi^\mp$ imply a $3.6\sigma$
direct \CP violation in the $\rho^+\pi^-$ mode, but not yet in
$\rho^-\pi^+$.

\begin{table}[t]
\caption{Experimental data for \CP averaged branching ratios (top,
in units of $10^{-6}$) and direct \CP asymmetries (bottom) for
$B\to\rho\pi$ \cite{HFAG,PDG}. Note that the direct $CP$-violating
quantities $\A_{\rho^\pm\pi^\mp}$ correspond to
$\A_{\rho\pi}^{\mp\pm}$ defined in \cite{CKMfitter}. }
\label{tab:Brhopi}
\begin{ruledtabular}
\begin{tabular}{c c c c c}
Mode & BaBar & Belle & CLEO & Average  \\
\hline
 $\ov B^0\to \rho^\pm\pi^\mp$ & $22.6\pm1.8\pm2.2$ & $29.1^{+5.0}_{-4.9}\pm4.0$ &
 $27.6^{+8.4}_{-7.4}\pm4.2$ & $24.0\pm2.5$ \\
 $\ov B^0\to\rho^0\pi^0$ &  $1.4\pm0.6\pm0.3<2.9$ & $5.1\pm1.6\pm0.8$ &
 $<5.5$ & $1.9\pm1.2$\footnotemark[1] \\
 $B^-\to \rho^-\pi^0$ & $10.9\pm1.9\pm1.9$ &
 $13.2\pm2.3^{+1.4}_{-1.9}$ & $<43$ & $12.0\pm2.0$\\
 $B^-\to \rho^0\pi^-$ & $9.5\pm1.1\pm0.9$ &
 $8.0^{+2.3}_{-2.0}\pm0.7$ & $10.4^{+3.3}_{-3.4}\pm2.1$ & $9.1\pm1.3$\\
 \hline
 $\A_{\rho^-\pi^+}$ & $-0.21\pm0.11\pm0.04$ & $-0.02\pm0.16^{+0.05}_{-0.02}$ &
 &  $-0.15\pm0.09$ \\
 $\A_{\rho^+\pi^-}$ & $-0.47\pm0.15\pm0.06$ & $-0.53\pm0.29^{+0.09}_{-0.04}$ &
 & $-0.47^{+0.13}_{-0.14}$ \\
 $\A_{\rho^-\pi^0}$ & $0.24\pm0.16\pm0.06$ &
 $0.06\pm0.19^{+0.04}_{-0.06}$ & & $0.16\pm0.13$ \\
 $\A_{\rho^0\pi^-}$ & $-0.19\pm0.11\pm0.02$ &
 & & $-0.19\pm0.11$ \\
\end{tabular}
\end{ruledtabular}
\footnotetext[1]{The error-bar of $\B(\ov B^0\to\rho^0\pi^0)$ is
scaled by an $S$ factor of 1.9. We have taken into account the
BaBar measurement to obtain the weighted average of the branching
ratio for this mode. Note that the average for $\B(\ov
B^0\to\rho^0\pi^0)$ is quoted to be $<2.9$ in \cite{HFAG}. }
\end{table}

The $B\to \rho\pi$ decay amplitudes have the general expressions
 \be \label{eq:ampBrhopi}
 A(\ov B^0\to\rho^+\pi^-) &=& \T_V+\P_V+{2\over 3}\P'^{\rm EW}_V, \non \\
 A(\ov B^0\to\rho^-\pi^+) &=& \T_P+\P_P+{2\over 3}\P'^{\rm EW}_P, \non \\
 A(\ov B^0\to\rho^0\pi^0) &=& -{1\over 2}(\C_V-\P_V+\P^{\rm EW}_V+{1\over 3}\P'^{\rm
 EW}_V+\C_P-\P_P+\P^{\rm EW}_P+{1\over 3}\P'^{\rm  EW}_P), \\
 A(B^-\to \rho^-\pi^0) &=& {1\over\sqrt{2}}(\T_P+\C_V+\P_P-\P_V+\P^{\rm EW}_V
 +{2\over 3}\P'^{\rm EW}_P+{1\over 3}\P'^{\rm EW}_V), \non \\
 A(B^-\to \rho^0\pi^-) &=& {1\over\sqrt{2}}(\T_V+\C_P+\P_V-\P_P+\P^{\rm
 EW}_P  +{2\over 3}\P'^{\rm EW}_V+{1\over 3}\P'^{\rm EW}_P). \non
 \en
They satisfy the strong isospin relation
 \be
 \sqrt{2}\left[ A(B^-\to
 \rho^-\pi^0)+A(B^-\to\rho^0\pi^-)+\sqrt{2}A(\ov
 B^0\to\rho^0\pi^0)\right]=A(\ov B^0\to\rho^+\pi^-)+A(\ov B^0\to
 \rho^-\pi^+)\non \\
 \en

In the factorization approach (the subscript ``SD" being dropped
for convenience),
 \be
 && \T_{V,P} = \kappa_{V,P}\lambda_ua_1, \qquad
 \C_{V,P} = \kappa_{V,P}\lambda_u a_2, \qquad
 \P_P=\kappa_P\Big[\lambda_u a_4^u+\lambda_c a_4^c\Big], \non \\
 && \P_V=\kappa_V\Big[\lambda_u(a_4^u+a_6^u
 r_\chi)+\lambda_c(a_4^c+a_6^c r_\chi)\Big],
 \qquad \P^{\rm EW}_{V,P} = {3\over
 2}\kappa_{V,P}(\lambda_u+\lambda_c)(-a_7+a_9), \non \\
 &&  \P'^{\rm
 EW}_V={3\over 2}\kappa_V\Big[\lambda_u(a_8^u r_\chi+a_{10}^u)+\lambda_c(a_8^c
 r_\chi+a_{10}^c)\Big], \qquad \P'^{\rm
 EW}_P={3\over 2}\kappa_P\Big[\lambda_u a_{10}^u+\lambda_c a_{10}^c\Big],
 \en
with
 \be
 \kappa_V=\sqrt{2}G_Ff_\pi A_0^{B\rho}(m_\pi^2)m_\rho(\vp\cdot p_B),
 \qquad \kappa_P=\sqrt{2}G_Ff_\rho F_1^{B\pi}(m_\rho^2)m_\rho(\vp\cdot p_B),
 \en
Short-distance contributions yield
 \be
 \B(\ov B^0\to\rho^-\pi^+)_{\rm SD} &=& 18.4\times 10^{-6}, \qquad
 \A_{\rho^-\pi^+}^{\rm SD}=-0.03, \non \\
 \B(\ov B^0\to\rho^+\pi^-)_{\rm SD} &=& 7.9\times 10^{-6}, \qquad~
 \A_{\rho^+\pi^-}^{\rm SD}=-0.01, \non \\
 \B(\ov B^0\to\rho^0\pi^0)_{\rm SD} &=& 5.9\times 10^{-7}, \qquad~
 \A_{\rho^0\pi^0}^{\rm SD}=0.01,  \non \\
 \B(B^-\to\rho^-\pi^0)_{\rm SD} &=& 12.8\times 10^{-6}, \qquad
 \A_{\rho^-\pi^0}^{\rm SD}=-0.04, \non \\
  \B(B^-\to\rho^0\pi^-)_{\rm SD} &=& 6.8\times 10^{-6}, \qquad~
 \A_{\rho^0\pi^-}^{\rm SD}=0.06\,.
 \en
Note that the central values of the branching ratios for
$\rho^-\pi^+,\rho^+\pi^-,\rho^0\pi^0$, $\rho^-\pi^0,\rho^0\pi^-$,
respectively, are predicted to be 21.2, 15.4, 0.4, 14.0, 11.9 in
units of $10^{-6}$ in \cite{BN}, which are larger than our results
by around a factor of 2 for $\rho^+\pi^-$ and 1.5 for
$\rho^0\pi^-$. This is ascribed to the fact that the form factors
$A_0^{B\rho}(0)$ and $F_1^{B\pi}(0)$ are 0.28 and 0.25,
respectively, in our case \cite{CCH}, while they are chosen to be
$0.37\pm0.06$ and $0.28\pm0.05$ in \cite{BN}. The central value of
the predicted  $\B(\ov
B^0\to\rho^\pm\pi^\mp)=(36.5^{+21.4}_{-17.6})\times 10^{-6}$
\cite{BN} is large compared to the experimental value of
$(24.0\pm2.5)\times 10^{-6}$, but it is accompanied by large
errors.

\begin{figure}[t]
\vspace{-1cm}
  \centerline{\psfig{figure=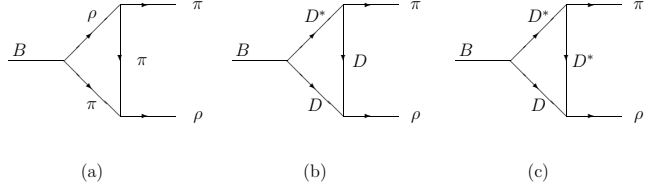,width=12cm}}
\vspace{-13cm}
    \caption[]{\small Long-distance $t$-channel rescattering contributions
    to $B\to \rho\pi$.  }\label{fig:Brhopi}
\end{figure}

The long-distance contributions are depicted in Fig.
\ref{fig:Brhopi}:
 \be
 \A bs(8a) &=& {G_F\over\sqrt{2}}\lambda_u\int_{-1}^1 {|\vec p_1|d\cos\theta\over 16\pi
 m_B}\,8g^2_{\rho\pi\pi}\,{F^2(t,m_\pi)\over
 t-m_\pi^2}m_\rho\Big[af_\pi A_0^{B\rho}(m_\pi^2)+bf_\rho
 F_1^{B\pi}(m_\rho^2)\Big]  \non \\
 &\times& \left(-p_2\cdot p_3+{
 (p_1\cdot p_2)(p_1\cdot p_3)\over m_\rho^2}\right)\,{E_2|\vec
 p_4|-E_4|\vec p_2|\cos\theta\over m_B|\vec p_4|}, \non \\
 \A bs(8b) &=& {G_F\over\sqrt{2}}\lambda_u\int_{-1}^1 {|\vec p_1|d\cos\theta\over 16\pi
 m_B}\,2\sqrt{2}\,g_{D^*D\pi}\,g_V\beta\,{F^2(t,m_D)\over
 t-m_D^2}m_{D^*} \non \\ &\times& \Big[cf_D A_0^{BD^*}(m_D^2)+d\,f_{D^*}
 F_1^{BD}(m_{D^*}^2)\Big]  \non \\
 &\times& \left(-p_2\cdot p_3+{
 (p_1\cdot p_2)(p_1\cdot p_3)\over m_\rho^2}\right)\,{E_2|\vec
 p_4|-E_4|\vec p_2|\cos\theta\over m_B|\vec p_4|}, \non \\
 \A bs(8c) &=& {G_F\over\sqrt{2}}\lambda_u\int_{-1}^1 {|\vec p_1|d\cos\theta\over 16\pi
 m_B}\,4\sqrt{2}\,g^2_{D^*D\pi}\,g_V\lambda\,{F^2(t,m_{D^*})\over
 t-m_{D^*}^2}m_{D^*}  \non \\
 &\times& \Big[cf_D A_0^{BD^*}(m_D^2)+d\,f_{D^*}
 F_1^{BD}(m_{D^*}^2)\Big]
 \Bigg(m_4^2p_1\cdot p_2-(p_2\cdot p_4)(p_1\cdot p_4)\non \\
 &+& {E_1|\vec
 p_4|+E_4|\vec p_1|\cos\theta\over m_B|\vec p_4|}\Big[(p_B\cdot p_2)
 (p_3\cdot p_4)-(p_B\cdot p_4)(p_2\cdot p_3)\Big] \Bigg),
 \en
where $a=b=1/2$ for $\rho^0\pi^0$, $a=1\,(1/\sqrt{2})$, $b=0$ for
$\rho^-\pi^+\,(\rho^0\pi^-)$, $a=0$, $b=1\,(1/\sqrt{2})$ for
$\rho^+\pi^-\,(\rho^-\pi^0)$, $c=d=1/2$ for $\rho^0\pi^0$,
$c=d=1/\sqrt{2}$ for $\rho^0\pi^-$, $c=d=-1/\sqrt{2}$ for
$\rho^-\pi^0$, $c=1,d=0$ for $\rho^-\pi^+$, and $c=0,d=1$ for
$\rho^+\pi^-$.

\begin{table}[t]
\caption{\label{tab:comparerhopi} Same as Table
\ref{tab:comparepiK} except for $B\to\rho\pi$ decays.}
\begin{ruledtabular}
\begin{tabular}{c r r r}
 Mode
   & Expt.
   & SD
   & SD+LD
   \\
\hline
 ${\mathcal B}(\ov B^0\to \rho^-\pi^+)$
   & $13.9^{+2.2}_{-2.1}$
   & 18.4
   & $18.8^{+0.3}_{-0.2}$
   \\
 ${\mathcal B}(\ov B^0\to \rho^+\pi^-)$
   & $10.1^{+2.1}_{-1.9}$
   & 7.9
   & $8.4\pm0.3$
   \\
 ${\mathcal B}(\ov B^0\to \rho^\pm\pi^\mp)$
   & $24.0\pm2.5$
   & 26.3
   & $27.3^{+0.6}_{-0.4}$
   \\
 ${\mathcal B}(\ov B^0\to\rho^0\pi^0)$
   & $1.9\pm1.2$
   & 0.6
   & $1.3^{+0.4}_{-0.3}$
   \\
 ${\mathcal B}(B^-\to \rho^-\pi^0)$
   & $12.0\pm2.0$
   & 12.9
   & $14.0^{+0.7}_{-0.4}$
   \\
 ${\mathcal B}(B^-\to \rho^0\pi^-)$
   & $9.1\pm1.3$
   & 6.8
   & $7.5^{+0.6}_{-0.3}$
   \\
 \hline
 $\A_{\rho^-\pi^+}$
   &  $-0.15\pm0.09$
   & $-0.03$
   & $-0.24\pm0.6$
   \\
 $\A_{\rho^+\pi^-}$
   & $-0.47^{+0.13}_{-0.14}$
   & $-0.01$
   & $-0.43\pm0.11$
   \\
 $\A_{\rho^0\pi^0}$
   &
   & $0.01$
   & $0.57^{+0.01}_{-0.03}$
   \\
 $\A_{\rho^-\pi^0}$
   & $0.16\pm0.13$
   & $-0.04$
   & $0.36\pm0.10$
   \\
 $\A_{\rho^0\pi^-}$
   & $-0.19\pm0.11$
   & $0.06$
   & $-0.56^{+0.14}_{-0.15}$
   \\
\end{tabular}
\end{ruledtabular}
\end{table}

A $\chi^2$ fit of the cutoff scales
$\Lambda_\pi,\Lambda_{D^{(*)}}$ or the cutoff parameters
$\eta_\pi$ and $\eta_{D^{(*)}}$to the measured branching ratios of
$B\to\rho\pi$ shows a flat $\chi^2$ without a minimum for
$\eta_D\lsim 2$. Therefore, we choose the criterion that FSI
contributions should accommodate the data of the $\rho^0\pi^0$ and
$\rho^0\pi^-$ without affecting $\rho^\pm\pi^\mp$ and
$\rho^-\pi^0$ significantly. Using $\eta_D=\eta_{D^*}=1.6$\,, the
branching ratios and direct \CP asymmetries after taking into
account long-distance rescattering effects are shown in Table
\ref{tab:comparerhopi}. (The results are rather insensitive to
$\eta_\pi$.) We see that, in contrast to the $\pi K$ case, the
$B\to\rho\pi$ decays are less sensitive to the cutoff scales. The
$\ov B^0\to\rho^0\pi^0$ rate is enhanced via rescattering  by a
factor of 2, which is consistent with the weighted average of
$(1.9\pm1.2)\times 10^{-6}$. However, it is important to clarify
the discrepancy between BaBar and Belle measurements for this
mode. It should be remarked that direct \CP violation in this mode
is significantly enhanced by FSI from around 1\% to 60\%. Naively,
this asymmetry ought to become accessible at $B$ factories with
$10^8$ $B\bar B$ pairs. As stressed before, there is $3.6\sigma$
direct \CP violation in the $\rho^+\pi^-$ mode. Our prediction of
$\A_{\rho^+\pi^-}$ agrees with the data in both magnitude and
sign, while the QCD factorization prediction
$(0.6^{+11.6}_{-11.8})\%$ \cite{BN} seems not consistent with
experiment.

\section{Polarization in $B\to\phi K^*,~\rho K^*$ decays}

\begin{table}[t]    \label{tab:BphiKV}
 \caption{ Experimental data for
\CP averaged branching ratios (in units of $10^{-6}$) and
polarization fractions for $B\to\phi K^*,\rho K^*,\rho\rho$
\cite{HFAG,PDG,BaBarVV,BelleVV}. The phases of the parallel and
perpendicular amplitudes relative to the longitudinal one for
$B\to\phi K^*$ are also included \cite{BaBarVV,BelleVV}. }
\begin{ruledtabular}
\begin{tabular}{c c c c c}
Mode & BaBar & Belle & CLEO & Average  \\
\hline
 $\ov B^0\to \phi \ov K^{*0}$
              & $9.2\pm0.9\pm0.5$
              & $10.0^{+1.6+0.7}_{-1.5-0.8}$
              & $11.5^{+4.5+1.8}_{-3.7-1.7}$
              & $9.5\pm0.9$
              \\
 $B^-\to\phi K^{*-}$
              & $12.7^{+2.2}_{-2.0}\pm1.1$
              & $6.7^{+2.1+0.7}_{-1.9-1.0}$
              & $10.6^{+6.4+1.8}_{-4.9-1.6}$
              & $9.7\pm1.5$
              \\
 $\ov B^0\to \rho^+ K^{*-}$
              & $16.3\pm5.4\pm2.3^{+0.0}_{-6.3}<24$
              &
              &
              & $<24$
              \\
 $B^-\to \rho^0 K^{*-}$
              & $10.6^{+3.0}_{-2.6}\pm2.4$
              &
              & $<74$
              & $10.6^{+3.8}_{-3.5}$
              \\
 $B^-\to \rho^- \ov K^{*0}$
              & $17.0\pm2.9\pm2.0^{+0.0}_{-1.9}$
              & $6.6\pm2.2\pm0.8$
              &
              & $9.6\pm4.7$\footnotemark[1]
              \\
 $B^-\to\rho^0\rho^-$
              & $22.5^{+5.7}_{-5.4}\pm5.8$
              & $31.7\pm7.1^{+3.8}_{-6.7}$
              &
              & $26.4^{+6.1}_{-6.4}$
              \\
 $\ov B^0\to \rho^+ \rho^-$
              & $30\pm4\pm5$
              &
              &
              & $30\pm6$
              \\
 $B^-\to \rho^- \omega$
              & $12.6^{+3.7}_{-3.3}\pm1.8$
              &
              &
              & $12.6^{+4.1}_{-3.8}$
              \\
 \hline
 $f_L(\phi \ov K^{*0})$
              & $0.52\pm0.05\pm0.02$
              & $0.52\pm0.07\pm0.05$
              &
              & $0.52\pm0.05$
              \\
 $f_\bot(\phi \ov K^{*0})$
              & $0.22\pm0.05\pm0.02$
              & $0.30\pm0.07\pm0.03$
              &
              & $0.25\pm0.04$
              \\
 $f_L(\phi K^{*-})$
              & $0.46\pm0.12\pm0.03$
              & $0.49\pm0.13\pm0.05$
              &
              & $0.47\pm0.09$
              \\
 $f_\bot(\phi K^{*-})$
              &
              & $0.12^{+0.11}_{-0.08}\pm0.03$
              &
              & $0.12^{+0.11}_{-0.09}$
              \\
 $\phi_\parallel(\phi K^{*0})$ (rad)\footnotemark[2]
              & $2.34^{+0.23}_{-0.20}\pm0.05$
              & $-2.30\pm0.28\pm0.04$
              &
              &
              \\
 $\phi_\bot(\phi K^{*0})$ (rad)\footnotemark[2]
              & $2.47\pm0.25\pm0.05$
              & $0.64\pm0.26\pm0.05$
              &
              &
              \\
 $\phi_\parallel(\phi K^{*+})$ (rad)\footnotemark[2]
              &
              & $-2.07\pm0.34\pm0.07$
              &
              & $-2.07\pm0.35$
              \\
 $\phi_\bot(\phi K^{*+})$ (rad)\footnotemark[2]
              &
              & $0.93^{+0.55}_{-0.39}\pm0.12$
              &
              & $0.93^{+0.56}_{-0.41}$
              \\
 \hline
 $f_L(\rho^0 K^{*-})$
              & $0.96^{+0.04}_{-0.15}\pm0.04$
              &
              &
              & $0.96^{+0.06}_{-0.16}$
              \\
 $f_L(\rho^- \ov K^{*0})$
              & $0.79\pm0.08\pm0.04\pm0.02$
              & $0.50\pm0.19^{+0.05}_{-0.07}$
              &
              & $0.74\pm0.08$
              \\
 \hline
 $f_L(\rho^0 \rho^-)$
              & $0.97^{+0.03}_{-0.07}\pm0.04$
              & $0.95\pm0.11\pm0.02$
              &
              & $0.96^{+0.04}_{-0.07}$
              \\
 $f_L(\rho^+ \rho^-)$
              & $0.99\pm0.03\pm0.04$
              &
              &
              & $0.99\pm0.05$
              \\
 $f_L(\rho^- \omega)$
              & $0.88^{+0.12}_{-0.15}\pm0.03$
              &
              &
              & $0.88^{+0.12}_{-0.15}$
              \\
\end{tabular}
\end{ruledtabular}
\footnotetext[1]{The error in $\B(B^-\to\rho^-\ov K^{*0})$ is
scaled by $S=2.4$. Our average is slightly different from the
value of $9.2\pm2.0$ quoted in \cite{HFAG}.}
 \footnotetext[2]{Experimental results of the relative strong
 phases are for $B^{(0,+)}\to \phi K^{*(0,+)}$ decays.}
\end{table}

The branching ratios and polarization fractions of charmless $B\to
VV$ decays have been measured for $\rho\rho,~\rho K^*$ and $\phi
K^*$ final states. In general, it is expected that they are
dominated by the longitudinal polarization states and respect the
scaling law, namely \cite{Kagan1} (see also footnote 1 on p.3)
 \be \label{eq:scaling}
1-f_L={\cal O}(m^2_V/m^2_B), \qquad f_\bot/f_\parallel=1+{\cal
O}(m_V/m_B)
 \en
with $f_L,f_\bot$ and $f_\parallel$ being the longitudinal,
perpendicular, and parallel polarization fractions, respectively.
However, in sharp contrast to the $\rho\rho$ case, the large
fraction of transverse polarization in $B\to\phi K^*$ decays
observed by BaBar and confirmed by Belle (see Table
\ref{tab:BphiKV}) is a surprise and poses an interesting challenge
for any theoretical interpretation. The aforementioned scaling law
remains true even when nonfactorizable graphs are included in QCD
factorization \cite{Kagan1}. Therefore, in order to obtain a large
transverse polarization in $B\to\phi K^*$, this scaling law valid
at short-distance interactions must be circumvented in one way or
another. One way is the Kagan's recent suggestion of sizable
penguin-induced annihilation contributions \cite{Kagan2}. Another
possibility is that perhaps a sizable transverse polarization can
be achieved via final-state rescattering. To be more specific, the
following two FSI processes are potentially important. First,
$B\to D^*D_s^*$ followed by a rescattering to $\phi K^*$ is not
subject to the scaling law and hence its large transverse
polarization can be conveyed to $\phi K^*$ via FSI. Second, the
FSI effect from $\ov B\to D^*D_s$ or $\ov B\to DD^*_s$ will
contribute only to the $A_\bot$ amplitude. In this section, we
shall carefully examine the long-distance rescattering effects on
$\phi K^*$ polarizations.

\subsection{Short-distance contributions to $\ov B\to\phi \ov K^*$}

In the factorization approach, the amplitude of the $\overline
B\to\phi \overline K {}^*$ decay is given by
 \be  \label{eq:phiKstSD}
 \la\phi(\lambda_\phi) \overline K {}^*(\lambda_{K^*})|H_{\rm eff}|\overline B\ra&=&\frac{G_{\rm F}}{\sqrt2}
 \sum_{q=u,c} V_{qb} V^*_{qs}~a^q_{\phi K^*}~\la\phi|(\bar s s)_V|0\ra\,\la\overline K {}^*|(\bar
 s b)_{V-A}|\overline B\ra
 \non\\
 &=&-i \alpha_{\phi K^*}~
 \vp^{*\mu}_\phi(\lambda_\phi)\vp^{*\nu}_{K^*}(\lambda_{K^*})\bigg(A_1^{BK^*}(m_\phi^2) g_{\mu\nu}
 \non\\
 &&-\frac{2
 A_2^{BK^*}(m_\phi^2)}{(m_B+m_{K^*})^2} p_{B\mu}
 p_{B\nu}-i\frac{2 V^{BK^*}(m_\phi^2)}{(m_B+m_{K^*})^2}\epsilon_{\mu\nu\alpha\beta}p_B^\alpha
 p_{K^*}^\beta\bigg),
 \en
where $\lambda_{\phi,K^*}$ are the corresponding helicities,
$a^q_{\phi K^*}\equiv a_3+a^q_4+a_5-(a_7+a_9+a^q_{10})/2$ and
$\alpha_{\phi K^*}\equiv G_{\rm F}m_\phi f_\phi(m_B+m_{K^*})
\sum_{q=u,c} V_{qb} V^*_{qs}~a^q_{\phi K^*}/\sqrt2$.
With $A_\lambda$ defined as $\la\phi(\lambda)\overline K
{}^*(\lambda)|H_{\rm eff}|\overline B\ra$, it is straightforward
to obtain~\cite{Dighe}
 \be
 A_0&=&-i\alpha_{\phi
 K^*}\bigg(-A_1^{BK^*}(m_\phi^2)\frac{m_B^2-m_\phi^2-m_{K^*}^2}{2 m_\phi
 m_{K^*}}+\frac{2
 A_2^{BK^*}(m_\phi^2)}{(m_B+m_{K^*})^2} \frac{m_B^2 p_c^2}{m_\phi m_{K^*}}\bigg),
 \non\\
 A_{\pm}&=&-i\alpha_{\phi
 K^*}\bigg(A_1^{BK^*}(m_\phi^2)\mp\frac{2 V^{BK^*}(m_\phi^2)}{(m_B+m_{K^*})^2}m_B p_c\bigg),
 \en
where $p_c$ is the center of mass momentum and we have used the
phase convention $\vp_\phi(\lambda)=\vp_{K^*}(-\lambda)$ for $\vec
p_{\phi,K^*}\to 0$. The longitudinal amplitude $A_0$ is sometimes
denoted as $A_L$, while the transverse amplitudes are defined by
 \be
 A_{\parallel}&=&\frac{A_++A_-}{\sqrt2}=-i\alpha_{\phi
 K^*}\sqrt2A_1^{BK^*}(m_\phi^2),
 \non\\
 A_{\bot}&=&\frac{A_+-A_-}{\sqrt2}=i\alpha_{\phi
 K^*}\frac{2\sqrt2 V^{BK^*}(m_\phi^2)}{(m_B+m_{K^*})^2}m_B p_c
 \label{eq:Atrans}
 \en
in the transversity basis. The decay rate can be expressed in
terms of these amplitudes as
 \be
 \Gamma=\frac{p_c}{8\pi m_B^2}(|A_0|^2+|A_+|^2+|A_-|^2)
       =\frac{p_c}{8\pi m_B^2}(|A_L|^2+|A_{\parallel}|^2+|A_{\bot}|^2),
 \en
and the polarization fractions are defined as
 \be
 f_\alpha\equiv \frac{\Gamma_\alpha}{\Gamma}
                     =\frac{|A_\alpha|^2}{|A_0|^2+|A_\parallel|^2+|A_\bot|^2},
 \label{eq:f}
 \en
with $\alpha=L,\parallel,\bot$.

The form factors $A_1, A_2, V$ usually are similar in size
\cite{CCH}. Using above equations with $a_i$ given in
(\ref{eq:ai}) \footnote{As shown explicitly in \cite{CYBVV} within
the QCD factorization approach, nonfactorizable corrections to
each partial-wave or helicity amplitude are in general not the
same; the effective parameters $a_i$ can vary for different
helicity amplitudes. However, since such an effect is too small to
account for the large transverse polarization~\cite{Kagan1}, for
simplicity we shall ignore the difference of $a_i$ in different
helicity amplitudes.}
and the form factors given in \cite{CCH}, one immediately finds
that the short-distance contributions yield (see also Table
\ref{tab:phiKst})
 \be
 f^{\rm SD}_L:f^{\rm SD}_\parallel:f^{\rm SD}_\bot=
    0.88\pm 0.06:0.07\pm0.03:0.05\pm0.03,\qquad
 1\simeq f^{\rm SD}_L\gg f^{\rm SD}_{\parallel,\bot},
 \label{eq:frel}
 \en
where the errors are estimated with 10\% uncertainties in form
factors. However, the above prediction is not borne out by the
$B\to\phi K^*$ data (see Table~\ref{tab:BphiKV}). It is easily
seen that Eq.~(\ref{eq:frel}) is closely related to the fact that
$m_B\gg m_{\phi,K^*}$. We do not expect that such a relation holds
in the case of $\overline B\to D^{(*)} D_s^{(*)}$ decays, which
can rescatter to the $\phi \overline K {}^*$ final state. It is
thus interesting to investigate the effect of FSI on these
polarization fractions. A similar analysis was also considered
recently in \cite{Colangelo04}.

Another possible shortcoming of the short-distance factorization
approach is that the central value of the predicted branching
ratio
 \be \label{eq:BRphiKstSD}
 \B(\ov B\to\phi \ov K^*)_{\rm SD}=(4.4\pm2.6)\times 10^{-6}
 \en
is too small by a factor of 2 compared to experiment. As pointed
out in \cite{CYBVV}, in the heavy quark limit, both vector mesons
in the charmless $B\to VV$ decays should have zero helicity and
the corresponding amplitude is governed by the form factor $A_0$.
Consequently, the predicted branching ratios are rather sensitive
to the form factor models of $A_0$ and can be easily different by
a factor of 2. Using the covariant light-front model for $B\to
K^*$ form factors \cite{CCH}, we are led to the above prediction
in Eq. (\ref{eq:BRphiKstSD}). A similar result is also obtained in
the BSW model \cite{CYBVV}. However, a much larger branching ratio
will be obtained if the form factors based on light-cone sum rules
\cite{Ball98,Ball03} are employed. At any rate, if the predicted
rate is too small, this is another important incentive for
considering long-distance effects.

\subsection{Long-distance contributions to $\ov B\to\phi \ov K^*$}

\begin{figure}[t]
  \centerline{\psfig{figure=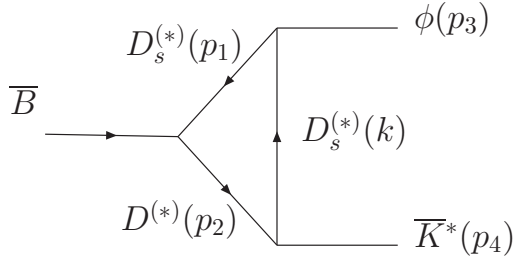,width=7cm}}
    \caption[]{\small Long-distance $t$-channel rescattering contributions
    to $\overline B\to \phi \overline K{}^*$. Note that flows in lines are along $b$ or $c$ quarks.  }
\label{fig:BphiKst}
\end{figure}

The $D^{(*)} D_s^{(*)}$ states from $\ov B$ decays can rescatter
to $\phi \overline K {}^*$ through the $t$-channel $D_s^{(*)}$
exchange in the triangle diagrams depicted in
Fig.~\ref{fig:BphiKst}. The effective Lagrangian for
$D^{(*)}D^{(*)}V$ vertices is given in Eq. (\ref{eq:LDDV}). Note
that as in Eq.~(\ref{eq:SU3breaking}), SU(3) or U(3) breaking
effects in strong couplings are assumed to be taken into account
by the relations
 \be \label{eq:gSU3}
 g_{D^{(*)} D^{(*)}_s K^*}=\sqrt\frac{m_{D^{(*)}_s}}{m_{D^{(*)}}}
 g_{D^{(*)}D^{(*)}\rho},\qquad
 g_{D_s^{(*)}D^{(*)}_s \phi}=\frac{m_{D^{(*)}_s}}{m_{D^{(*)}}}
 g_{D^{(*)}D^{(*)}\rho},
 \en
and likewise for $f_{D^{(*)} D^{(*)}_s K^*}$ and
$f_{D_s^{(*)}D^{(*)}_s \phi}$.

There are totally eight different FSI diagrams in
Fig.~\ref{fig:BphiKst}. The absorptive part contribution of
$\overline B\to D_s D\to\phi \overline K^*$ amplitude via the
$D_s$ exchange is given by
 \be \label{eq:DsDDspre}
 \A bs\,(D_sD;D_s) &=& {1\over 2}\int {d^3\vec p_1\over (2\pi)^32E_1}\,{d^3 \vec p_2\over
(2\pi)^3 2E_2}\,(2\pi)^4\delta^4(p_B-p_1-p_2)A(\ov B^0\to D_s D)
\non \\
 &&\times  (2i) g_{D_s D_s\phi} {F^2(t,m_{D_s})\over
 t-m_{D_s}^2}(-2i)g_{D D_s K^*}\,
 (\vp^*_3\cdot p_1) (\vp^*_4\cdot p_2).
 \en
We need to use some identities to recast the above amplitude into
the standard expression given in Eq.~(\ref{eq:phiKstSD}). To
proceed, we note that after performing the integration, the
integrals
 \be
 \int {d^3\vec p_1\over (2\pi)^32E_1}\,{d^3 \vec p_2\over
(2\pi)^3
2E_2}\,(2\pi)^4\delta^4(p_B-p_1-p_2)f(t)\times\{p_{1\mu},\,p_{1\mu}
p_{1\nu},\,p_{1\mu} p_{1\nu} p_{1\alpha}\}
 \en
can be expressed only in terms of the external momenta $p_3, p_4$
with suitable Lorentz and permutation structures. Therefore, under
the integration we have
  \be
 p_{1\mu}
       &\doteq& P_\mu A_1^{(1)}+q_\mu A_2^{(1)},
 \non\\
 p_{1\mu} p_{1\nu}
       &\doteq& g_{\mu\nu} A_1^{(2)}+P_\mu P_\nu A_2^{(2)}+(P_\mu
                q_\nu+ q_\mu P_\nu) A^{(2)}_3+q_\mu q_\nu A^{(2)}_4,
 \non\\
 p_{1\mu} p_{1\nu} p_{1\alpha}
       &\doteq& (g_{\mu\nu} P_\alpha+g_{\mu\alpha} P_\nu+g_{\nu\alpha} P_\mu) A_1^{(3)}
               +(g_{\mu\nu} q_\alpha+g_{\mu\alpha} q_\nu+g_{\nu\alpha} q_\mu) A_2^{(3)}
 \non \\
       &&       +P_\mu P_\nu P_\alpha A_3^{(3)}
                +(P_\mu P_\nu q_\alpha+ P_\mu q_\nu P_\alpha+q_\mu P_\nu P_\alpha) A^{(3)}_4
 \non \\
       &&       +(q_\mu q_\nu P_\alpha+ q_\mu P_\nu q_\alpha+P_\mu q_\nu q_\alpha)
                 A^{(3)}_5
                +q_\mu q_\nu q_\alpha  A^{(3)}_6,
 \label{eq:p1}
 \en
where $P=p_3+p_4$, $q=p_3-p_4$ and
$A^{(i)}_j=A^{(i)}_j(t,m_B^2,m_1^2,m_2^2,m_3^2,m_4^2)$ have the
analytic forms given in Appendix A. With the aid of
Eq.~(\ref{eq:p1}), we finally obtain
 \be \label{eq:DsDDs}
 \A bs\,(D_sD;D_s) &=& {1\over 2}\int {d^3\vec p_1\over (2\pi)^32E_1}\,{d^3 \vec p_2\over
(2\pi)^3 2E_2}\,(2\pi)^4\delta^4(p_B-p_1-p_2)A(\ov B^0\to D_s D)
\non \\
 &&\times  4~g_{D_s D_s \phi} {F^2(t,m_{D_s})\over
 t-m_{D_s}^2}g_{D D_s K^* }
\non\\
 &&\times\Big\{\vp^*_3\cdot \vp^*_4(-A^{(2)}_1)-(\vp^*_3\cdot P)(\vp^*_4\cdot
 P)
 (-A^{(1)}_1+A^{(1)}_2+A^{(2)}_2-A^{(2)}_4)\Big\}.
 \en
Likewise, the absorptive part of the $\overline B\to D_s D\to\phi
\overline K^*$ amplitude via the $D^*_s$ exchange is given by
\footnote{To avoid using too many dummy indices, we define
$[A,B,C,D]\equiv\epsilon_{\alpha\beta\gamma\delta}A^\alpha B^\beta
C^\gamma D^\delta$,
$[A,B,C,{}_\mu]\equiv\epsilon_{\alpha\beta\gamma\mu}A^\alpha
B^\beta C^\gamma $ and so on for our convenience.}
 \be \label{eq:DsDDsst}
 \A bs\,(D_sD;D^*_s) &=& {1\over 2}\int {d^3\vec p_1\over (2\pi)^32E_1}\,{d^3 \vec p_2\over
(2\pi)^3 2E_2}\,(2\pi)^4\delta^4(p_B-p_1-p_2)A(\ov B^0\to D_s D)
\non \\
 &&\times (-4) f_{D_s D^*_s \phi} {F^2(t,m_{D_s})\over
 t-m_{D_s^*}^2}\,4f_{D D_s^* K^*}
  [p_3,\vp^*_3,p_1,{}^\mu]\left(-g_{\mu\nu}+\frac{k_\mu k_\nu}{m_{D_s^*}^2}\right)[p_4,\vp^*_4,p_2,{}^\nu]
  \non\\
&=& {1\over 2}\int {d^3\vec p_1\over (2\pi)^32E_1}\,{d^3 \vec
p_2\over (2\pi)^3 2E_2}\,(2\pi)^4\delta^4(p_B-p_1-p_2)A(\ov B^0\to
D_s D)
\non \\
 &&\times 16 f_{D_s D^*_s \phi} {F^2(t,m_{D_s^*})\over
 t-m_{D_s^*}^2}f_{D D_s^* K^*}
 \non\\
 &&\times\Bigg\{\vp^*_3\cdot\vp^*_4\Big\{2 A^{(2)}_1 p_3\cdot p_4-[(p_3\cdot
 p_4)^2-m_3^2 m_4^2](A^{(1)}_1-A^{(1)}_2-A^{(2)}_2+A^{(2)}_4)\Big\}
 \non\\
 &&-\vp^*_3\cdot P\vp^*_4\cdot P\Big[2 A^{(2)}_1-p_3\cdot
 p_4(A^{(1)}_1-A^{(1)}_2-A^{(2)}_2+A^{(2)}_4)\Big]\Bigg\},
 \en
where $k=p_1-p_3$. It can be easily seen that the FSI contribution
from the $\ov B\to D D_s$ decay will affect both $A_L$ and
$A_\parallel$ amplitudes of the $\ov B\to \phi \ov K {}^*$ decay.

The absorptive part contributions of $\overline B\to D^*_s
D\to\phi \overline K^*$ amplitudes via $D_s$ and $D_s^*$ exchanges
are given by
 \be \label{eq:DsstD}
 \A bs\,(D^*_sD;D_s) &=& {1\over 2}\int {d^3\vec p_1\over (2\pi)^32E_1}\,{d^3 \vec p_2\over
  (2\pi)^3 2E_2}\,(2\pi)^4\delta^4(p_B-p_1-p_2)\frac{A(\ov B^0\to
  D^*_s D)}{2\vp^*_1\cdot P}
\non \\
 &&\times  \sum_{\lambda_1}{2\vp^*_1\cdot P}\,(-4i) f_{D^*_s D_s \phi} {F^2(t,m_{D_s})\over
 t-m_{D_s}^2}(-2i)g_{D D_s K^*}\,
 [p_3,\vp^*_3,p_1,\vp_1] p_2\cdot \vp^*_4
 \non\\
 &=&{1\over 2}\int {d^3\vec p_1\over (2\pi)^32E_1}\,{d^3 \vec p_2\over
(2\pi)^3 2E_2}\,(2\pi)^4\delta^4(p_B-p_1-p_2)\frac{A(\ov B^0\to
D^*_s D)}{2\vp^*_1\cdot P}
\non \\
 &&\times (-i)[\vp_3^*,\vp_4^*,p_3,p_4]\bigg\{-16i\,f_{D^*_s D_s \phi} {F^2(t,m_{D_s})\over
 t-m_{D_s}^2}g_{D D_s K^*}\, A^{(2)}_1\bigg\},
 \non\\
 \A bs\,(D^*_sD;D^*_s) &=& {1\over 2}\int {d^3\vec p_1\over (2\pi)^32E_1}\,{d^3 \vec p_2\over
(2\pi)^3 2E_2}\,(2\pi)^4\delta^4(p_B-p_1-p_2)\frac{A(\ov B^0\to
D^*_s D)}{2\vp^*_1\cdot P}
\non \\
 &&\times  \sum_{\lambda_1}{2\vp^*_1\cdot P}\,(-4i) f_{D^*_s D^*_s \phi} {F^2(t,m_{D^*_s})\over
 t-m_{D^*_s}^2}4i\,f_{D D_s^* K^*}\,
 \non\\
 &&\times\vp_1^\rho(\sigma g_{\rho\mu} p_1\cdot\vp^*_3+p_{3\rho}\vp_{3\mu}^*-\vp_{3\rho}^*p_{3\mu})
          \bigg(-g^{\mu\nu}+\frac{k^\mu k^\nu}{m_{D^*_s}^2}\bigg)[p_4,\vp^*_4,p_2,{}^\nu]
 \non\\
 &=&{1\over 2}\int {d^3\vec p_1\over (2\pi)^32E_1}\,{d^3 \vec p_2\over
(2\pi)^3 2E_2}\,(2\pi)^4\delta^4(p_B-p_1-p_2)\frac{A(\ov B^0\to
D^*_s D)}{2\vp^*_1\cdot P}
\non \\
 &&\times (-i)[\vp_3^*,\vp_4^*,p_3,p_4]~(-32i)\,f_{D^*_s D^*_s \phi} {F^2(t,m_{D^*_s})\over
 t-m_{D^*_s}^2}f_{D D_s^* K^*}
 \bigg\{A^{(2)}_1\bigg[\sigma+(1-\sigma)\frac{P\cdot p_1}{m_1^2}\bigg]
 \non\\
 &&
 +(1-A^{(1)}_1-A^{(1)}_2)\bigg(P\cdot p_3-\frac {P\cdot p_1 p_1\cdot p_3}{m_1^2}\bigg)\bigg\},
  \en
where the dependence of the polarization vector in $A(\ov B^0\to
D^*_s D)$ has been extracted out explicitly and $\sigma\equiv
g_{D^*D^*V}/(2f_{D^*D^*V})$.

Similarly, the absorptive parts of $\overline B\to D_s D^*\to\phi
\overline K^*$ amplitudes via $D_s$ and $D^*_s$ exchanges are
given by
 \be \label{eq:DsDst}
 \A bs\,(D_sD^*;D_s) &=& {1\over 2}\int {d^3\vec p_1\over (2\pi)^32E_1}\,{d^3 \vec p_2\over
(2\pi)^3 2E_2}\,(2\pi)^4\delta^4(p_B-p_1-p_2)\frac{A(\ov B^0\to
D_s D^*)}{2\vp^*_2\cdot P}
\non \\
 &&\times  \sum_{\lambda_2}{2\vp^*_2\cdot P}\,2i g_{D_s D_s \phi} {F^2(t,m_{D_s})\over
 t-m_{D_s}^2}\,(-4i)f_{D^* D_s K^*}\,
 [p_4,\vp^*_4,p_2,\vp_2] p_1\cdot \vp^*_3
 \non\\
 &=&{1\over 2}\int {d^3\vec p_1\over (2\pi)^32E_1}\,{d^3 \vec p_2\over
(2\pi)^3 2E_2}\,(2\pi)^4\delta^4(p_B-p_1-p_2)\frac{A(\ov B^0\to
D_s D^*)}{2\vp^*_2\cdot P}
\non \\
 &&\times (-i)[\vp_3^*,\vp_4^*,p_3,p_4]\bigg\{16i\,g_{D_s D_s \phi} {F^2(t,m_{D_s})\over
 t-m_{D_s}^2}f_{D^* D_s K^*}\, A^{(2)}_1\bigg\},
 \non\\
 \A bs\,(D_sD^*;D^*_s) &=& {1\over 2}\int {d^3\vec p_1\over (2\pi)^32E_1}\,{d^3 \vec p_2\over
(2\pi)^3 2E_2}\,(2\pi)^4\delta^4(p_B-p_1-p_2)\frac{A(\ov B^0\to
D_s D^*)}{2\vp^*_2\cdot P}
\non \\
 &&\times  \sum_{\lambda_2}{2\vp^*_2\cdot P}\,4i\, f_{D_s D^*_s \phi} {F^2(t,m_{D^*_s})\over
 t-m_{D^*_s}^2}(4i)f_{D^* D_s^* K^*}\,
 \non\\
 &&\times[p_3,\vp^*_3,p_1,{}_\mu]\bigg(-g^{\mu\nu}+\frac{k^\mu k^\nu}{m_{D^*_s}^2}\bigg)
  (\sigma g_{\nu\alpha} p_2\cdot\vp^*_4-p_{4\nu}\vp_{4\alpha}^*+\vp_{4\nu}^*p_{4\alpha})
          \vp_2^\alpha
 \non\\
 &=&{1\over 2}\int {d^3\vec p_1\over (2\pi)^32E_1}\,{d^3 \vec p_2\over
(2\pi)^3 2E_2}\,(2\pi)^4\delta^4(p_B-p_1-p_2)\frac{A(\ov B^0\to
D_s D^*)}{2\vp^*_2\cdot P}
\non \\
 &&\times (-i)[\vp_3^*,\vp_4^*,p_3,p_4]~32i\,f_{D_s D^*_s \phi} {F^2(t,m_{D^*_s})\over
 t-m_{D^*_s}^2}f_{D^* D_s^* K^*}
 \bigg\{A^{(2)}_1\bigg[\sigma+(1-\sigma)\frac{P\cdot p_2}{m_2^2}\bigg]
 \non\\
 &&
 +(A^{(1)}_1-A^{(1)}_2)\bigg(P\cdot p_4-\frac {P\cdot p_2 p_2\cdot p_4}{m_2^2}\bigg)\bigg\},
  \en
where again the dependence of the polarization vector in $A(\ov
B^0\to D_s D^*)$ has been extracted out explicitly. Note that
through rescattering FSI, both $\ov B\to D^*_s D$ and $\ov B\to
D_s D^*$ will affect only the $A_{\bot}$ term of the $\ov B\to\phi
\ov K {}^*$ decay amplitude.

To consider the FSI effect from the decay $\overline B\to D^*_s
D^*$, we denote the decay amplitude as
 \be
 A(\overline B\to D^*_s(p_1,\lambda_1) D^*(p_2,\lambda_2))=\vp^{*\mu}_1\vp^{*\nu}_2(a\,
 g_{\mu\nu}+bP_\mu P_\nu+ic\,[{}_\mu,{}_\nu,P,p_2]).
 \en
The absorptive part of the $\overline B\to D^*_s D^*\to\phi
\overline K^*$ amplitude via a $t$-channel $D_s$ exchange has the
expression
 \be \label{eq:DsstDstDs}
 \A bs\,(D^*_s D^*;D_s) &=& {1\over 2}\int {d^3\vec p_1\over (2\pi)^32E_1}\,{d^3 \vec p_2\over
 (2\pi)^3 2E_2}\,(2\pi)^4\delta^4(p_B-p_1-p_2)(a\,g_{\mu\nu}+bP_\mu
 P_\nu+ic[{}_\mu,{}_\nu,P,p_2])
\non \\
 &&\times  \sum_{\lambda_1,\lambda_2}{\vp^{*\mu}_1\vp^{*\nu}_2}\,(-4i) f_{D^*_s D_s \phi} {F^2(t,m_{D_s})\over
 t-m_{D_s}^2}\,(-4i)f_{D^* D_s K^*}\,
 [p_3,\vp^*_3,p_1,\vp_1] [p_4,\vp^*_4,p_2,\vp_2]
 \non\\
 &=&{1\over 2}\int {d^3\vec p_1\over (2\pi)^32E_1}\,{d^3 \vec p_2\over
(2\pi)^3 2E_2}\,(2\pi)^4\delta^4(p_B-p_1-p_2)16 f_{D^*_s D_s \phi}
{F^2(t,m_{D_s})\over t-m_{D_s}^2} f_{D^* D_s K^*}
\non \\
 &&\times\Bigg\{\vp^*_3\cdot\vp^*_4\{2 a\, p_3\cdot p_4 A^{(2)}_1\!\!
 -[(p_3\cdot p_4)^2\!\!-m_3^2 m_4^2][a(A^{(1)}_1\!\!-A^{(1)}_2\!\!-A^{(2)}_2\!\!+A^{(2)}_4)+b
 A^{(2)}_1]\}
 \non\\
 &&-\vp^*_3\cdot P\vp^*_4\cdot P\{2 a\, A^{(2)}_1
 -p_3\cdot p_4[a(A^{(1)}_1-A^{(1)}_2-A^{(2)}_2+A^{(2)}_4)+b
 A^{(2)}_1]\}
 \non\\
 &&-i[\vp_3^*,\vp_4^*,p_3,p_4]\,2c
 \Big\{P\cdot p_3(A^{(2)}_1-A^{(3)}_1-A^{(3)}_2)+P\cdot p_4 (A^{(3)}_1-A^{(3)}_2)-p_3\cdot p_4 A^{(3)}_2
 \non\\
 &&+[(p_3\cdot p_4)^2-m_3^2 m_4^2](A^{(2)}_3-A^{(2)}_4-A^{(3)}_4+A^{(3)}_6)\Big\}\Bigg\}.
  \en
We see that the above FSI contributes to all three polarization
components. The contribution from the $D^*_s$ exchange is
 \be \label{eq:DsstDstDsst}
 \A bs\,(D^*_s D^*;D^*_s) &=& {1\over 2}\int {d^3\vec p_1\over (2\pi)^32E_1}\,{d^3 \vec p_2\over
(2\pi)^3 2E_2}\,(2\pi)^4\delta^4(p_B-p_1-p_2)(a\,
 g_{\delta\beta}+bP_\delta P_\beta+ic[{}_\delta,{}_\beta,P,p_2])
\non \\
 &&\times  \sum_{\lambda_1,\lambda_2}{\vp^{*\delta}_1\vp^{*\beta}_2}\,
 (-4i) f_{D^*_s D^*_s \phi} {F^2(t,m_{D^*_s})\over t-m_{D^*_s}^2}(4i)f_{D^* D^*_s K^*}\,
 \vp_1^\rho(\sigma g_{\rho\mu} p_1\cdot\vp^*_3\!\!+p_{3\rho}\vp_{3\mu}^*\!\!-\vp_{3\rho}^*p_{3\mu})
\non\\
      &&\times \bigg(-g^{\mu\nu}+\frac{k^\mu k^\nu}{m_{D^*_s}^2}\bigg)
        (\sigma g_{\nu\alpha} p_2\cdot\vp^*_4-p_{4\nu}\vp_{4\alpha}^*+\vp_{3\nu}^*p_{3\alpha})
          \vp_2^\alpha
 \non\\
 &=&{1\over 2}\int {d^3\vec p_1\over (2\pi)^32E_1}\,{d^3 \vec p_2\over
(2\pi)^3 2E_2}\,(2\pi)^4\delta^4(p_B-p_1-p_2)\,16 f_{D^*_s D^*_s
\phi} {F^2(t,m_{D^*_s})\over t-m_{D^*_s}^2} f_{D^* D^*_s K^*}
\non \\
 &&\times\Bigg\{\vp^*_3\cdot\vp^*_4
 \bigg\{\bigg[-1+\frac{(1-\sigma)^2}{m_{D^*_s}^2}A^{(2)}_1\bigg]
 \bigg[a\,p_3\cdot p_4+b p_3\cdot p_2 p_1\cdot p_4
 \non\\
 &&+\frac{a+b p_1\cdot
 p_2}{m_1^2 m_2^2}(-m_2^2 p_3\cdot p_1 p_1\cdot p_4-m_1^2 p_3\cdot
 p_2 p_2\cdot p_4+p_3\cdot p_1 p_1\cdot p_2 p_2\cdot p_4)\bigg]
 \non\\
 &&+a\bigg(-p_3\cdot p_4+\frac{p_3\cdot k k\cdot
 p_4}{m_{D_s^*}^2}\bigg)
 +A^{(2)}_1\bigg\{\sigma^2\bigg[2a-b
 p_1\cdot p_2+\frac{a+b p_1\cdot
 p_2}{m_1^2 m_2^2}(p_1\cdot p_2)^2\bigg]
 \non\\
 &&+\sigma\bigg[b-\frac{a+b p_1\cdot
 p_2}{m_1^2 m_2^2} p_1\cdot p_2\bigg]P^2
 \non\\
 &&+(\sigma -1)\frac{p_4\cdot k}{m_{D^*_s}^2}\bigg[b p_2\cdot p_3+\frac{a+b p_1\cdot
 p_2}{m_1^2 m_2^2}(m_1^2 p_2\cdot p_3-m_2^2 p_1\cdot p_3-p_2\cdot
 p_1 p_1\cdot p_3)\bigg]
 \non\\
 &&+(\sigma -1)\frac{p_3\cdot k}{m_{D^*_s}^2}\bigg[-b p_1\cdot p_4+\frac{a+b p_1\cdot
 p_2}{m_1^2 m_2^2}(m_1^2 p_2\cdot p_4-m_2^2 p_1\cdot p_4+p_1\cdot
 p_2 p_2\cdot p_4)\bigg]
 \non\\
 &&+\bigg(p_3\cdot p_4-\frac{p_3\cdot k k\cdot
 p_4}{m_{D^*_s}^2}\bigg)\bigg[b+\frac{a+b p_1\cdot
 p_2}{m_1^2 m_2^2}(m_1^2+m_2^2+p_1\cdot p_2)\bigg]
 \bigg\}\bigg\}
 \non\\
 &&-\vp^*_3\cdot P\vp^*_4\cdot P
 \bigg\{\frac{(1-\sigma)^2}{m_{D^*_s}^2}(A^{(1)}_1-A^{(1)}_2-A^{(2)}_2+A^{(2)}_4)
 \bigg[a\,p_3\cdot p_4+b p_3\cdot p_2 p_1\cdot p_4
 \non\\
 &&+\frac{a+b p_1\cdot p_2}{m_1^2 m_2^2}(-m_2^2 p_3\cdot p_1 p_1\cdot p_4-m_1^2 p_3\cdot
 p_2 p_2\cdot p_4+p_3\cdot p_1 p_1\cdot p_2 p_2\cdot p_4)\bigg]
 \non\\
 &&+\sigma^2(A^{(1)}_1-A^{(1)}_2-A^{(2)}_2+A^{(2)}_4)\bigg[2a-b
 p_1\cdot p_2+\frac{(p_1\cdot p_2)^2}{m_1^2 m_2^2}(a+b p_1\cdot
 p_2)\bigg]
 \non\\
 &&-\sigma\bigg[b-\frac{a+b p_1\cdot
 p_2}{m_1^2 m_2^2} p_1\cdot p_2\bigg][(A^{(2)}_2-A^{(2)}_4)P^2+(1-2 A^{(1)}_1) p_1\cdot p_3
 \non\\
 &&-(A^{(1)}_1-A^{(1)}_2)(p_1\cdot p_4+p_2\cdot p_3)]
 -\bigg[1-(1-\sigma)(A^{(1)}_1-A^{(1)}_2)\frac{p_4\cdot k}{m_{D^*_s}^2}\bigg]
 \non\\
 &&\times
 \bigg[a+\frac{a+b p_1\cdot p_2}{m_1^2 m_2^2}(-m_1^2 p_2\cdot
 p_3+p_2\cdot p_1 p_1\cdot p_3)\bigg]
 \non\\
 &&-\bigg[1+(1-\sigma)(1-A^{(1)}_1-A^{(1)}_2)\frac{p_3\cdot k}{m_{D^*_s}^2}\bigg]
 \bigg(a+b p_1\cdot p_4-\frac{a+b p_1\cdot p_2}{m_2^2} p_2\cdot
 p_4\bigg)
 \non\\
 &&-\bigg[A^{(1)}_1+A^{(1)}_2-(1-\sigma)(A^{(2)}_2-A^{(2)}_4)\frac{p_4\cdot
 k}{m_{D^*_s}^2}\bigg] \non \\
 && \times \bigg[b p_2\cdot p_3
 +\frac{a+b p_1\cdot p_2}{m_1^2 m_2^2}(m_1^2 p_2\cdot p_3-m_2^2 p_1\cdot p_3-p_2\cdot p_1 p_1\cdot p_3)\bigg]
 \non\\
 &&+\bigg[A^{(1)}_1-A^{(1)}_2+(1-\sigma)\frac{p_3\cdot k}{m_{D_s^*}^2}
 (A^{(1)}_1-A^{(1)}_2-A^{(2)}_2+A^{(2)}_4)\bigg)
 \non\\
 &&\times\bigg[b p_1\cdot p_4
 -\frac{a+b p_1\cdot p_2}{m_1^2 m_2^2}(m_1^2 p_2\cdot p_4-m_2^2 p_1\cdot p_4+p_1\cdot p_2 p_2\cdot p_4)\bigg]
 \non\\
 &&+\bigg(p_3\cdot p_4-\frac{p_3\cdot k k\cdot p_4}{m_{D_s^*}^2}\bigg)
 \Big\{b(A^{(1)}_1+A^{(1)}_2-A^{(2)}_2+A^{(2)}_4)
 \non\\
 &&+\frac{a+b p_1\cdot p_2}{m_1^2 m_2^2}
 [p_1\cdot p_2(A^{(1)}_1-A^{(1)}_2-A^{(2)}_2+A^{(2)}_4)
 -m_1^2(1-2 A^{(1)}_1+A^{(2)}_2-A^{(2)}_4)
 \non\\
 &&-m_2^2(A^{(2)}_2-A^{(2)}_4)]\Big\}\bigg\}
 -i[\vp_3^*,\vp_4^*,p_3,p_4]\,2c
 \bigg\{\bigg(p_3\cdot p_4-\frac{p_3\cdot k k\cdot p_4}{m_{D_s^*}^2}\bigg)A^{(1)}_2
 \non\\
 &&
 -\bigg[2\sigma-(1-\sigma)\frac{q\cdot k}{2m_{D^*_s}^2}\bigg]A^{(2)}_1\bigg\}\Bigg\},
 \en
with $P=p_3+p_4$, $q=p_3-p_4$ and $k=p_1-p_3=p_4-p_2$ as before.
Clearly the above rescattering amplitude contributes to all three
polarization components $A_{L,\parallel,\bot}$.

\subsection{Numerical results}

In order to perform a numerical study of  the analytic results
obtained in the previous subsection, we need to specify the
short-distance $A(\ov B\to D_s^{(*)} D^{(*)})$ amplitudes. In the
factorization approach, we have
 \be
 A(\ov B\to D_s D)_{SD}&=&i\frac{G_{\rm F}}{\sqrt2} V_{cb} V^*_{cs} a_1
 f_{D_s}(m_B^2-m_D^2) F_0^{BD}(m_{D_s}^2),
 \non\\
 A(\ov B\to D^*_s D)_{SD} &=&\frac{G_{\rm F}}{\sqrt2} V_{cb} V^*_{cs} a_1
 f_{D^*_s} m_{D_s^*} F_1^{BD}(m_{D^*_s}^2)(2\vp^*_{D_s^*}\cdot p_B),
 \non\\
 A(\ov B\to D_s D^*)_{SD} &=&\frac{G_{\rm F}}{\sqrt2} V_{cb} V^*_{cs} a_1
 f_{D_s} m_{D^*} A_0^{BD^*}(m_{D_s}^2)(2\vp^*_{D^*}\cdot p_B),
 \non\\
 A(\ov B\to D^*_s D^*)_{SD} &=& -i\frac{G_{\rm F}}{\sqrt2} V_{cb} V^*_{cs} a_1
 f_{D^*_s} m_{D_s^*} (m_B+m_{D^*})\vp^{*\mu}_{D_s^*}\vp^{*\nu}_{D^*}
 \bigg[A_1^{BD^*}(m_{D^*_s}^2) g_{\mu\nu}
 \non\\
 &&-\frac{2 A_2^{BD^*}(m_{D^*_s}^2)}{(m_B+m_{D^*})^2}p_{B\mu} p_{B\nu}
 -i\frac{2 V^{BD^*}(m_{D^*_s}^2)}{(m_B+m_{D^*})^2}
 \epsilon_{\mu\nu\alpha\beta}p_B^\alpha p_{D^*}^\beta \bigg].
 \en
Writing $A=A^{SD}+i \a bs A^{LD}$ with $\a bs A^{LD}$ being given
in the previous subsection and using the form factors given in
\cite{CCH}, we obtain the numerical results for $B\to\phi K^*$
amplitudes as exhibited in Table~\ref{tab:phiKst}. The branching
ratio after the inclusion of FSIs is
 \be
 {\mathcal B}(B\to\phi K^*)=(9.2^{+4.3}_{-2.7})\times 10^{-6}
 \en
for $\Lambda_{D_s,D^{*}_s}=m_{D_s,D^{*}_s}+\eta\,\Lambda_{\rm
QCD}$ with $\eta=0.80$ and $15\%$ error in $\Lambda_{\rm QCD}$.
Several remarks are in order. (i) The polarization amplitudes are
very sensitive to the cutoffs $\Lambda_{D_s,D^*_s}$.  (ii) We have
assumed a monopole behavior [i.e. $n=1$ in Eq. (\ref{eq:FF})] for
the form factor $F(t,m_{D_s})$ and a dipole form (i.e. $n=2$) for
$F(t,m_{D_s^*})$.\footnote{For $B\to D\pi,\,K\pi,\pi\pi$ decays,
assuming a dipole form for the form factor appearing in the
$V_{\rm exc}VP$ or $V_{\rm exc}VV$ vertex with $V_{\rm exc}$ being
the exchanged vector particle [see e.g. Figs. \ref{fig:BDpi}(e),
\ref{fig:BDpi}(f), \ref{fig:BKpi}(d) and \ref{fig:Bpipi}(e)] and a
monopole behavior for all other form factors, we found that the
best $\chi^2$ fit to the measured decay rates is very similar to
the one with the monopole form for all the form factors. Only the
cutoff $\Lambda$ or the parameter $\eta$ appearing in Eq.
(\ref{eq:Lambda}) is slightly modified. The resultant branching
ratios and \CP asymmetries remain almost the same. The same
exercise also indicates that if the dipole behavior is assumed for
all the form factors, then the $\chi^2$ value becomes too poor to
be acceptable. This is the another good reason why we usually
choose $n=1$ in Eq. (\ref{eq:FF}) for the form factors except for
the one appearing in $V_{\rm exc}VP$ or $V_{\rm exc}VV$ vertex.}
If the monopole form for $F(t,m_{D_s^*})$ is utilized, the
parameter $\eta$ will become unnaturally small in order to avoid
too large FSI contributions. (iii) the interference between
short-distance and long-distance contributions is mild. This is
because $B\to\phi K^*$ and $B\to D^{(*)}_s D^{(*)}$ amplitudes
should have the similar weak phase as $|V_{cb}V^*_{cs}|\gg
|V_{ub}V_{us}^*|$. \footnote{The weak phase of $SD$ $\ov
B\to\phi\ov K^*$ and $\ov B\to D_s^{(*)} D^{(*)}$ is not exactly
the same. Moreover, our $\ov B\to\phi\ov K^*$ $SD$ amplitude is
not purely imaginary. It carries a phase as shown in
Table~\ref{tab:phiKst} mainly because of the phase difference in
$a_4^u$ and $a_4^c$ in our case [cf. Eq.~(\ref{eq:ai})].}
As a result, the long-distance contribution is essentially
``orthogonal" to the short-distance one. (iv) The calculated
strong phases shown in Table~\ref{tab:phiKst} are in agreement
with experiment for $\phi_\parallel$ but not for
$\phi_\bot$.\footnote{In Table \ref{tab:BphiKV}, the experimental
results of $\phi_\parallel$ and $\phi_\bot$ are for $B\to \phi
K^*$ decays. The Belle results for ($\phi_\parallel,\phi_\bot)$
should be transformed to ($-\phi_\parallel,\pi-\phi_\bot)$. This
phase transformation will not modify the experimentally measured
distributions (see \cite{BaBarVV1} for details) but it will flip
the relative sign between $\phi_\parallel$, $\phi_\bot$ and lead
to $|A_+|>|A_-|$ for $B\to\phi K^*$ as expected from the
factorization approach. Since our calculations are for $\ov B\to
\phi\ov K^{*}$ decays, in Table \ref{tab:phiKst} we have
transformed BaBar and Belle results from $\phi_\bot$ to
$\phi_\bot-\pi$ so that $|A_+|<|A_-|$ in $\ov B\to \phi \ov
K^{*}$. Note that in the absence of FSIs,
$\phi_\parallel=\phi_\bot=\pi$ for $B\to \phi K^*$ and
$\phi_\parallel=\pi$, $\phi_\bot=0$ for $\ov B\to \phi \ov K^*$.}
(v) There are large cancellations among various long-distance
contributions, e.g. between  $A(D^*_s D;D^{(*)}_s)$ and $A(D_s
D^*; D^{(*)}_s)$.

\begin{table}[t]
\caption{Short-distance ($SD$) and long-distance ($LD$)
contributions to the decay amplitudes (in unit of $10^{-8}$\,GeV)
and the polarization fractions for $\ov B\to\phi \ov K^*$. $LD$
contributions are calculated with
$\Lambda_{D_s,D^{*}_s}=m_{D_s,D^{*}_s}+\eta \Lambda_{\rm QCD}$,
where $\eta=0.8$ and errors come from a $15\%$ variation on
$\Lambda_{\rm QCD}$.}
 \label{tab:phiKst}
 \begin{ruledtabular}
 \begin{tabular}{ c c c c}
    & longitudinal ($L$)
    & parallel ($\parallel$)
    & perpendicular ($\bot$)
 \\
    \hline
 $A^{SD}$
    & $0.63-1.98i$
    & $-0.18+0.55i$
    & $0.15-0.46i$
    \\
 $A^{SD+LD}$
    & $-0.57^{+0.33}_{-0.39}-(2.03\pm0.01)i$
    & $2.28^{+0.77}_{-0.67}+(0.64\pm 0.03)i$
    & $0.32\pm0.05-(0.44\pm 0.00)i$
    \\
    \hline
 $A(D_s D;D_s)$
    & $-1.10^{+0.30}_{-0.34}-(0.04\pm0.01)i$
    & $0.07\pm0.02+0.00i$
    & 0
    \\
 $A(D_s D;D^*_s)$
    & $0.00$
    & $-0.02\pm0.01-0.00i$
    & 0
    \\
 $A(D^*_s D;D_s)$
    & 0
    & 0
    & $-0.19\pm0.05-0.01i$
    \\
 $A(D^*_s D;D^*_s)$
    & 0
    & 0
    & $-0.01-0.00i$
    \\
 $A(D_s D^*;D_s)$
    & 0
    & 0
    & $0.17\pm0.05+0.01i$
    \\
   $A(D_s D^*;D^*_s)$
    & 0
    & 0
    & $0.01+0.00i$
    \\
  $A(D^*_s D^*;D_s)$
    & $-0.06\pm0.02-0.00i$
    & $2.35^{+0.73}_{-0.63}+(0.09^{+0.03}_{-0.02})i$
    & $0.21^{+0.07}_{-0.06}+0.01i$
    \\
  $A(D^*_s D^*;D_s^*)$
    & $-0.04^{+0.02}_{-0.03}-0.00i$
    & $0.06^{+0.04}_{-0.03}+0.00i$
    & $-0.01\pm0.01-0.00i$
    \\
     \hline
  $f^{SD}$
    &0.88
    &0.07
    &0.05
    \\
 $f^{SD+LD}$
    &$0.43^{+0.13}_{-0.09}$
    &$0.54^{+0.10}_{-0.14}$
    &$0.03\pm 0.01$
    \\
 $f^{\rm expt}$
    &$0.51\pm0.04$~\footnotemark[1]
    &$0.27\pm0.06$~\footnotemark[1]
    &$0.22\pm0.04$~\footnotemark[1]
    \\  \hline
 $\arg(A_i A^*_L)^{SD}$
    & --
    &$\pi$
    &0
    \\
 $\arg(A_i A^*_L)^{SD+LD}$
    & --
    &$2.12^{+0.11}_{-0.07}$
    &$0.89^{+0.21}_{-0.24}$
    \\
 $\arg(A_i A^*_L)^{\rm BaBar}$
    & --
    &$2.34^{+0.23}_{-0.20}\pm0.05$
    &$-0.67\pm0.25\pm0.05$~\footnotemark[2]
    \\
 $\arg(A_i A^*_L)^{\rm Belle}$
    & --
    &$2.21\pm0.22\pm0.05$~\footnotemark[2]
    &$-0.72\pm0.21\pm0.06$~\footnotemark[2]
\end{tabular}
\end{ruledtabular}
\footnotetext[1]{$f^{\rm expt}$ is the experimental average of
polarization fractions for $B^-\to\phi K^{*-}$ and $\ov B^0\to\phi
\ov K^{*0}$.}
 \footnotetext[2]{See footnote 18 for details.}
\end{table}

The cancellation in $\overline B\to D_s^* D\to\phi K^*$ and
$\overline B\to D_s D^*\to\phi K^*$ contributions can be
understood from SU(3) symmetry, \CP conjugation~\footnote{The
combined symmetry is reminiscent of CPS symmetry which is used
extensively in kaon decay matrix elements on the lattice
\cite{CPS,xghe}.} and the similarity in the size of source
amplitudes. To see this, let $\la\phi(\bar s s) K^*(\bar q s)
|T_{\rm st}|D_s^*(\bar c s) D(\bar q c)\ra$ be the $D_s^* D\to\phi
K^*$ rescattering matrix element, which is represented in the
right-hand part of Fig.~\ref{fig:BphiKst}. After changing $s\to q$
in $D_s^*(\bar c s)$ and $\phi(\bar s s)$ (on the upper right part
of Fig.~\ref{fig:BphiKst}), and $\bar q\to \bar s$ in $D(\bar q
c)$ and $K^*(\bar q s)$ (on the lower right part of of the same
figure) under the SU(3) symmetry transformation, the above matrix
element becomes $\la K^*(\bar s q) \phi(\bar s s)|T_{\rm
st}|D^*(\bar c q) D_s(\bar s c)\ra$. Applying charge conjugation
after the SU(3) transformation, we are led to
 \be
 \la \phi(\bar s s) K^*(\bar q s)|T_{\rm st}|D_s^*(\bar c s) D(\bar q c)\ra
 =-\la K^*(\bar q s)
\phi(\bar s s)|T_{\rm st}|D^*(\bar q c) D_s(\bar c s)\ra,
 \en
where the overall negative sign arises from the transformation of
vector particles under charge conjugation. In the $\ov B\to D^*_s
D\to\phi K^*$ rescattering amplitude, the above matrix element is
sandwiched by $\vp^*(p_1,\lambda)\cdot p_B$ from $\ov B\to D^*_s
D$ amplitudes and $\epsilon_{\mu\nu\rho\sigma}
\vp^\mu(p_3,\lambda')\vp^\nu(p_4,\lambda'')p_3^\rho p_4^\sigma$ to
project out the $A_\bot$ term. These factors bring the initial and
final states into $p$-wave configurations. Applying parity
transformation will interchange the particle momentum in the $B$
rest frame. Such an interchange does not yield a further sign flip
as the initial and final states are both in the $p$-wave
configuration. To be more specific, we have
 \be
 {\bf P}\sum_\lambda|D^*(p_1,\lambda)
 D(p_2)\ra~\vp(p_1,\lambda)^\mu p_{B\mu}=\sum_\lambda
 |D^*(p_2,-\lambda) D(p_1)\ra~[-\vp_\mu(p_2,-\lambda) p_B^\mu],
 \en
where the identity $\vp_\mu(\vec p,\lambda)=-\vp^\mu(-\vec
p,-\lambda)$ with $p_1^\mu=(E,\vec p)=p_{2\mu}$ in the $B$ rest
frame for the $m_1=m_2$ case and $p_1\leftrightarrow p_2$ under
the parity transformation ({\bf P}) on initial state particles
have been used. For the parity transformation of the final state,
we have
 \be
 &&{\bf
 P}\sum_{\lambda',\lambda''}|K^*(p_3,\lambda')\phi(p_4,\lambda'')\ra~
 \epsilon_{\mu\nu\rho\sigma}
 \vp^\mu(p_3,\lambda')\vp^\nu(p_4,\lambda'')p_3^\rho p_4^\sigma
 \non\\
 &&=\sum_{\lambda',\lambda''}|K^*(p_4,-\lambda')\phi(p_3,-\lambda'')\ra~
 \epsilon_{\mu\nu\rho\sigma}
 (-)^2\vp_\mu(p_4,-\lambda')\vp_\nu(p_3,-\lambda'')p_{4\rho}
 p_{3\sigma}\non\\
 &&=-\sum_{\lambda',\lambda''}|K^*(p_4,\lambda'')\phi(p_3,\lambda')\ra~
 \epsilon_{\mu\nu\rho\sigma}
 \vp^\mu(p_3,\lambda')\vp^\nu(p_4,\lambda'')p_3^\rho p_4^\sigma,
 \en
where similar identities and indices relabelling have been used.
We thus have
 \be
 &&\sum_{\lambda,\lambda'\lambda''}\epsilon_{\mu\nu\rho\sigma}
 \vp^\mu(p_3,\lambda')\vp^\nu(p_4,\lambda'')p_3^\rho p_4^\sigma
 ~\la \phi(p_3,\lambda') K^*(p_4,\lambda'')|T_{\rm st}|D_s^*(p_1,\lambda)
 D(p_2)\ra~\vp^*(p_1,\lambda)\cdot p_B
 \non\\
 &&=-\sum_{\lambda,\lambda'\lambda''}\epsilon_{\mu\nu\rho\sigma}
 \vp^\mu(p_3,\lambda')\vp^\nu(p_4,\lambda'')p_3^\rho p_4^\sigma
 ~\la K^*(p_4,\lambda'') \phi(p_3,\lambda') |T_{\rm st}|
 D^*(p_2,\lambda)D_s(p_1)\ra~\vp^*(p_2,\lambda)\cdot p_B
 \non\\
 \en
from  SU(3) [U(3)] symmetry and \CP conjugation.  The right-hand
side of the above equation is proportional to the $\ov B\to D_s
D^*\to\phi K^*$ rescattering matrix element with an additional
negative sign, which is the source of cancellation. This
cancellation turns out to be quite effective as the $\ov B\to
D^*_s D$ and $\ov B\to D_s D^*$ amplitudes are numerically
similar, i.e. $f_{D^*_s} m_{D^*_s} F_1^{BD}(m^2_{D^*_s})\simeq
f_{D_s} m_{D^*} A_0^{BD^*}(m_{D_s}^2)$. Note that SU(3) breaking
does not really help avoid the cancellation. Since the number of
$s$ quarks involved in $\ov B\to D_s^* D\to\phi\ov K^*$ and $\ov
B\to D_s D^*\to\phi \ov K^*$ is the same, SU(3) breaking affects
both amplitudes in a similar manner and the net effect cancels.

Because of the above-mentioned cancellation and the fact that
$\bar B\to D_s^* D^*$ does not give rise to a large $A_\bot$, we
cannot obtain large $f_\bot(\ov B\to\phi \ov K^*)$ through FSI.
The latter point can be seen from the fact that the polarization
fraction of $\ov B\to D_s^* D^*$ is $f_L:f_\parallel:f_\bot\simeq
0.51:0.41:0.08$. Although the $D_s^* D^*$ intermediate state has a
smaller $A_L$, it does not have a large $A_\bot$. The suppression
in $A_\bot$ can be understood from the smallness of the $p_c/m_B$
factor as shown in Eq.~(\ref{eq:Atrans}).
In principle, one could obtain a larger $A_\bot^{LD}(B\to \phi
K^*)$ by applying a larger $\Lambda_{D_s^*}$ to get a larger
$A(D^*_s D^*;D^*_s)$. However, such a large $A(D^*_s D^*;D^*_s)$
will also enhance $A_L^{LD}$, giving a too large $\ov B\to\phi\ov
K^*$ decay rate and does not really bring up $f_\bot$.
%

\subsection{$B^-\to K^{*-}\rho^0 $ decays}

In the factorization approach, the amplitude of the $\overline
B\to K^{*-}\rho^0 $ decay is given by
 \be
 \la K^{*-}(\lambda_{K^*}) \rho^0 (\lambda_{\rho})|H_{\rm eff}|B^-\ra
 &=&
 \frac{G_{\rm F}}{\sqrt2}
 \sum_{q=u,c} V_{qb} V^*_{qs}~a^q_{K^*\rho}~\la \rho|(\bar u u)_V|0\ra\,\la K^{*-}|(\bar
 s b)_{V-A}|B^-\ra
 \non\\
 &&+\frac{G_{\rm F}}{\sqrt2}
 \sum_{q=u,c} V_{qb} V^*_{qs}~b^q_{K^*\rho}~\la K^{*-}|(\bar s u)_V|0\ra\,\la \rho^0|(\bar
 u b)_{V-A}|B^-\ra
 \non\\
 &=&-i~
 \vp^{*\mu}_{K^*}(\lambda_{K^*})\vp^{*\nu}_\rho(\lambda_\rho)
 \bigg\{[\alpha_{K^*\rho} A_1^{BK^*}(m_\rho^2)+\beta_{K^*\rho} A_1^{B\rho}(m_{K^*}^2)]
 g_{\mu\nu}
 \non\\
 &&-2\bigg[\alpha_{K^*\rho}\frac{
 A_2^{BK^*}(m_\rho^2)}{(m_B+m_{K^*})^2}+\beta_{K^*\rho}\frac{
 A_2^{B\rho}(m_{K^*}^2)}{(m_B+m_\rho)^2}\bigg] p_{B\mu}
 p_{B\nu}
 \non\\
 &&-2i\bigg[-\alpha_{K^*\rho}\frac{ V^{BK^*}(m_\phi^2)}{(m_B+m_{K^*})^2}
           +\beta_{K^*\rho}\frac{ V^{B\rho}(m_{K^*}^2)}{(m_B+m_\rho)^2}\bigg]
 \epsilon_{\mu\nu\alpha\beta}p_B^\alpha
 p_{\rho}^\beta\bigg\},
 \non\\
  \la \ov K^{*0}(\lambda_{K^*}) \rho^- (\lambda_{\rho})|H_{\rm eff}|B^-\ra
 &=&
 \frac{G_{\rm F}}{\sqrt2}
 \sum_{q=u,c} V_{qb} V^*_{qs}~c^q_{K^*\rho}~\la \ov K^{*0}|(\bar s d)_V|0\ra\,\la \rho^-|(\bar
 d b)_{V-A}|B^-\ra
 \non\\
 &=&-i~\gamma_{K^*\rho}
 \vp^{*\mu}_{K^*}(\lambda_{K^*})\vp^{*\nu}_\rho(\lambda_\rho)
 \bigg\{ A_1^{B\rho}(m_{K^*}^2)
 g_{\mu\nu}
 \non\\
 &&-2\frac{A_2^{B\rho}(m_{K^*}^2)}{(m_B+m_\rho)^2} p_{B\mu} p_{B\nu}
 -2i\frac{ V^{B\rho}(m_{K^*}^2)}{(m_B+m_\rho)^2} \epsilon_{\mu\nu\alpha\beta}p_B^\alpha
 p_{\rho}^\beta\bigg\},
 \label{eq:KstrhoSD}
 \en
where
 \be
 a^u_{K^*\rho}&=&a_2+\frac{3}{2}(a_7+a_9),\quad
 a^c_{K^*\rho}=\frac{3}{2}(a_7+a_9),\quad
 b^u_{K^*\rho}=a_1+a^u_4+a^u_{10},\quad
 b^c_{K^*\rho}=a^c_4+a^c_{10},
 \non\\
c^u_{K^*\rho}&=&a^u_4-\frac{1}{2}a^u_{10},\quad
 c^c_{K^*\rho}=a^c_4-\frac{1}{2}a^c_{10},
 \non\\
 \alpha_{K^*\rho}&=&\frac{G_{\rm F}}{2} f_\rho m_\rho (m_B+m_{K^*})\sum_{q=u,c} V_{qb}
 V^*_{qs}~a^q_{K^*\rho},
 \quad
 \beta_{K^*\rho}=\frac{G_{\rm F}}{2} f_{K^*} m_{K^*}
 (m_B+m_\rho)\sum_{q=u,c} V_{qb} V^*_{qs}~b^q_{K^*\rho}, \non\\
 \gamma_{K^*\rho}&=&\frac{G_{\rm F}}{\sqrt2} f_{K^*} m_{K^*}
(m_B+m_\rho)\sum_{q=u,c} V_{qb} V^*_{qs}~c^q_{K^*\rho}.\non
 \en

\begin{table}[t]
\caption{Same as Table \ref{tab:phiKst} except for $B^-\to
K^{*-}\rho^0$ and $\Lambda_{D,D^{*}}=m_{D,D^{*}}+\eta\Lambda_{\rm
QCD}$.}
 \label{tab:Kstrho}
 \begin{ruledtabular}
 \begin{tabular}{ c c c c}
    & longitudinal ($L$)
    & parallel ($\parallel$)
    & perpendicular ($\bot$)
 \\
    \hline
 $A^{SD}$
    & $1.18-1.69i$
    & $-0.26+0.37i$
    & $0.16-0.10i$
    \\
 $A^{SD+LD}$
    & $0.11^{+0.30}_{-0.35}-(1.73^{+0.02}_{-0.01})i$
    & $1.43^{+0.53}_{-0.46}+(0.43\pm 0.02)i$
    & $0.25\pm0.03-0.10i$
    \\
    \hline
 $A(D_s D;D)$
    & $-1.01^{+0.27}_{-0.31}-(0.04\pm0.01)i$
    & $0.04\pm0.01+0.00i$
    & 0
    \\
 $A(D_s D;D^*)$
    & $0.00$
    & $-0.01\pm0.01-0.00i$
    & 0
    \\
 $A(D^*_s D;D)$
    & 0
    & 0
    & $-0.13^{+0.03}_{-0.04}-0.00i$
    \\
 $A(D^*_s D;D^*)$
    & 0
    & 0
    & $0.00$
    \\
 $A(D_s D^*;D)$
    & 0
    & 0
    & $0.11\pm0.03+0.00i$
    \\
   $A(D_s D^*;D^*)$
    & 0
    & 0
    & $0.00$
    \\
  $A(D^*_s D^*;D)$
    & $-0.03\pm0.01-0.00i$
    & $1.62^{+0.50}_{-0.44}+(0.06\pm0.02)i$
    & $0.12^{+0.04}_{-0.03}+0.00i$
    \\
  $A(D^*_s D^*;D^*)$
    & $-0.04\pm0.02-0.00i$
    & $0.04^{+0.03}_{-0.02}+0.00i$
    & $-0.01^{+0.00}_{-0.01}-0.00i$
    \\
     \hline
  $f^{SD}$
    &0.95
    &0.04
    &0.01
    \\
 $f^{SD+LD}$
    &$0.57^{+0.16}_{-0.14}$
    &$0.42^{+0.14}_{-0.16}$
    &$0.01$
    \\
 $f^{\rm expt}$
    &$0.96^{+0.04}_{-0.16}$
     \end{tabular}
\end{ruledtabular}
\end{table}

Likewise, we have
 \be
 A_0(K^{*-}\rho^0)&=&-i\bigg(-[\alpha_{K^*\rho} A_1^{BK^*}(m_\rho^2)+\beta_{K^*\rho} A_1^{B\rho}(m_{K^*}^2)]
 \frac{m_B^2-m_{K^*}^2-m_\rho^2}{2 m_{K^*} m_\rho}
 \non\\
 &&+2\bigg[\alpha_{K^*\rho}\frac{
 A_2^{BK^*}(m_\rho^2)}{(m_B+m_{K^*})^2}+\beta_{K^*\rho}\frac{
 A_2^{B\rho}(m_{K^*}^2)}{(m_B+m_\rho)^2}\bigg] \frac{m_B^2 p_c^2}{m_{K^*} m_\rho}\bigg),
 \non\\
 A_{\parallel}(K^{*-}\rho^0)&=&-i\sqrt2 [\alpha_{K^*\rho} A_1^{BK^*}(m_\rho^2)+\beta_{K^*\rho} A_1^{B\rho}(m_{K^*}^2)],
 \non\\
 A_{\bot}(K^{*-}\rho^0)&=&2\sqrt2i \bigg[-\alpha_{K^*\rho}\frac{ V^{BK^*}(m_\phi^2)}{(m_B+m_{K^*})^2}
           +\beta_{K^*\rho}\frac{ V^{B\rho}(m_{K^*}^2)}{(m_B+m_\rho)^2}\bigg]m_B
           p_c,
 \non\\
 A_0(\ov K^{*0}\rho^-)&=&-i\gamma_{K^*\rho}\bigg(-A_1^{B\rho}(m_{K^*}^2)\frac{m_B^2-m_{K^*}^2-m_\rho^2}{2 m_{K^*} m_\rho}
 +\frac{2A_2^{B\rho}(m_{K^*}^2)}{(m_B+m_\rho)^2} \frac{m_B^2 p_c^2}{m_{K^*} m_\rho}\bigg),
 \non\\
 A_{\parallel}(\ov K^{*0}\rho^-)&=&-i\gamma_{K^*\rho}\sqrt2 A_1^{B\rho}(m_{K^*}^2),
 \quad
 A_{\bot}(\ov K^{*0}\rho^-)=i\gamma_{K^*\rho} \frac{2\sqrt2 V^{B\rho}(m_{K^*}^2)}{(m_B+m_\rho)^2}m_B
           p_c.
 \label{eq:AiKstrho}
 \en
Similar to the calculation for $B\to \phi K^*$,  we obtain
 \be
 {\mathcal B}^{SD}(B^-\to K^{*-}\rho^0) &=& (4.1\pm2.2)\times 10^{-6},
  \non \\
 {\mathcal B}^{SD}(B^-\to \ov K^{*0}\rho^-) &=& (4.3\pm2.9)\times 10^{-6},
 \non \\
 f^{\rm SD}_L:f^{\rm SD}_\parallel:f^{\rm SD}_\bot(K^{*-}\rho^0) &=&
    0.95\pm 0.02:0.04\pm0.02:0.01\pm0.00,
 \non\\
 f^{\rm SD}_L:f^{\rm SD}_\parallel:f^{\rm SD}_\bot(\ov K^{*0}\rho^-) &=&
    0.92\pm 0.05:0.04\pm0.02:0.04\pm0.01,
 \label{eq:KstrhoSD}
 \en
where the errors are estimated from 10\% uncertainties in form
factors.

\begin{table}[t]
\caption{Same as Table \ref{tab:phiKst} except for $B^-\to \ov
K^{*0}\rho^-$.}
 \label{tab:Kstrho1}
 \begin{ruledtabular}
 \begin{tabular}{ c c c c}
    & longitudinal ($L$)
    & parallel ($\parallel$)
    & perpendicular ($\bot$)
 \\
    \hline
 $A^{SD}$
    & $0.43-2.04i$
    & $-0.10+0.45i$
    & $0.09-0.41i$
    \\
 $A^{SD+LD}$
    & $-1.09^{+0.42}_{-0.49}-(2.09\pm0.02)i$
    & $2.29^{+0.75}_{-0.65}+(0.54^{+0.03}_{-0.02})i$
    & $0.22\pm0.04-0.41i$
    \\
    \hline
 $A(D_s D;D)$
    & $-1.42^{+0.39}_{-0.44}-(0.05\pm0.02)i$
    & $0.06\pm0.02+0.00i$
    & 0
    \\
 $A(D_s D;D^*)$
    & $0.00$
    & $-0.02\pm0.01-0.00i$
    & 0
    \\
 $A(D^*_s D;D)$
    & 0
    & 0
    & $-0.18\pm0.05-0.00i$
    \\
 $A(D^*_s D;D^*)$
    & 0
    & 0
    & $-0.01-0.00i$
    \\
 $A(D_s D^*;D)$
    & 0
    & 0
    & $0.16^{+0.05}_{-0.04}+0.00i$
    \\
   $A(D_s D^*;D^*)$
    & 0
    & 0
    & $0.01+0.00i$
    \\
  $A(D^*_s D^*;D)$
    & $-0.04\pm0.01-0.00i$
    & $2.29^{+0.70}_{-0.62}+(0.09^{+0.03}_{-0.02})i$
    & $0.17^{+0.06}_{-0.05}+0.01i$
    \\
  $A(D^*_s D^*;D^*)$
    & $-0.05^{+0.02}_{-0.04}-0.00i$
    & $0.06^{+0.04}_{-0.03}+0.00i$
    & $-0.01\pm0.01-0.00i$
    \\
    \hline
  $f^{SD}$
    &0.92
    &0.04
    &0.04
    \\
 $f^{SD+LD}$
    &$0.49^{+0.11}_{-0.08}$
    &$0.49^{+0.08}_{-0.12}$
    &$0.02\pm0.01$
    \\
 $f^{\rm expt}$
    &$0.74\pm0.08$
\end{tabular}
\end{ruledtabular}
\end{table}

The formulae for long-distance contributions to $B^-\to
K^{*-}\rho^0$ decays are similar to that in $B\to\phi K^*$ with
the intermediate state $D^{(*)}_s$ and strong couplings
$f(g)_{D_s^{(*)} D_s^{(*)}\phi}$ and $f(g)_{D^{(*)} D^{(*)}_s
K^*}$ replaced by $D^{(*)}$, $f(g)_{D_s^{(*)} D^{(*)} K^*}$ and
$f(g)_{D^{(*)}D^{(*)}\rho}/\sqrt2$, respectively. Likewise, the
$B^-\to\ov K^{*0}\rho^-$ rescattering amplitudes can be obtained
by a similar replacement except for an additional factor of
$1/\sqrt2$ associated with $f(g)_{D^{(*)}D^{(*)}\rho}$.

Numerical results for $B^-\to K^{*-}\rho^0$ and $B^-\to
K^{*0}\rho^-$ are shown in Tables~\ref{tab:Kstrho} and
\ref{tab:Kstrho1}, respectively. Using
$\Lambda_{D^{(*)}}=m_{D^{(*)}}+\eta\Lambda_{\rm QCD}$ with
$\eta=0.8$, we obtain
 \be
 && {\mathcal B}^{SD+LD}(K^{*-}\rho^0)=(4.8^{+1.8}_{-0.9})\times 10^{-6},
\qquad
    {\mathcal B}^{SD+LD}(\ov K^{*0}\rho^-)=(10.3^{+4.9}_{-3.1})\times 10^{-6},
    \non \\
 && \A^{SD}(K^{*-}\rho^0)=0.30,  \qquad\qquad\qquad\qquad~~
    \A^{SD}(\ov K^{*0}\rho^-)=0.11, \non \\
 && \A^{SD+LD}(K^{*-}\rho^0)= 0.02^{+0.05}_{-0.02}, \qquad\quad
    \A^{SD+LD}(\ov K^{*0}\rho^-)\simeq0.00\,.
 \en
Note that the long-distance contributions are similar to that of
the $\phi \ov K^*$ case up to some SU(3) breaking and the $LD$
contributions to $K^{*-}\rho^0$ and $\ov K^{*0}\rho^-$ are related
through the above-mentioned factor of $\sqrt2$ from the
$D^{(*)}D^{(*)}\rho$ vertices. Before the inclusion of FSIs,
$K^{*-}\rho^0$ and $\ov K^{*0}\rho^-$  have similar rates as the
former receives additional tree contributions from $\T$ and $\C$
topologies. However, the LD contribution to the former is
suppressed at the amplitude level by a factor of $\sqrt{2}$ and
this accounts for the disparity in rates between these two modes.
The cancellation between $\ov B\to PV\to VV$ and $\ov B\to VP\to
VV$ amplitudes remains. Since the short distance $B^-\to
K^{*-}\rho^0$ amplitude has a non-vanishing weak phase, we have
interferences between $SD$ and $LD$ contributions resulting in
quite sensitive direct $CP$. Comparing to the data shown in
Table~\ref{tab:BphiKV}, we note that (i) the $K^{*-}\rho^0$ and
$\ov K^{*0}\rho^-$ rates are consistent with the data within
errors and (ii) the predicted $f_L(\ov K^{*0}\rho^-)$ agrees well
with experiment, while $f_L(K^{*-}\rho^0)$ is lower than the BaBar
data. This should be clarified by future improved experiments.

\subsection{FSI contributions from other channels}

For LD contributions to the $\ov B\to \phi \ov K^{(*)}$ decay, it
is also possible to have $B\to J/\psi K^{(*)}\to \phi K^*$ via a
$t$-channel $\eta_s$ (a pseudoscalar with the $s\bar s$ quark
content) exchange and $B\to K^{(*)} J/\psi\to \phi K^*$ via a
$t$-channel $K$ exchange. However, it does not seem that it can
fully explain the polarization anomaly for two reasons: (i) The
\CP and SU(3) [CPS] symmetry argument also leads to the
cancellation in the $B\to J/\psi K \to \phi K^*$ via a $t$-channel
$\eta_s$ exchange and $B\to K J/\psi\to \phi K^*$ via a
$t$-channel $K$ exchange contribution. (ii) The $B\to J/\psi K^*$
decay does not generate a large enough $A_\bot$ source. The
polarization fractions of $B\to J/\psi K^*$ are
$f_L:f_\parallel:f_\bot\simeq 0.47:0.35:0.17$, which differ not
much from that of $\ov B\to D_s^* D^*$.
On the other hand, it is known that SU(3) breaking and the
electromagnetic interaction take place in $J/\psi$ decays (in
particular, in the case of $J/\psi\to\ov K^* K,\,K^*\ov K$ and
$\phi\eta^{(\prime)}$ decays of interest) at the level of $\sim
20\%$ and $\sim 10\%$~\cite{J/psi}, respectively. The
above-mentioned cancellation may not be very effective.
Hence, we should look at the FSI contributions from this channel
more carefully.

The Lagrangian relevant for $J/\psi\to PV$ and $\phi\to PP$ decays
is given by
 \be
 {\cal L}=ig_{VPP} {\rm Tr}(V^\mu P \lrpartial_\mu P)+\frac{g_{J/\psi
 VP}}{\sqrt2} \epsilon^{\mu\nu\alpha\beta} \partial_\mu (J/\psi)_\nu
 {\rm Tr}(\partial_\alpha V_\beta P),
 \en
where $V (P)$ is the vector (pseudoscalar) nonet matrix. We have
$g_{\phi KK}=4.8$ and $g_{J/\psi K^* K}=2.6\times
10^{-3}$~GeV$^{-1}$ fitted from the data~\cite{PDG}. The
$g_{J\psi\phi\eta_s}$ coupling is estimated to be $0.4\, g_{J/\psi
K^* K}$, where the deviation comes from SU(3) breaking and the
contribution of electromagnetic interactions~\cite{J/psi}, and we
estimate $g_{K^*K\eta_s}\simeq g_{\phi KK}$.

It is straightforward to obtain
 \be \label{eq:DsstD}
 \A bs\,(J/\psi \ov K;\eta_s) &=& {1\over 2}\int {d^3\vec p_1\over (2\pi)^32E_1}\,{d^3 \vec p_2\over
(2\pi)^3 2E_2}\,(2\pi)^4\delta^4(p_B-p_1-p_2)\frac{A(\ov B^0\to
J/\psi \ov K)}{2\vp^*_1\cdot P}
\non \\
 &&\times (-i)[\vp_3^*,\vp_4^*,p_3,p_4]\bigg\{-2\sqrt2 i\,g_{J/\psi \phi\eta_s} {F^2(t,m_{\eta_s})\over
 t-m_{\eta_s}^2} g_{K^* K\eta}\, A^{(2)}_1\bigg\},
 \non\\
 \A bs\,(\ov K/ J\psi;K) &=& {1\over 2}\int {d^3\vec p_1\over (2\pi)^32E_1}\,{d^3 \vec p_2\over
(2\pi)^3 2E_2}\,(2\pi)^4\delta^4(p_B-p_1-p_2)\frac{A(\ov B^0\to
\ov K J/\psi)}{2\vp^*_2\cdot P}
\non \\
 &&\times (-i)[\vp_3^*,\vp_4^*,p_3,p_4]\bigg\{2\sqrt2 i\,g_{\phi KK} {F^2(t,m_{K})\over
 t-m_{K}^2}f_{J/\psi K^* K}\, A^{(2)}_1\bigg\},
\en with
 \be
 A(\ov B\to J/\psi \ov K)&=&\frac{G_{\rm F}}{\sqrt2} V_{cb} V^*_{cs}
 a_2
 f_{J/\psi} m_{J/\psi} F_1^{BK}(m_{J/\psi}^2)(2\vp^*_{J/\psi}\cdot
 p_B).
 \en
We obtain
 \be
 A^{LD}_\bot(J/\psi \ov K;\eta_s)=2.4\times 10^{-11}\,{\rm
 GeV},\quad
 A^{LD}_\bot(J/\psi \ov K; K)=-6.9\times 10^{-11}\,{\rm GeV},
 \en
where use of $a_2=0.28$, $f_{J/\psi}=405$ MeV, $m_{\eta_s}\simeq
800$~MeV and $F(t,m_{K,\eta_s})=1$ has been made. Although the
cancellation is incomplete as expected, the FSI contributions
(even without the form factor suppression) from the $\ov B\to
J/\psi \ov K$ decay are too small to explain the data (see
Tables~\ref{tab:BphiKV} and \ref{tab:phiKst}). The smallness of
these contributions is due to the smallness of the OZI suppressed
$g_{J/\psi PV}$ coupling.

One possibility of having a large $f_\bot$ is to circumvent the
cancellation in $\ov B\to VP\to VV$ and $\ov B\to PV\to VV$
rescatterings. For example, we may consider $B\to SA\to\phi K^*$
contributions, where $S$ and $A$ denotes scalar and axial vector
mesons, respectively. We need $SA$ instead of $VS,PA$ intermediate
states to match the $(\phi K^*)_{p-wave}$ quantum numbers. The
$B\to SA$ and $B\to AS$ amplitudes may not be similar (note that
$f_{D_{s0}^*}\ll f_{D_0^*})$ and the charge conjugation properties
of $A(^1P_1)$ and $A(^3P_1)$ are different.

To show this, we consider the FSI contributions to $f_\bot(VV)$
from the even-parity charmed meson (denoted by $D^{**}$)
intermediate states. The Lagrangian (${\cal L}_{D^{**}D^{**}V})$
for the interactions of the charmed meson $J^P_j=(0^+,1^+)_{1/2}$
multiplets and vector mesons is similar to (\ref{eq:LDDV}) for the
$s$-wave charm meson case. In chiral and heavy quark limits, the
effective Lagrangian can be expressed compactly in terms of
superfields \cite{Casalbuoni}
 \be
 {\cal L}_{D^{**}D^{**}V} =&& i\beta_1\la S_b v^\mu(V_\mu-\rho_\mu)_{ba}\ov S_a\ra+i\ov\lambda_1\la
 S_b\sigma^{\mu\nu}F_{\mu\nu}(\rho)_{ba}\ov S_a\ra,
 \en
with the superfield $S={1+v\!\!\!/\over
2}(D^\mu_1\gamma_\mu\gamma_5-D_0^*)$, which is similar to the
superfield $H$ except for the $\gamma_5$ factor and the phase in
front of $D_0^*$. Hence, the Lagrangian in terms of the component
fields can be obtained from Eq. (\ref{eq:LDDV}) with $g_{DDV}$,
$f_{D^*DV}$, $f_{D^*D^*V}$, $g_{D^*D^*V}$, $\beta$, $\lambda$,
$D^*$-field and $D$-field replaced by $g_{D^*_0 D^*_0 V}$, $f_{D_1
D^*_0V}$, $f_{D_1D_1V}$, $g_{D_1D_1V}$, $\beta_1$, $\ov\lambda_1$,
$D_1$-field and $-i D_0^*$-field, respectively, and with an
overall negative sign on $\cal L$.

\begin{table}[t]
\caption{Short-distance ($SD$) and long-distance ($LD$)
contributions from $D^{**}_s D^{**}$ intermediate states to the
decay amplitudes (in unit of $10^{-8}$\,GeV) and the polarization
fractions for $B\to\phi K^*$. $LD$ contributions are calculated
with
$\Lambda_{D^*_{s0},D_{s1}}=m_{D^*_{s0},D_{s1}}+\eta\Lambda_{\rm
QCD}$ and $\eta=1.5$.}
 \label{tab:pwave}
 \begin{ruledtabular}
 \begin{tabular}{ c c c c}
    & longitudinal ($L$)
    & parallel ($\parallel$)
    & perpendicular ($\bot$)
 \\
    \hline
 $A^{SD}$
   & $0.63-1.98i$
    & $-0.18+0.55i$
    & $0.15-0.46i$
    \\
 $A^{SD+LD}$
    & $-1.65-2.07i$
    & $0.50+0.58i$
    & $1.29-0.42i$
    \\
    \hline
 $A(D_{s0}^* D_0^*;D^*_{s0})$
    & $-0.33-0.01i$
    & $0.00+0.00i$
    & 0
    \\
 $A(D^*_{s0} D^*_0;D_{s1})$
    & $0.03+0.00i$
    & $-1.99-0.07i$
    & 0
    \\
 $A(D_{s1} D^*_0;D^*_{s0})$
    & 0
    & 0
    & $0.07+0.00i$
    \\
 $A(D_{s1} D^*_0;D_{s1})$
    & 0
    & 0
    & $1.15+0.04i$
    \\
 $A(D^*_{s0} D_1;D_{s0})$
    & 0
    & 0
    & $0.02+0.00i$
    \\
   $A(D_{s0}^* D_1;D_{s1})$
    & 0
    & 0
    & $0.26+0.01i$
    \\
  $A(D_{s1} D_1;D^*_{s0})$
    & $-0.01-0.00i$
    & $0.83+0.03i$
    & $0.14+0.01i$
    \\
  $A(D_{s1} D_1;D_{s1})$
    & $-1.97-0.07i$
    & $1.83+0.07i$
    & $-0.50-0.02i$
    \\
     \hline
  $f^{SD}$
    &0.88
    &0.07
    &0.05
    \\
 $f^{SD+LD}$
    &$0.74$
    &$0.06$
    &$0.19$
    \\
 $f^{\rm expt}$
    &$0.52\pm0.05$
    &$0.23\pm0.06$
    &$0.25\pm0.04$
\end{tabular}
\end{ruledtabular}
\end{table}

The absorptive part contributions of $\overline B\to D^{**}_{s}
D^{**}\to\phi \overline K^*$ via $D_{s0}^*$ and $D_{s1}$ exchanges
are given by
 \be \label{eq:Ds**D**}
 \A bs\,(D^*_{s0} D_0^*;\mbox{$D_{s0}^*$ or $D_{s1}$}) &=& \A bs\,(D_s D;\mbox{$D_s$ or $D_s^*$})\quad
      \mbox{with $D^*_{(s)}\to D_{(s)1}$ and $D_{(s)}\to D_{(s)0}^*$},
 \non\\
  \A bs\,(D_{s1} D_0^*;\mbox{$D_{s0}^*$ or $D_{s1}$}) &=& -i \A bs\,(D^*_s D;\mbox{$D_s$ or $D_s^*$})\quad
      \mbox{with $D^*_{(s)}\to D_{(s)1}$ and $D_{(s)}\to D_{(s)0}^*$},
 \non\\
 \A bs\,(D^*_{s0} D_1;\mbox{$D_{s0}^*$ or $D_{s1}$}) &=& i\A bs\,(D_sD^*;\mbox{$D_s$ or $D_s^*$})\quad
       \mbox{with $D^*_{(s)}\to D_{(s)1}$ and $D_{(s)}\to D_{(s)0}^*$},
 \non\\
 \A bs\,(D_{s1} D_1;\mbox{$D_{s0}^*$ or $D_{s1}$}) &=& \A bs\,(D_s^* D^*;\mbox{$D_s$ or $D_s^*$})\quad
       \mbox{with $D^*_{(s)}\to D_{(s)1}$ and $D_{(s)}\to
       D_{(s)0}^*$}, \non \\
  \en
and the $\ov B \to D^{**}_{s} D^{**}$ factorization amplitudes are
  \be
  A(\ov B\to D^*_{s0} D_0^*)&=&i\frac{G_{\rm F}}{\sqrt2} V_{cb} V^*_{cs} a_1
 f_{D_{s0}^*} (m_B^2-m_{D_0^*}^2) F_0^{BD_0^*}(m_{D^*_{s0}}^2),
 \non\\
  A(\ov B\to D_{s1} D_0^*)&=&-i\frac{G_{\rm F}}{\sqrt2} V_{cb} V^*_{cs} a_1
 f_{D_{s1}} m_{D_{s1}} F_1^{BD_0^*}(m_{D_{s1}}^2)(2\vp^*_{D_{s1}}\cdot p_B),
 \non\\
 A(\ov B\to D_{s0}^* D_1)&=&-i\frac{G_{\rm F}}{\sqrt2} V_{cb} V^*_{cs} a_1
 f_{D^*_{s0}} m_{D_1} V_0^{BD_1}(m_{D^*_{s0}}^2)(2\vp^*_{D_1}\cdot p_B),
 \non\\
 A(\ov B\to D^*_s D^*)&=&i\frac{G_{\rm F}}{\sqrt2} V_{cb} V^*_{cs} a_1
 f_{D_{s1}} m_{D_{s1}} (m_B-m_{D_1})\vp^{*\mu}_{D_{s1}}\vp^{*\nu}_{D_1}
 \bigg[V_1^{BD_1}(m_{D_{s1}}^2) g_{\mu\nu}
 \non\\
 &&-\frac{2 V_2^{BD_1}(m_{D_{s1}}^2)}{(m_B-m_{D^*})^2}p_{B\mu} p_{B\nu}
 -i\frac{2 A^{BD_1}(m_{D_{s1}}^2)}{(m_B-m_{D^*})^2}
 \epsilon_{\mu\nu\alpha\beta}p_B^\alpha p_{D_1}^\beta \bigg].
 \en
with decay constants and form factors given in \cite{CCH}. Note
that the $-i$ and $i$ factors appearing in (\ref{eq:Ds**D**})
originate from the phases differences of $D_{(s)0}^*$ and
$D_{(s)}$ in the superfields $S$ and $H$, respectively. From these
equations, it is now evident that the contributions from
$\overline B\to D_{s1} D_0^*\to\phi \overline K^*$ and $\overline
B\to D^*_{s0} D_1\to\phi \overline K^*$ rescatterings add up
instead of cancellation as noticed before.
To estimate the numerical significance of these contributions, we
shall use $\beta_1=1(\simeq \beta)$, $\ov\lambda_1=1\, {\rm
GeV}^{-1}(\simeq\lambda)$ and
$\Lambda_{D^*_{s0},D_{s1}}=m_{D^*_{s0},D_{s1}}+\eta \Lambda_{\rm
QCD}$ with $\eta=1.5$ and the monopole form for the form factors
$F(t,m_{D_s^*})$ and $F(t, m_{D_s1})$. Numerical results are shown
in Table~\ref{tab:pwave}. The branching ratio is
$\B^{SD+LD}(B\to\phi K^*)=8.4\times 10^{-6}$. It is interesting to
notice that most of the FSI contributions from the intermediate
$D^{**}_s D^{**}$ states add up in $A_\bot$ giving
$f_\bot\simeq0.2$, which is close to data.
Note that we only consider the contribution from $D_{s1/2}
D_{1/2}$ intermediate states. A more detailed study including
other multiplets such as $D_{3/2}$ along this line is worth
pursuing.

\subsection{Comparison with other works}

Although the analysis of the $\phi K^*$ polarization anomaly by
Colangelo, De Fazio and Pham (CDP) \cite{Colangelo04} is very
similar to ours, their results are quite different form ours in
some respects (see also \cite{Ladisa} for a different analysis
based on the Regge theory). First, in our case, the interference
between short-distance and long-distance contributions is mild
because the absorptive long-distance contribution is essentially
``orthogonal" to the short-distance one (see Table
\ref{tab:Kstrho}). By fixing $\Lambda=2.3$ GeV, to be compared
with our choice of
$\Lambda_{D_s,D^*_s}=m_{D_s,D^*_s}+\eta\Lambda_{\rm QCD}$ with
$\eta=0.8$, CDP introduced a parameter $r$ so that
$A=A_{SD}+rA_{LD}$ with $r=0$ corresponding to the absence of
rescattering. It appears from Fig. 3 of \cite{Colangelo04} that
there is a drastic SD and LD interference behavior in the CDP
calculation, namely, the resulting branching ratio is not
symmetric with respect to the reflection of $r$: $r\to -r$. It is
not clear to us why the absorptive LD contribution obtained by CDP
is not orthogonal to the SD one at all. Second, instead of
employing a small $r$ factor of order 0.08 to reduce the FSI
contribution, we use the form factors with different momentum
dependence for the exchanged particles $D_s$ and $D^*_s$. As a
result, $f_L\approx 1/2$ can be accommodated in our work with a
reasonable cutoff. Third, we have shown large cancellations
occurring in the processes $\overline B\to D_s^* D\to\phi\ov K^*$
and $\overline B\to D_s D^*\to\phi\ov K^*$ and this can be
understood as a consequence of \CP and SU(3) [CPS] symmetry. We
are not able to check the aforementioned cancellation in the CDP
work as no details are offered to the individual LD amplitudes.
Anyway, $f_\bot$ is very small (of order 3\%) in our case, while
CDP obtained $f_\bot\approx 0.15$. Fourth, CDP adapt a different
phase convention for pseudoscalar mesons. Considerable efforts
have been made in keeping the consistency of phase conventions and
reminders on phase subtleties (see, for example, footnote 2 on
p.11) are given in this work. On the other hand, as far as the
analytic expressions are concerned we do not find any
inconsistency in the phase convention chosen in the CDP work.
Fifth, we apply the $U(3)$ symmetry for the $\phi$ meson, while
CDP seem to consider only the $\omega_8$ component of it. Since in
both approaches $\phi$ always couples to a $D_s^{(*)} D_s^{(*)}$
pair (see Fig.~\ref{fig:BphiKst} in this work and Fig.~2 in
\cite{Colangelo04}), only the $s\bar s$ component of $\phi$ is
relevant. Hence, the FSI amplitudes in CDP should be identical to
ours up to an overall factor of the Clebsch-Gordan coefficient,
$-2/\sqrt6$. We do not expect any qualitative difference arising
from the different treatments of the $\phi$ wave function. In
particular, the cancellation argument presented in Sec.~VII.C
should be also applicable to their case.

Working in the context of QCD factorization, Kagan \cite{Kagan1}
has recently argued that the lower value of the longitudinal
polarization fraction can be accommodated in the Standard Model
provided that the penguin-induced annihilation contributions are
taken into account. In general, the annihilation amplitudes
induced by the $(V-A)(V-A)$ operators are subject to helicity
suppression. Indeed, they are formally power suppressed by order
$(\Lambda_{\rm QCD}/m_b)^2$ \cite{BBNS}. However, the $(S-P)(S+P)$
penguin operator induced annihilation is no longer subject to
helicity suppression and could be substantially enhanced. In QCD
factorization, the penguin-induced annihilation amplitude contains
some infrared divergent integral, say $X_A$, arising from the
logarithmic endpoint divergence, and the logarithmic divergence
squared $X_A^2$. Modelling the logarithmic divergence by
\cite{BBNS}
 \be
 X_A=(1+\rho_A e^{i\phi_A})\ln {m_B\over \Lambda_h},
 \en
with $\Lambda_h$ being a hadron scale of order 500 MeV and
$\rho_A\leq 1$, Kagan has shown that the $\phi K^*$ polarization
measurements can be accommodated by fitting to the annihilation
parameter $\rho_A$ (see Fig. 4 of \cite{Kagan1}). Note that the
penguin-induced annihilation terms being formally of order
$1/m_b^2$ do respect the scaling law stated in Eq.
(\ref{eq:scaling}), but they can be ${\cal O}(1)$ numerically.

Assuming Kagan's reasoning \cite{Kagan2} is correct, it means that
although QCD factorization does not predict a lower value of the
$\phi K^*$ longitudinal polarization, one may be able to find a
value for $\rho_A$ within an acceptable range to be able to
accommodate the data.

It is worth mentioning also that the annihilation amplitudes in
$VV$ modes are calculable in the so-called perturbative QCD
approach for hadronic $B$ decays where the end-point divergence is
regulated by the parton's transverse momentum. In particular, the
$\ov B\to\phi\bar K^*$ decay is studied in \cite{Chen:2002pz},
giving $f_L:f_\parallel:f_\bot= 0.75:0.13:0.11$ and ${\mathcal
B}(B^-\to\phi K^{*-})=16.0\times10^{-6}$. Although $f_L$ is
reduced after the inclusion of nonfactorizable and annihilation
contributions, $f_{\parallel,\bot}$ are only of order $10\%$. The
annihilation contribution indeed helps but it is not sufficient.

A common feature of the final-state rescattering or a large
penguin-induced annihilation contribution as a mechanism for
explaining the observed $\phi K^*$ anomaly is that the same
mechanism will also lead to a large transverse polarization in the
$\rho K^*$ modes, which is borne out by experiment in the decay
$B^-\to\rho^- \ov K^{*0}$ but not in $B^-\to\rho^0K^{*-}$ (see
Table \ref{tab:BphiKV}). This has to be clarified experimentally.

An alternative suggestion for the solution of the $\phi K^*$
anomaly was advocated in \cite{Hou04} that a energetic transverse
gluon from the $b\to s g$ chromodipole operator keeps most of its
quantum numbers except color when it somehow penetrates through
the $B$ meson surface and descends to a transversely polarized
$\phi$ meson. Sizable transverse components of the $\ov B\to
\phi\bar K^*$ decay can be accommodated by having
$f_\parallel>f_\perp$ (see also Table \ref{tab:phiKst}). Since the
gluon is a flavor singlet, this mechanism can distinguish $\phi$
from $\rho$, hence it affects $\ov B\to\phi \ov K^{*},\omega \ov
K^{*}$ but not $\ov B\to \ov K^{*}\rho$. The $\ov B\to\omega\ov
K^*$ is predicted to be of order $4\times 10^{-6}$ with large
transverse components. On the other hand, a recent estimation
based on PQCD~\cite{Li04} seems to indicate the above-mentioned
contribution is too small to explain the $\phi \ov K^*$ anomaly.
Moreover, the predicted absence of the transverse polarization in
$B\to\rho K^*$ is not consistent with experiment at least for
$\rho^-\ov K^{*0}$ and $\rho^+K^{*-}$ modes.

\section{Discussion and conclusion}
In this work we have studied the effects of final-state
interactions on the hadronic $B$ decay rates and their impact on
direct \CP violation. Such effects are modelled as soft
final-state rescattering of some leading intermediate two-body
channels and can be classified according to the topological quark
diagrams. It amounts to considering the one-particle-exchange
processes at the hadron level for long-distance rescattering
effects. Our main results are as follows:

\begin{enumerate}

\item The color-suppressed neutral modes such as $B^0\to
D^0\pi^0,\pi^0\pi^0,\rho^0\pi^0,K^0\pi^0$ can be substantially
enhanced by long-distance rescattering effects, whereas the
color-allowed modes are not significantly affected by FSIs.

\item All measured color-suppressed charmful decays of $\ov B^0$
into $D^{(*)0}\pi^0,D^0\eta,D^0\omega$ and $D^0\rho^0$ are
significantly larger than theoretical expectations based on naive
factorization. The rescattering from $B\to\{ D\pi,D^*\rho\}\to
D\pi$ contributes to the color-suppressed $W$-emission and
$W$-exchange topologies and accounts for the observed enhancement
of the $D^0\pi^0$ mode without arbitrarily assigning the ratio of
$a_2/a_1$ a large magnitude and strong phase as done in many
previous works.

\item The branching ratios of all penguin-dominated $B\to \pi K$
decays are enhanced via final-state rescattering effects by
(30-40)\%. \CP asymmetry in the $K^- \pi^+$ mode is predicted by
the short-distance approach to be $\sim 4\%$. It is modified by
final-state rescattering to the level of $14\%$ with a sign flip,
in good agreement with the world average of $-0.11\pm0.02$. Given
the establishment of direct \CP violation in $\ov B^0\to
K^-\pi^+$, a model-independent relation assuming SU(3) implies
that the $\pi^+\pi^-$ mode has direct \CP asymmetry of order 40\%
with a positive sign. An isospin sum-rule relation involving the
branching ratios and \CP asymmetries of the four $B\to K\pi$ modes
indicates the presence of electroweak contributions and/or some
New Physics. Better measurements are needed to put this on a
firmer footing.

\item If only the absorptive part of final-state rescattering is
considered, it cannot explain the observed enhancement of the
$\pi^0\pi^0$ rate, whereas the predicted $\pi^+\pi^-$ will be too
large by a factor of 2 compared to experiment. The dispersive part
of rescattering from $D\bar D\to \pi\pi$ and $\pi\pi\to\pi\pi$ via
meson annihilation (i.e. the so-called vertical $W$-loop diagram
$\V$) which interferes destructively with the short-distance
amplitude of $B^0\to\pi^+\pi^-$ can reduce $\pi^+\pi^-$ and
enhance $\pi^0\pi^0$ substantially.

\item For tree-dominated $B\to\pi\pi$ decays, it is known that the
predicted direct \CP asymmetry in the $\pi^+\pi^-$ mode by the
short-distance approach is small with a negative sign. We have
shown that its sign is flipped by final-state rescattering and its
magnitude is substantially enhanced. Direct \CP violation in
$B\to\pi^0\pi^0$ is predicted to have a sign opposite to that of
$\pi^+\pi^-$, in contrast to the predictions based on perturbative
QCD (PQCD). Hence, even a sign measurement of direct \CP asymmetry
in $\pi^0\pi^0$ can be used to discriminate between the FSI and
PQCD approaches for \CP violation.

\item \CP partial rate asymmetry in $B^\pm \to \pi^\pm \pi^0$ is
very small in the SM and remains so even after the inclusion of
FSIs, with a magnitude less than 1 percent. Since \CP asymmetry is
caused by the electroweak penguin but not the QCD penguin (up to
isospin violation), this is a good mode for searching New Physics.

\item For tree-dominated $B\to \rho \pi$ decays, we showed that
(i) the color-suppressed $\rho^0 \pi^0$ mode is slightly enhanced
by rescattering effects to the order of $1.3\times 10^{-6}$, which
is consistent with the weighted average of the experimental
values. However, the discrepancy between BaBar and Belle for this
mode should be clarified soon. It should be stressed that direct
\CP violation in this mode is significantly enhanced by FSI from
around 1\% to 60\%. (ii) Direct \CP violation in the $\rho^+
\pi^-$ mode is greatly enhanced by FSI from the naive expectation
of $\sim -0.01$ from the short-distance approach to the level of
$-0.42$, in agreement with BaBar and Belle.

\item As for the intriguing $B\to\phi K^*$ polarization anomaly,
the longitudinal polarization fraction can be significantly
reduced by the rescattering contribution from the intermediate
$DD_s$ states. However, no sizable perpendicular polarization is
found owing mainly to the large cancellations among various
contributions from intermediate $D_s D^*$ and $D^*_s D$ states.
Consequently, our result for the perpendicular polarization
fraction is different from a recent similar analysis in
\cite{Colangelo04}. Final-state rescattering effects from this
particular set of states seem not be able to fully account for the
polarization anomaly observed in $B\to \phi K^*$ and FSI from
other intermediate states and/or some other mechanism e.g. the
penguin-induced annihilation~\cite{Kagan2}, may have to be
invoked. Given the fact that both Belle and BaBar observe large
phases in various polarization amplitudes (see Table
\ref{tab:BphiKV}), FSI may still provide a plausible explanation.
In any case, our conclusion is that the small value of the
longitudinal polarization in $VV$ modes cannot be regarded as a
clean signal for New Physics.

\end{enumerate}

Needless to say, the calculation of final-state rescattering
effects in hadronic $B$ decays is rather complicated and very much
involved and hence it suffers from several possible theoretical
uncertainties. Though most of them have been discussed before, it
is useful to make a summary here:

(i) The strong form factor $F(t)$ for the off-shell effects of the
exchanged particle. It is parameterized as in Eq. (\ref{eq:FF}) by
introducing a cutoff scale $\Lambda$. Moreover, we write
$\Lambda=m_{\rm exc}+\eta\Lambda_{\rm QCD}$ [cf. Eq.
(\ref{eq:Lambda})], where the parameter $\eta$ is expected to be
of order unity and can be determined from the measured rates. For
a given exchanged particle, the cutoff varies from process to
process. For example, $\eta_D=2.1,~0.69,~1.6$ for $B\to
D\pi,K\pi,\rho\pi$ decays, respectively. To see the sensitivity on
the cutoff, we have allowed 15\% error in the QCD scale
$\Lambda_{\rm QCD}$. Another important uncertainty arises from the
momentum dependence of the form factor $F(t)$. Normally it is
assumed to be of the monopole form. However, the analysis of the
$\phi K^*$ decays prefers to a dipole behavior for the form factor
$F(t,m_{D_s^*})$ appearing in the $V_{\rm exc}VP$ or $V_{\rm
exc}VV$ vertex with the exchanged vector particle $V_{\rm exc}$
and a monopole dependence for other form factors. A more rigorous
study of the momentum dependence of the strong form factor is
needed. It turns out that \CP asymmetries and the decay rates
especially for the color-suppressed modes are in general sensitive
to the cutoff scale. This constitutes the major theoretical
uncertainties for long-distance rescattering effects.

(ii) Form factors and decay constants. Model predictions for the
form factors can vary as much as 30\% which in turn imply large
uncertainties in the branching ratios. In particular, the $B\to
VV$ decays are fairly sensitive to the {\it difference} between
the form factors $A_1$ and $A_2$. For form factor transitions in
this work we rely on the covariant light-front model which is
favored by the current data.

(iii) Strong couplings of heavy mesons and their SU(3) breaking.
We have applied chiral and heavy quark symmetries to relate
various strong couplings of heavy mesons with the light
pseudoscalar or vector mesons. It is not clear how important are
the chiral and $1/m_Q$ corrections. For SU(3) breaking effects, we
have assumed that they are taken into account in the relations
given in Eqs. (\ref{eq:SU3breaking}) and (\ref{eq:gSU3}).

(iv) The real part of the long-distance contribution which can be
obtained from the dispersion relation (\ref{eq:dispersive}).
Unlike the absorptive part, the dispersive contribution suffers
from the large uncertainties due to some possible subtractions and
the complication from integrations. For this reason, we have
ignored it so far. However, in order to resolve the discrepancy
between theory and experiment for $B^0\to\pi^+\pi^-$ and
$B^0\to\pi^0\pi^0$, we have argued that it is the dispersive part
of long-distance rescattering of $D\bar D\to \pi\pi$ and
$\pi\pi\to\pi\pi$ via meson annihilation that accounts for the
suppression of the $\pi^+\pi^-$ mode and the enhancement of
$\pi^0\pi^0$.

(v) CKM matrix elements and $\gamma$. Direct \CP asymmetry is
proportional to $\sin\gamma$. All numbers in the present work are
generated by using $\gamma=60^\circ$.

(vi) Intermediate multi-body contributions which have not been
considered thus far and may have cancellations with the
contributions from two-body channels.

\vspace{0.35cm} Finally, it is worth remarking that the
final-state rescattering effects from intermediate charmed mesons
as elaborated in the present work has some similarity to the
long-distance ``charming penguin" effects advocated in the
literature \cite{charmpenguin,Isola2003,charmpenguin1}. The
relevance of this long-distance effect has been conjectured to be
justified in the so-called soft collinear effective theory
\cite{Bauer}. Indeed, we have pointed out that some of the charmed
meson loop diagrams in the decays e.g. $B\to \pi\pi, \pi K,\phi
K^*$ will manifest as the long-distance $c\bar c$ penguins.
However, we have also considered FSIs free of charming penguins.
For example, long-distance rescattering effects in $B\to D\pi$ and
in $B^-\to\pi^-\pi^0$ (also $\rho^-\rho^0$) decays have nothing to
do with the charming penguin effects. In other words, our
systematic approach for FSIs goes beyond the long-distance
charming penguin mechanism.

\vskip 2.0cm \acknowledgments We are grateful to Andrei V. Gritsan
for helpful discussions on $B$ to $VV$ data. This research was
supported in part by the National Science Council of R.O.C. under
Grant Nos. NSC93-2112-M-001-043, NSC93-2112-M-001-053 and by the
U.S. DOE contract No. DE-AC02-98CH10886(BNL).

\newpage
\appendix
\section{Useful formula}

Under the integration, the covariant integrals
 \be
 \int {d^3\vec p_1\over (2\pi)^32E_1}\,{d^3 \vec p_2\over
(2\pi)^3
2E_2}\,(2\pi)^4\delta^4(p_B-p_1-p_2)f(t)\times\{p_{1\mu},\,p_{1\mu}
p_{1\nu},\,p_{1\mu} p_{1\nu} p_{1\alpha}\}
 \en
can only be expressed by external momenta $p_3, p_4$ with suitable
Lorentz and permutation structures. Hence we can express these
$p_{1\mu},\,p_{1\mu} p_{1\nu},\,p_{1\mu} p_{1\nu} p_{1\alpha}$ in
terms for $P=p_3+p_4$ and $q=p_3-p_4$ as in Eq.~(\ref{eq:p1})
within the integration. By contracting the left-hand side of
Eq.~(\ref{eq:p1}) with $P_{\mu}$, $q_{\mu}$ and $g_{\mu\nu}$, we
are able to solve for these $A^{(i)}_j$ and obtain
 \be
 \left(
 \begin{array}{c}
 A^{(1)}_1
 \\
 A^{(1)}_2
 \end{array}
 \right)
 &=&
 \left(
 \begin{array}{cc}
 P^2         &P\cdot q
 \\
 P\cdot q    &q^2
 \end{array}
 \right)^{-1}
 \cdot
 \left(
 \begin{array}{c}
 P\cdot p_1
 \\
 q\cdot p_1
 \end{array}
 \right),
 \non\\
 \left(
 \begin{array}{c}
 A^{(2)}_1
 \\
 A^{(2)}_2
 \\
 A^{(2)}_3
 \\
 A^{(2)}_4
 \end{array}
 \right)
 &=&
 \left(
 \begin{array}{cccc}
 4         &P^2          &2P\cdot q       &q^2
 \\
 P^2       &(P^2)^2      &2 P^2 P\cdot q  &(P\cdot q)^2
 \\
 2P\cdot q &2 P^2 P\cdot q  &2P^2 q^2+(P\cdot q)^2 &2P\cdot q q^2
 \\
 q^2       &(P\cdot q)^2  &2P\cdot q q^2    &q^2
 \end{array}
 \right)^{-1}
 \cdot
 \left(
 \begin{array}{c}
 p_1^2
 \\
 (P\cdot p_1)^2
 \\
 2P\cdot p_1 p_1\cdot q
 \\
 (q\cdot p_1)^2
 \end{array}
 \right),
 \non\\
 \left(
 \begin{array}{c}
 A^{(3)}_1
 \\
 A^{(3)}_2
 \\
 A^{(3)}_3
 \\
 A^{(3)}_4
 \\
 A^{(3)}_5
 \\
 A^{(3)}_6
 \end{array}
 \right)
 &=&
 \left(
 {\footnotesize
 \begin{array}{cccccc}
 18 P^2
     &18 P\cdot q
     &3(P^2)^2
     &9 P^2 P\cdot q
     &3q^2 P^2+6(P\cdot q)^2
     &3q^2 P\cdot q
 \\
     &18q^2
     &3 P^2 P\cdot q
     &3P^2 q^2+6(P\cdot q)^2
     &9q^2 P\cdot q
     &3(q^2)^2
 \\
     &
     &(P^2)^3
     &3(P^2)^2P\cdot q
     &3(P\cdot q)^2 P^2
     &(P\cdot q)^3
 \\
     &
     &
     &3(P^2)^2q^2+6P^2(P\cdot q)^2
     &3(P\cdot q)^3+6P^2 q^2 P\cdot q
     &3(P\cdot q)^2 q^2
     \\
     &%
     &%
     &%
     &3(q^2)^2P^2+6q^2(P\cdot q)^2
     &3(q^2)^2P\cdot q
     \\
     &
     &
     &
     &
     &(q^2)^3
     \\
 \end{array}
 }
 \right)^{-1}
 \non\\
 &&\cdot
 \left(
 \begin{array}{c}
 3m_1^2 P\cdot p_1
 \\
 3 m_1^2 q\cdot p_1
 \\
 (P\cdot p_1)^3
 \\
 3(P\cdot p_1)^2 p_1\cdot q
 \\
 3(q\cdot p_1)^2 p_1\cdot P
 \\
 (q\cdot p_1)^3
 \end{array}
 \right),
 \en
where the lower left part of the symmetric inverse matrix for the
$A^{(3)}_j$ case is not shown explicitly. Note that the $t$
dependence of $A^{(i)}_j$ is in the column matrices in the
right-hand side of the above equations and $t$ always appears in
the numerators of $A^{(i)}_j$ in a polynomial form.

\section{Theoretical input parameters}
In this Appendix we summarize the input parameters used in the
present paper. For decay constants we use
 \be
 f_\pi=132\,{\rm MeV}, \quad f_K=160\,{\rm MeV}, \quad
 f_D=200\,{\rm MeV}, \quad f_{D^*}=230\,{\rm MeV}, \quad
 f_{a_1}=-205\,{\rm MeV}.
 \en
Note that a preliminary CLEO measurement of the semileptonic decay
$D^+\to\mu^+\nu$ yields $f_{D^+}=(202\pm41\pm17)$ MeV
\cite{CLEOfD}. For form factors we follow the covariant
light-front approach \cite{CCH}. For purpose of comparison, we
list some of the form factors at $q^2=0$ used in the main text:
 \be
 F_0^{B\pi}(0)=0.25, \quad F_0^{BK}(0)=0.35, \quad
 A_0^{B\rho}(0)=0.28,\quad A_0^{BK^*}(0)=0.31
 \en
to be compared with
 \be
 F_0^{B\pi}(0)=0.28\pm0.05, \quad F_0^{BK}(0)=0.34\pm0.05, \quad
 A_0^{B\rho}(0)=0.37\pm0.06,\quad A_0^{BK^*}(0)=0.45\pm0.07 \non
 \\
 \en
employed in \cite{BN} and
 \be
 F_0^{B\pi}(0)=0.33, \quad F_0^{BK}(0)=0.38, \quad
 A_0^{B\rho}(0)=0.28,\quad A_0^{BK^*}(0)=0.32
 \en
used by Bauer, Stech and Wirbel \cite{BSW}. The most recent
light-cone sum rule analysis yields \cite{Ball}
 \be
 F_0^{B\pi}(0)=0.258\pm0.031, \qquad
 F_0^{BK}(0)=0.331+0.041+0.025\delta_{a_1}
 \en
with $\delta_{a_1}$ being defined in \cite{Ball}. The
$B^-\to\pi^-\pi^0$ data favors a smaller $F^{B\pi}(0)$ of order
0.25, while $B\to\rho\pi$ measurement prefers to small form
factors for $B\to\rho$ transition.

For the quark mixing matrix elements, we use $A=0.801$ and
$\lambda=0.2265$ \cite{CKMfitter} in the Wolfenstein
parametrization of the quark mixing angles \cite{Wolfenstein}. The
other two Wolfenstein parameters $\rho$ and $\eta$ obey the
relations $\bar\eta=R\sin\gamma$ and $\bar\rho=R\cos\gamma$, where
$R=|V_{ud}V^*_{ub}/(V_{cd}V^*_{cb})|$,
$\bar\rho=(1-\lambda^2/2)\rho$ and $\bar\eta=(1-\lambda^2/2)\eta$.
We take $\gamma=60^\circ$ and $R=0.39$ in this work for
calculations. For current quark masses, we use $m_b(m_b)=4.4$ GeV,
$m_c(m_b)=1.3$ GeV, $m_s(2.1\,{\rm GeV})=90$ MeV and
$m_q/m_s=0.044$.

The physical strong coupling constants are
 \be
g_{\rho\pi\pi} &=& 6.05\pm0.02, \qquad g_{K^*K\pi}=4.6, \qquad
g_{D^*D\pi} =17.9\pm0.3\pm1.9, \non \\
g_V &=& 5.8, \qquad \beta=0.9, \qquad \lambda=0.56\,{\rm
GeV}^{-1},
 \en
where  $g_V$, $\beta$ and $\lambda$ (not to be confused with the
parameter $\lambda$ appearing in the parametrization of quark
mixing angles) are the parameters in the effective chiral
Lagrangian describing the interactions of heavy mesons with low
momentum light vector mesons.


\end{document}